\documentclass[12pt]{article}

\usepackage[vcentermath]{youngtab}
\usepackage{color}

\def\be{\begin{equation}}
\def\ee{\end{equation}}
\def\ba{\begin{eqnarray}}
\def\ea{\end{eqnarray}}

\def\ps{\langle}
\def\pd{\rangle}
\def\de{\right}
\def\si{\left}
\def\ap{\a {\rm '}}

\def\qb{{\bar q}}
\def\p{\partial}
\def\nb{\nonumber}

\def\beq{\begin{equation}}
\def\eeq{\end{equation}}
\def\bea{\begin{eqnarray}}
\def\eea{\end{eqnarray}}

\def\e{\epsilon}
\def\h{\eta}

\def\z{\zeta}
\def\m{\mu}
\def\n{\nu}
\def\r{\rho}
\def\s{\sigma}

\def\t{\tau}
\def\f{\varphi}
\def\a{\alpha}
\def\b{\beta}
\def\d{\delta}
\def\l{\lambda}
\def\g{\gamma}

\def\D{\Delta}
\def\w{\omega}

\usepackage[centertags]{amsmath}
\usepackage{amsfonts} 
\usepackage{amssymb}
\usepackage{amsthm}
\numberwithin{equation}{section} 
\def\ii{{\rm i}}
\newcommand{\ex}[1]{{\rm e}^{#1}} 

\begin{document}
\begin{titlepage}
\hfill \hbox{CERN-PH-TH/2013-227}
\vskip 0.1cm
\hfill \hbox{NORDITA-2013-81}
\vskip 0.1cm
\hfill \hbox{QMUL-PH-13-10}
\vskip 1.0cm
\begin{flushright}
\end{flushright}
\vskip 1.0cm
\begin{center}
{\Large \bf  Microscopic unitary description of tidal excitations in high-energy string-brane collisions}
 
  \vskip 1.0cm {\large Giuseppe
D'Appollonio$^{a}$, Paolo Di Vecchia$^{b, c}$,
Rodolfo Russo$^{d, e}$, \\
Gabriele Veneziano$^{f, g}$ } \\[0.7cm]
{\it $^a$ Dipartimento di Fisica, Universit\`a di Cagliari and
INFN\\ Cittadella
Universitaria, 09042 Monserrato, Italy}\\
{\it $^b$ The Niels Bohr Institute, University of Copenhagen, Blegdamsvej 17, \\
DK-2100 Copenhagen, Denmark}\\
{\it $^c$ Nordita, KTH Royal Institute of Technology and Stockholm University, \\Roslagstullsbacken 23, SE-10691 Stockholm, Sweden}\\
{\it $^d$ Queen Mary University of London, Mile End Road, E1 4NS London, United Kingdom}\\
{\it $^e$ Laboratoire de Physique Th\'eorique de L'Ecole Normale Sup\'erieure\\
24 rue Lhomond, 75231 Paris cedex, France}\\
{\it $^f$ Coll\`ege de France, 11 place M. Berthelot, 75005 Paris, France}\\
{\it $^g$Theory Division, CERN, CH-1211 Geneva 23, Switzerland}
\end{center}
\begin{abstract}
The eikonal operator was originally introduced  to describe the effect of tidal excitations on higher-genus elastic 
string amplitudes at high energy. In this paper we provide a precise interpretation for this operator through the 
explicit tree-level calculation of  generic inelastic transitions between closed strings as they scatter  
off a stack of parallel D$p$-branes. We perform this analysis both in the light-cone gauge, using the Green-Schwarz 
vertex, and in the covariant formalism, using  the Reggeon vertex operator. We also present a detailed 
discussion of the high-energy behaviour of the covariant string amplitudes, showing how to take  into account the energy 
factors that enhance the contribution of the longitudinally polarized massive states in a simple way. 

\end{abstract}

\end{titlepage}

\tableofcontents

\section{Introduction}
\label{intro}

Since about 25 years transplanckian-energy gravitational scattering has been the target of numerous investigations.
The original thrust was focused on the  scattering among point-like particles \cite{'tHooft:1987rb,  Muzinich:1987in} or light strings \cite{Amati:1987wq}
 (see also \cite{Sundborg:1988tb}). One of its main goals was to  understand how an effective curved geometry originates from studying collisions in 
flat $D$-dimensional space-time. A more ambitious aim was to find out whether and  how unitarity of the 
$S$-matrix is preserved in all regimes within a consistent quantum-gravity framework.

Indeed, at sufficiently high energy and at any finite order in perturbation theory, (partial-wave) unitarity bounds are already violated even at  large impact parameters. 
This problem was neatly solved \cite{Amati:1987wq,  Muzinich:1987in} by an all-loop  resummation of the leading high-energy contributions: in the point particle case these exponentiate (in impact parameter space) leading to an elastic-unitarity-preserving eikonal 
 $S$-matrix at leading order in the small parameter $G_D\sqrt{s}b^{3-D}$, where $D$ is the number of non-compact space-time dimensions and $G_D$ is the Newton constant in $D$ dimensions
\begin{equation} \label{d1}
 S(E,b) = 1 + 2 \ii \delta(s,b) + \dots \rightarrow \exp\left( 2 \ii \delta(s,b)  \right) \ ,   \hspace{1cm} \delta(s,b) \sim \hbar^{-1} G_D s b^{4-D} \ .
\end{equation}
This small parameter controls the typical gravitational deflection angle, $\theta \sim (R_s/b)^{D-3}$~\footnote{Here $R_s^{D-3} = \frac{8 \pi \Gamma(\frac{D-1}{2})G_D \sqrt{s} }{(D-2) \pi^{\frac{D-1}{2}} }$ is  the Schwarzschild radius  for a mass $\sqrt{s}$ in $D$ non-compact dimensions.}.  In the case of string-string collisions the above $c$-number factorization and exponentiation 
fail below a certain impact parameter $b_D \gg  l_s \equiv \sqrt{2 \alpha'  \hbar}$
\footnote{We use units in which $c=1$ and, in the following sections, we shall  also set $\hbar =1$ thus identifying $l^2_s$ with $2 \alpha'$.}. They can only be recovered  \cite{Amati:1987wq, Amati:1987uf}  at the price of promoting the conventional eikonal phase   to  an  eikonal operator
\begin{equation}
  \label{deltahat}
\delta(s,b) \rightarrow \hat{\delta} (s,b) \sim  \int_0^{2 \pi}  \int_0^{2 \pi}\frac{d\sigma_u}{2 \pi} \frac{d\sigma_d}{2 \pi}: \delta(s, b + \hat{X}_u(\sigma_u,  \tau=0) - \hat{X}_d(\sigma_d, \tau=0)) : \ ,
\end{equation}
where $\hat{X}_u(\s, \t)$ and $\hat{X}_d(\s, \t)$ are independent free bosonic fields in two dimensions containing only oscillation modes.  
The intuitive meaning of this operator  \cite{Amati:1987wq, Amati:1987uf} is that the gravitons responsible for the transplanckian scattering 
are exchanged between two arbitrary points on the two colliding strings at a Lorentz contracted instant $\tau =0$. 
More physically \cite{Giddings:2006vu}, the eikonal operator should  ensure inelastic unitarity in the regime in which tidal forces induce substantial excitations of the incoming strings. However, in order to fully understand how unitarity works, and the precise microscopic nature of the transitions induced by such tidal forces, it would be necessary to consider, at tree level, the individual inelastic transitions whose shadow is collectively taken into account by the elastic loop amplitudes. This issue will be one of the main objectives of this paper, although we will discuss it
explicitly not in the context of string-string collisions but in the slightly different context of 
string-brane collisions that we shall  illustrate in a moment.

Understanding how unitarity is preserved at arbitrary impact parameter in string-string collisions has proven to  be a much more difficult task. Nice progress was made \cite{Amati:1987uf, Amati:1988tn, Veneziano:2004er}  in the so-called stringy regime ($l_s  > R_S , b$)  in which  
one does not expect black-hole formation to occur. In the opposite regime ($R_s  > l_s, b$), 
in which black-hole formation is expected to occur on the basis of classical collapse criteria 
\cite{Penrose1974U,Eardley:2002re,Kohlprath:2002yh,Giddings:2004xy}, 
only a crude approximation was attempted \cite{Amati:2007ak}. While this approximation could reproduce semi-quantitatively the expected critical points for gravitational collapse \cite{Amati:2007ak, Marchesini:2008yh,  Veneziano:2008zb, Veneziano:2008xa}, it has failed,  so far  \cite{Ciafaloni:2008dg, Ciafaloni:2011de}, to explain how unitarity is preserved beyond such critical points (i.e. in the supposed collapse regime).

In order to address such questions
in an easier context we have recently turned our attention  \cite{D'Appollonio:2010ae} to the collision of a light closed string off a stack of $N$ parallel D$p$-branes (at large $N$ and small string coupling), where the effective metric, rather than being produced by the collision itself, should be, up to possible corrections, the known  classical one generated by the branes. When the volume of the branes 
is compactified on a $p$-dimensional torus, one obtains a point-like $1/2$-BPS object in $9-p$ non-compact spatial dimensions
with a mass proportional to the D$p$-brane tension and thus  very large at weak string coupling.
The study of the high-energy scattering of a fundamental string on this kind of target,
although it does not represent a standard black hole with a macroscopic horizon, is non trivial 
and the construction of an explicitly unitary string $S$-matrix 
very interesting. The $S$-matrix
does in fact allow to test the regime of validity of the classical gravity solution as an effective description of the D$p$-branes
and to explore the small $b$ regime in which the energy of the incoming string should dissipate through the creation of many open string excitations on the branes.

With these motivations in mind, 
in~\cite{D'Appollonio:2010ae} we have computed the high-energy (Regge) limit  of the elastic scattering of a massless closed string state, belonging to the NS-NS sector of type II string theory, on a maximally supersymmetric D$p$-brane system. It turns out that also in this case  the 
tree-level amplitude, the disk diagram, diverges with the energy of the incoming string violating unitarity at sufficiently high energy. 
Neglecting string-size effects unitarity is again recovered by summing the  contributions to the elastic amplitude coming from   
surfaces with any number of boundaries. The sum of these terms exponentiates into a phase, 
as in the previous case of a string-string collision, 
 \begin{equation}
 S(E,b) = 1 + 2 \ii \delta(s,b) + \dots \rightarrow \exp\left( 2 \ii \delta(s,b)  \right) \  , 
\hspace{1cm}  \delta(s,b) \sim \frac{E~b}{\hbar} ~ \left(  \frac{R_p}{b}\right) ^{7-p} \ , 
\end{equation} 
where $R_p^{7-p} \sim
g N l_s^{7-p}$ is the characteristic scale of the geometry produced by the branes. 

Taking into account string size effects, the eikonal phase becomes once more an  eikonal operator 
which  contains the bosonic oscillators of the superstring corresponding to the $8-p$ directions transverse to the world-volume of the D$p$-brane and to the momentum of the fast-moving string
  \be
  \label{deltahat2} S(s, b) = {\rm e}^{2 \ii \hat{\delta} (s,b)} \ , \hspace{1cm}
\hat{\delta} (s,b) =  \int_0^{2 \pi} \frac{d\sigma}{2 \pi}: \delta(s, b + \hat{X}(\sigma, \tau=0) ) : \ . 
\ee
All calculations are done in flat Minkowski space-time  with suitable conditions  imposed on the closed strings by  the presence of the D$p$-branes, but the final results show the curved space structure generated by the presence of the D$p$-branes. In particular, in the zero-slope
limit $\alpha' \rightarrow 0$ 
one reproduces  the eikonal computed with a  curved space formalism in field theory.

Present  derivations of the eikonal operator are somewhat indirect in that the existence and nature of such an operator is argued on the basis of the imaginary part of a higher-loop elastic amplitude which, in principle, only provides some inclusive sum over intermediate excited string states, rather than a precise microscopic description of each produced string. 
In previous work the eikonal operator was indeed mainly used to study the absorption of the elastic channel due to the excitation of the massive
modes of the string and to identify the average excitation mass \cite{Amati:1987uf}. To this aim all that is required
is the algebra satisfied by the modes of the bosonic fields $\hat{X}$. 

The information on the high-energy string dynamics encoded in the eikonal operator is by far more detailed. Its
matrix elements give in fact the asymptotic behaviour at high energy of the transition amplitudes between four
(in the case of a string-string collision) or two (in the case of  a string-brane collision) arbitrary string states,
resumming to all orders the perturbative series. 
For the evaluation of these matrix elements it is necessary to give a
precise definition of the eikonal operator by specifying the Hilbert
space on which it acts, an issue that has never been clarified in the
previous literature on the subject. 

Although the eikonal operator is the result of an all-loop resummation, to complete its definition
it is enough to consider string amplitudes at tree level. 
To identify the correct Hilbert space it is in fact sufficient to study the eikonal phase, 
the  operator $\hat \d(s, b)$, which gives the  asymptotic behaviour at high energy of the
tree-level two-point (four-point) amplitudes between arbitrary string states. 
In this paper we will present two independent derivations of the eikonal phase, one in the light-cone gauge 
and the other fully covariant, thus
giving a precise meaning to the operators $\hat{X}$ in Eqs.~\eqref{deltahat}
and~\eqref{deltahat2}. 

We will first 
derive $\hat \d(s, b)$ by quantizing the string in a light-cone gauge adapted to the kinematics of the high-energy scattering, 
that is with the spatial direction of the light-cone chosen along the direction of the large
momentum.  This derivation is based on the Green-Schwarz three-string vertex
and shows that  the free fields $\hat{X}$ can be identified with the transverse
string coordinates in the light-cone gauge.

That the eikonal operator can be interpreted as acting on the space of the physical states of the
string quantized in a specific light-cone gauge is of course not unexpected. It is quite natural given the kinematics
of the high-energy scattering in the Regge limit, which is characterized by the presence
of one privileged direction of large momentum, and it is also clearly suggested 
by the original papers~\cite{Amati:1987wq,Amati:1987uf}. Moreover both
the eikonal operator for string-string collisions and for string-brane collisions
can be derived (at large impact parameters and 
to first order in the string corrections, i.e. in $l_s^2/b^2$) by quantizing  in the light-cone gauge
the string sigma-model for an effective curved background. 
In the first case, \cite{Veneziano:1988aj}, \cite{deVega:1988ts,Giddings:2007bw}
the relevant background
is  the Aichelburg-Sexl metric \cite {Aichelburg:1970dh}, the 
shock-wave generated in first approximation by one of the two colliding strings,
in the second case~\cite{D'Appollonio:2010ae} it is the  Penrose limit of the extremal $p$-brane solutions
of Type II supergravity.

Once the Hilbert space has been identified, one can proceed to consider the implications of the eikonal operator
for the high-energy string dynamics. 
The simple way in which the dependence on the string coordinates $\hat{X}$ enters in $\hat \d(s, b)$, as a shift of the impact
parameter, and the absence of the worldsheet fermions $\psi$ lead to interesting selection rules for the possible transitions. 
The class of states
that can be reached from any given initial state in a high-energy collision can be readily identified. 
The form of the inelastic amplitudes, which can only involve the external polarizations and the momentum transferred,
is also very constrained. Since the matrix elements are evaluated in the light-cone gauge, 
the initial and final state of a transition as well as the tensors that appear in the amplitudes derived from the eikonal operator
are characterized only by their transformation properties with respect to the transverse $SO(8)$ group. 

The natural question  then arises about what is the covariant dynamics responsible for the simple 
properties of the eikonal operator in the light-cone gauge. This represents
 the other main topic of our paper. To answer this question one needs to recall
that in the Regge limit the string amplitudes are dominated by the exchange in the $t$-channel of the states of the leading
Regge trajectory, which carry,  for a given mass, the highest spin. The effect of the exchange
of the whole leading
Regge trajectory can be summarized by the exchange of a single effective
string state, the Reggeon\footnote{We refer to this effective string state and to the corresponding vertex operator as the Reggeon,
since it describes the high-energy dynamics in the Regge limit \cite{Ademollo:1989ag, Ademollo:1990sd}.  It 
is also called the Pomeron, especially in the context of the string/gauge duality \cite{Brower:2006ea}.}
\cite{Ademollo:1989ag, Ademollo:1990sd, Brower:2006ea}. The covariant dynamics captured by the eikonal
operator is precisely the exponentiation of the Reggeon exchange at tree level.

The Reggeon vertex operator considerably simplifies the derivation of the Regge limit of the covariant string amplitudes
and gives their  high-energy behaviour directly in a neat and factorized form. For instance a four-point function reduces to the product of the
Reggeon propagator and the
three-point couplings of the external states to the Reggeon. Similarly a two-point
function in the D$p$-brane background reduces to the product of the Reggeon tadpole
and the holomorphic and antiholomorphic
part of the three-point couplings of the external states to the Reggeon.

This very specific factorization in the $t$-channel, which isolates a single coupling to a process-independent intermediate
state, is the first major simplification that occurs in the covariant dynamics. 
The covariant equivalent of the selection rules and of the high-energy amplitudes
given by the eikonal operator in the light-cone gauge are then to be found in the properties of 
the three-point couplings 
of the covariant states to the Reggeon. 

The form of these couplings is restricted to be a contraction of the polarization tensors with the metric, 
the momentum transfer and the longitudinal polarization vectors of the massive states. The longitudinal polarization vectors  
appear in the asymptotic
behaviour at high energy since their components increase with the energy of the massive state
(as in the well-known problem of unitarity-violating amplitudes in the standard model in the absence of  or for a very heavy   Higgs boson). 
It is of course essential in order to derive the correct high-energy behaviour of a string 
amplitude which involves massive states to take the factors of the energy carried by the longitudinal 
polarizations into proper account. 

We will review the derivation of the Reggeon vertex and present a detailed 
discussion of the evaluation of string amplitudes with massive states in the high-energy limit. The
structure of these amplitudes is interesting on its own and may help to clarify the dynamics of the
massive string spectrum, whose typical states transform as traceless irreducible tensors of mixed symmetry,
the generic representations of the Lorentz group.  

The Reggeon vertex will allow us to provide a very simple and fully covariant derivation of the eikonal phase.
To achieve this we will choose a basis of physical states 
 adapted to the kinematics of the high-energy scattering, the basis  of the DDF operators~\cite{DelGiudice:1971fp}.
Although only the $SO(8)$ symmetry group of the
space transverse to the collision axis is manifestly realized in this basis, it has the advantage that
all the physical states can be easily enumerated and their couplings to the Reggeon become elementary.
We will show that when expressed in this basis the tree-level scattering matrix
in the Regge limit can be written in a compact operator form which coincides with 
the   operator $\hat \d(s, b)$. This covariant  derivation of the eikonal
phase thus leads to  the identification of the modes of the 
free fields $\hat{X}$ in  Eqs.~\eqref{deltahat}
and~\eqref{deltahat2} with the bosonic DDF operators. 

The two interpretations of the eikonal operator presented in this paper,
either as an operator written in the light-cone gauge or as a covariant operator
written in terms of the DDF basis, are of course connected since there is a direct correspondence
between the DDF operators of the covariant string and the physical states in the light-cone gauge.
The simple matrix elements of the eikonal phase as an operator in the light-cone gauge are indeed precisely the
simple couplings of the DDF operators to the Reggeon.

It is interesting to understand in more detail the link between the matrix elements of the
eikonal phase and the scattering amplitudes 
written in the basis of the covariant string states. The former contain only tensors 
with well-defined transformation properties with respect  to the
$SO(8)$ symmetry group of the transverse directions while the latter are expressed in terms of ten-dimensional
tensors and of physical polarizations characterized by their transformation properties
with respect to the little group $SO(9)$.  The first step to relate the
covariant amplitudes with the matrix elements of the eikonal operator
is clearly to decompose the covariant tensors with respect to the transverse
$SO(8)$. Each covariant
amplitude thus gives rise to several subamplitudes labeled by
representations of $SO(8)$ and expressed in terms of tensors living in
the space transverse to the collision axis.

The restricted number of linearly independent 
amplitudes allowed at high energy results in the decoupling  of a large number
of covariant states. At every mass level, for a given $SO(8)$ representation one can find the linear combinations of $SO(8)$
components of the covariant states that do not couple to the Reggeon and the  linear combinations which are produced
with a specific form of the amplitude. 
We will show that, once written in this high-energy basis, the covariant amplitudes precisely match the matrix elements of the eikonal phase.

The discussion in this paper will be based on the string-brane system but
the analysis for the case of string-string collisions would be very similar.  
Our main examples will thus be two-point inelastic disk amplitudes involving massive strings. In most cases
we will consider explicitly only transitions from an initial massless state.

A comparison between disk amplitudes  and  matrix elements
of the eikonal operator 
was attempted in two interesting previous papers  \cite{Black:2011ep, Bianchi:2011se} for the states of the leading Regge trajectory. 
The results obtained in these works for the Regge limit of the scattering amplitudes 
are however correct only when the non-vanishing components of the polarization tensors 
are restricted to the transverse directions,
since the energy factors that enhance the contribution of the longitudinal polarizations were not taken into account. 

The rest of the paper is organized as follows. Section \ref{kinsec} is devoted to the kinematics of 
the high-energy scattering of a string on a collection of $N$ D$p$-branes.  In section \ref{gssec} we show how the  operator 
$\hat \d(s, b)$ can be derived  by
 a systematic study of the transition amplitudes at tree level in the light-cone GS formalism, provided the light-cone direction is judiciously chosen.
We also anticipate that the result can be rewritten in a  covariant form using the bosonic DDF operators.
In section \ref{eikonal}  we analyse in detail the structure of the inelastic amplitudes implied by the light-cone eikonal phase,
illustrating the general case with the transitions from the massless NS-NS sector to the first two massive
levels of the superstring.  

In section \ref{covariant}  we turn to a fully covariant derivation of the transition amplitudes in the Regge limit.
We begin with a review of the Reggeon vertex, where we emphasize the importance of taking into account
the longitudinal polarizations when counting the powers of the energy and describe the essential steps 
necessary to evaluate the high-energy limit of the massive string amplitudes. We then explicitly derive 
the Regge limit of the inelastic amplitudes  from the massless NS-NS sector to the first two massive
levels of the superstring. Finally we discuss our covariant derivation of the eikonal phase, which exploits
the simple couplings of the DDF operators to the Reggeon.

We give in this section a complete and detailed description of the transitions to the second massive level for two main
reasons. The first is that it is at this level that the first examples of holomorphic string states transforming
as tensors of mixed symmetry appear. The dynamics of these states, which are generic in the string spectrum,
has not been explored much in the past, although they may provide some useful lessons on the
symmetries of string theory and on the consistent interactions of fields of higher spin (for more details
and references see for instance \cite{Campoleoni:2008jq}).
The second reason is that these amplitudes neatly display all the important features of the relation between the covariant and
the light-cone calculations.

This is the subject of section \ref{covariant-eikonal}, which is devoted to the comparison between 
 the scattering amplitudes with covariant external states and the matrix elements of the eikonal phase.
We shall proceed in both 
directions. To relate the covariant amplitudes with the matrix elements of the eikonal phase it is
sufficient to decompose the covariant tensors with respect to the transverse $SO(8)$ and to perform
a change of basis to a high-energy basis of states, characterized by their couplings to the Reggeon. 
To relate the matrix elements of the eikonal phase with the covariant amplitudes, it is sufficient
to identify the linear combination of covariant states that corresponds to a given light-cone state.
This problem is well-known and is usually addressed by studying the action of the generators of the full Lorentz group
on the light-cone states. We shall address it using the DDF operators which
make the connection between the light-cone and the covariant states somewhat more transparent. 
In section \ref{Conclusions} we summarize our results and draw our main conclusions.

Some additional details can be found in a few appendices.
In Appendix \ref{app:conv} we state our conventions for describing the physical states 
of the RNS superstring in the old covariant quantization.
In Appendix \ref{app:kine} we collect some formulae for  the kinematics of  the high energy string-brane scattering 
 in a convenient reference frame. In Appendix \ref{app:DDF} we review  the DDF operators and discuss their main properties.  
In Appendix \ref{app:Young} we discuss some properties of the polarization tensors of the massive states, in particular
of those of mixed symmetry,  and give
the explicit expressions for the decomposition of the polarizations of 
the covariant states in the second massive level
with respect to the transverse $SO(8)$.

\section{Kinematics of the eikonal scattering}
\label{kinsec}

The process that we will analyse in this paper
is the scattering of a perturbative closed string state
(the probe) from a stack of D$p$-branes (the target) at high energy $E$
and fixed momentum transfer $t$.
In this section we then begin  by discussing the kinematics which  is relevant for a string-brane collision.

The D$p$-branes are static and aligned along 
the first $p$ space-like directions.
We indicate the (incoming) momenta of the two
states by the $SO(1,9)$ vectors $p_r^\mu~(r = 1,2)$, with $p_r^2=-m_r^2$, and
their spatial part by $\vec{p}_r$ (without loss of generality we
assume $\vec{p}_r$ to be
orthogonal to the D$p$-branes). We also set $\hat p_r = \vec{p}_r/| \vec{p}_r|$.
The Regge limit we are interested in corresponds to taking the energies of the two closed strings very large while keeping the
momentum transfer $q =  p_{1} + p_{2}$ fixed and typically small (corresponding to a large impact parameter for the collision\footnote{We recall from \cite{Amati:1987wq, Amati:1987uf} that, in eikonal approximation, the typical transverse momentum carried by an individual graviton is of order $\hbar b^{-1}$,  the overall momentum transfer in the collision $q \sim \theta E$ being shared among many gravitons.}).

The identification of the left
and  right moving parts of the closed strings absorbed or emitted
by the D$p$-brane is described by a diagonal matrix $R^\mu_{~\nu}$
\begin{equation}
R^\mu_{~\nu}  =  \delta^\mu_{~\nu} \ , \hspace{0.4cm} \mu , \nu = 0,\ldots, p
\ , \hspace{1.4cm} R^\mu_{~\nu}  = -  \delta^\mu_{~\nu} \ , \hspace{0.4cm} \mu , \nu = p+1,\ldots,9 \ .
\label{MMD}
\end{equation}
The two kinematic (Mandelstam-like) invariants characterizing
this process can be chosen as follows
\begin{equation}
  \label{eq:s-t}
  t= -(p_1 + p_2)^2\ , \hspace{2cm}
  s= -\frac{1}{4} (p_1 + R p_1)^2= -\frac{1}{4} (p_2 + R p_2)^2\equiv E^2 \ ,
\end{equation}
where in the second equation we used momentum conservation along
the Neumann directions and $E > 0$ will denote, hereafter, the common energy of the  incoming and outgoing closed string.
Scalar products among the external momenta and the reflection
matrix can be expressed in terms of the variables $s$, $t$  and the masses of the external states
\be 2 p_1 p_2 = - t + m_1^2 + m_2^2 \ , \hspace{1cm} 
p_r R p_r = - 2s + m_r^2 \ , \hspace{1cm} 
2 p_1 R p_2 = 4s +  t - m_1^2 - m_2^2  \ . 
\label{kine}
\ee
The physical polarizations of a massive string state can be described by introducing a basis
of nine space-like polarization vectors.   
For a massive state with a non-zero space-like momentum $\vec{p}_r$ we
first define the longitudinal polarization vector $v_r$
\begin{equation}
  \label{eq:vi}
  v_r^\mu = -\frac{m_r}{|\vec{p}_r|} \, \hat{t}^\mu + \frac{E_r}{|\vec{p}_r|}
\,   \frac{p_r^\mu}{m_r} \ = \ \frac{|\vec{p}_r|}{m_r}  \, \hat{t}^\mu + \frac{E_r}{m_r} \, \hat{p}_r^{\,\mu}
\ , \hspace{1cm}  v_r p_r = 0 \ , \hspace{1cm} v_r^2 = 1 \; \ ,
\end{equation}
where $\hat{t}$ is the unit vector in the time direction\footnote{Some additional formulae for the kinematics
in a convenient  reference frame are collected in
  Appendix~\ref{app:kine}.}.
The remaining physical polarizations are given by eight unit
vectors transverse both to $p_r$ and to $v_r$.

Since at high energy the longitudinal polarizations of
the massive states play a special role, in the following sections we will
 express the covariant amplitudes in terms of the basis of physical polarizations that can be
attached, in the way just described, to each state taking part in the scattering process.
In order to do this it is sufficient to decompose
every tensor contracted with the polarization of one of the external states 
along the basis of physical  polarizations pertaining to that state. 
For instance, when the
momentum $q$ transferred to the D$p$-branes is contracted with the
polarization of the second external state, one can use the
decomposition
\begin{equation}
q^{\rho} \equiv p_{1}^{\rho} + p_{2}^{\rho} = \frac{E (t+m_2^2-m_1^2)}{2m_2
  \sqrt{E^2-m_2^2}} v^{\rho}_{2} +  \frac{t+m_2^2-m_1^2}{2m_2^2} p_{2}^{\rho}  +
{\bar{q}}^{\,\rho} ~,
\label{q-v}
\end{equation}
where ${\bar{q}}$ is perpendicular to both $v_2$ and $p_2$. Notice
that the difference between $q$ and $\bar{q}$ is small, $(q-\bar q)^2 \sim t/s$ as it can been
seen by using~\eqref{eq:vi} and \eqref{q-v}.
We define $q_9 = q-\bar{q}$, since  in the frame introduced in
Appendix~\ref{app:kine} this vector is aligned with the ninth direction. 

Similar decompositions hold for the
Minkowski metric $\eta$ and the reflection matrix $R$ 
\begin{align}
\label{eq:Minma}
  \eta^{\rho\sigma} & = -\frac{p_2^\rho}{m_2}  \frac{p_2^\sigma}{m_2} +
  \sum_{i=1}^8 \hat{w}_i^\rho \hat{w}_i^\sigma + v_2^\rho v_2^\sigma
  \equiv -\frac{p_2^\rho}{m_2}  \frac{p_2^\sigma}{m_2} + \hat\eta^{\rho\sigma}\;,
   \\  \label{eq:Dma}
  R^{\rho\sigma} & = - \frac{2 E^2 - m_2^2}{m_2^2} \left(\frac{p_2^\rho}{m_2}
    \frac{p_2^\sigma}{m_2} + v_2^\rho v_2^\sigma\right) + \sum_{i=1}^p  \hat{w}_i^\rho
  \hat{w}_i^\sigma 
\\ \nonumber 
  & - \sum_{i=p+1}^8 \hat{w}_i^\rho \hat{w}_i^\sigma - \frac{2 E\sqrt{E^2 -
    m_2^2}}{m_2^2} \left(\frac{p_2^\rho}{m_2} v^\sigma+ v^\rho \frac{p_2^\sigma}{m_2}
\right) ~,
\end{align}
where $\hat{w}_i$, with $i=1,\ldots,8$, is a set of unit vectors
spanning the space perpendicular to $p_2$ and $v_2$, while $\hat\eta$
is the metric in the space transverse to $p_2$.

Our main example in this paper will be the inelastic process where a massless string state of
momentum $p_1$ is excited by the tidal forces of the
D$p$-brane\footnote{We will limit ourselves to the case $p\leq 6$ so
  as to have an asymptotically flat region.}  gravitational
field and emerges as a massive string state of momentum $p_2$, with
$p_2^2=-m^2$. 
Setting $v_2=v$, $m_1=0$ and $m_2=m$, the
high-energy limit of~\eqref{q-v} reads
\begin{equation}
  \label{eq:q-vHE}
  \bar{q}= q - \frac{m}{2} v \left(1+\frac{t}{m^2}\right) -
  \frac{t+m^2}{2m^2} p_2 + {\cal O}(1/E^2)~,
\end{equation}
and from~\eqref{eq:Dma} we have
\begin{equation}
  \label{eq:p2D}
  p_2 \frac{\eta+R}{2} =   -p_1 \frac{\eta+R}{2} = 
  \frac{E^2}{m^2} p_2 + \frac{E \sqrt{E^2-m^2}}{m} v~.
\end{equation}
The polarizations of the massless state can be written in terms of
vectors $\epsilon_k$ that satisfy both a transversality  and
a light-cone gauge constraint 
\begin{equation}
\epsilon_{k} p_{1} = \epsilon_{k}  e^{+} 
=0 \ , \hspace{2cm} k=1 \dots 8  \ .
\label{emuKv}
\end{equation}
At high energy it is natural to identify the light-cone vectors
with the large components of the external momenta in the frame where
the D-branes are at rest, thus connecting the gauge choice with physical
quantities in the problem. We then define the light-cone
vectors as follows
\begin{equation}
  \label{eq:epm}
  \sqrt{2} (e^{-})^{\mu}   = \lim_{E_1\to \infty} 
  \frac{{p}_1^{\,\mu}}{E_1} = -\lim_{E_2\to \infty} 
  \frac{{p}_2^{\,\mu}}{E_2} \ , \hspace{1cm}
  (e^{+})^{\mu}   = (-(e^{-})^0, \vec{e}^{\,-}) ~.
\end{equation}
The direction $e^-$ defines the large relative boost between the perturbative states and the D-branes, while $e^+$ is the complementary null direction satisfying $e^+ e^-=1$.
When the polarizations $\epsilon_k$'s are contracted with $q$ and $v$ we have
\begin{equation}
  \label{eq:epsexp}
  \epsilon_k q = \epsilon_k p_2 \sim \bar{q}^k \ , \hspace{2cm}
  \epsilon_k v  \sim - \frac{\bar{q}^k}{m}~,
\end{equation}
where we neglected terms of order ${\cal
  O}(1/E^2)$ in the large $E$ expansion. These relations will be useful in section \ref{covariant}
in order to express the covariant amplitudes in terms of tensors with non-trivial components only in the
space transverse to the collision axis.

\section{The eikonal phase from the light-cone GS vertex}
\label{gssec}

The eikonal operator $\exp(2 i \hat{\delta}(s, b))$
was introduced in~\cite{Amati:1987wq,Amati:1987uf} to provide a manifestly unitary description
of high-energy string-string collisions below a critical impact parameter $b_D$ at which string excitations due to tidal forces become important,
as briefly reviewed in the introduction. 
In this paper we shall focus on the eikonal operator for the high-energy 
scattering of a string on a collection of $N$ D$p$-branes \cite{D'Appollonio:2010ae},  the process described in 
the previous section, but
a very similar analysis could be performed  for the former process. 

Exponentiation of the eikonal phase operator $ \hat{\delta}(s, b)$ was proven in~\cite{Amati:1987wq,Amati:1987uf}  by considering, at each loop order, the leading terms in a high-energy expansion. It provides the so-called 
leading eikonal operator, giving, in the present context, the leading term  in an expansion in powers of $\frac{R_p}{b}$, where $R_p$ is the scale of the curved
D$p$-brane background 
\be R_p^{7-p} = g N \frac{(2 \pi \sqrt{\ap})^{7-p}}{(7-p) V_{S^{8-p}}} \ , 
\hspace{1cm} V_{S^n} = \frac{2 \pi^{\frac{n+1}{2}}}{\Gamma(\frac{n+1}{2})} \ .  \label{Rp}
\ee
On the other hand, the eikonal operator is supposed to resum all string ($\alpha'/b^2$) corrections to the leading eikonal phase. 
One can also argue \cite{Amati:1988tn} that the leading eikonal operator 
gives the correct description of the high-energy scattering
for any value of the impact parameter provided the  string coupling $g$ is sufficiently weak, i.e. when $R_p$ (or $R_s$ in the case of string-string collisions) is smaller than the string length scale $l_s$.

A similar exponentiation is conjectured to occur also at non-leading order in $\frac{R_p}{b}$, but the exact structure of  $ \hat{\delta}(s, b)$ 
for that case is not known\footnote{ In \cite{D'Appollonio:2010ae} we have computed the next-to-leading correction to the eikonal  phase in the field theory limit. 
We plan to come back to its modification due to string-size corrections in a forthcoming paper.}.
For this reason, here we will limit our attention to the leading eikonal operator whose phase is given by the
operator~\cite{D'Appollonio:2010ae}
\be 
\hat \d(s, b) = \frac{1}{4 E} \int\limits_0^{2\pi} d\s \, : {\cal A}(s, b +\hat{X}) : \ ,  \hspace{1cm} 
 {\cal A}(s,b) = \int \frac{d^{8-p}\bar q}{(2\pi)^{8-p}}  {\rm e}^{\ii \bar q b} {\cal A}(s, \bar q) \ .
\label{eikgs1} 
 \ee
Here ${\cal A}(s,b)$ is the Fourier transform in impact parameter space of the disc amplitude in the Regge limit
\be 
 {\cal A}(s, \bar q) = \frac{R_{p}^{7-p}
       \pi^{\frac{9-p}{2}}}{\Gamma \left(\frac{7-p}{2}
       \right)} \Gamma\left( -\frac{\alpha' t}{4}\right) \ex{ - \ii
       \pi\frac{\alpha' t}{4}} (\alpha' s)^{1+\frac{\alpha' t}{4}} \ , \label{lep} \ee
${\bar{q}}^2 = - t$ and $\hat{X}$ is a free bosonic field, without zero modes and evaluated at $\t = 0$, with components
only along the $8-p$ spatial directions transverse to the collision axis and to the brane
\be \hat{X}^i = \ii \sqrt{\frac{\ap}{2}} \sum_{n \ne 0} \si ( \frac{a^i_n}{n} {\rm e}^{\ii n \s} +  \frac{\bar a^i_n}{n} e^{-i n \s} \de )
\ , \hspace{1cm} [a^i_n, a^j_m] = n  \d^{ij} \d_{n+m,0} \ . \label{gsmodes}\ee
In order to fully understand how inelastic unitarity works, it is necessary to identify precisely the Hilbert space on which this operator acts.
The aim of this section is to show  that, not unexpectedly,  the leading eikonal operator
acts on the space of the physical states of the string quantized in 
the light-cone gauge, with the spatial direction of the light-cone chosen
along the direction of the large momentum appearing in the process. The free fields $\hat{X}^i$ in Eq.~\eqref{gsmodes} should then be identified with 
the transverse string coordinates and the modes $a^i_n$, $\bar a^i_n$ with the light-cone oscillators.  
In section \ref{covariant} we will derive the eikonal phase from the covariant dynamics and we will show that
the modes $a^i_n$ are then naturally identified with the bosonic DDF operators. 
The two interpretations are of course equivalent, as a consequence of the one-to-one correspondence 
between the DDF operators and the physical states in the light-cone gauge.

The idea of the derivation presented in this section is simple.
We will use the  light-cone 3-string vertex~\cite{Green:1982tc,Green:1983hw}
in the Green-Schwarz formalism, which
encodes the interaction among three generic string states,  to write an operator that generates all the 
 tree-level  amplitudes
in the D$p$-brane background with two arbitrary string states. In order to do this
we first need to specialise the GS vertex to the case where only one of the three states
is off-shell, while the remaining two are arbitrary on-shell states.  

We then take the high-energy Regge limit, sending the energy $E$ of the two external states
to infinity while keeping  finite the momentum transfer $t$,  the momentum squared 
of the state exchanged between the string probe and the target. As we will see, provided
the light-cone direction is aligned to the direction of the large momentum, the structure of the GS vertex
considerably simplifies in the limit. 
The final step consists in contracting the
off-shell leg with the closed string propagator and with
the boundary state that describes the coupling of an arbitrary string
state to a collection of D$p$-branes.

In order to simplify the analysis in this section, we will consider the limit of large impact parameter
$b$ or, equivalently, of  vanishing momentum transfer $t$. This corresponds to exchanging only the graviton
between the probe and the target, since, on one hand, only the massless states contribute  for small momentum transfers,
  and, furthermore, states with the highest spin dominate at high
energy. 
We will show that the final result coincides with the phase of the eikonal operator in Eq.~\eqref{lep}
in the  limit of large impact parameter
$b$.  This is sufficient to identify the oscillators that appear in the eikonal operator as 
the bosonic string modes in the light-cone gauge. In any case
the covariant derivation
of the eikonal phase discussed in section \ref{covariant} will be valid for arbitrary value of the momentum transfer
and not only for massless exchanges.

It is also worth noticing that the
calculation is split into two independent parts, of which the first - the GS vertex - 
captures the emission of an off-shell graviton from the high energy probe
while the second describes how the graviton
propagates and then interacts with the target. 
The same factorized form will be displayed 
by the covariant amplitudes, with the
GS vertex  replaced by the three-point couplings of the external states to the Reggeon
and the information about the target  carried by the Reggeon tadpole in the given background.

Let us now start our first derivation of the eikonal phase
by recalling the GS light-cone vertex.
In this approach, the left moving part of the light-cone states is
described by a set of bosonic oscillators $A^i_n$ transforming as a
vector of $SO(8)$ and a set of fermionic oscillators $Q_n^a$
transforming as a spinor of $SO(8)$. These operators refer to (and
depend from) the choice of the light-cone vectors (\ref{eq:epm}) but
we omit for simplicity a label referring to that choice.  The left
moving part of the spectrum is obtained by acting with the raising
operators $A^i_{-n}$ and $Q^a_{-n}$ on a degenerate massless ground
state which we indicate with $|i\rangle$ and $|\dot{a}\rangle$. The
first ket represents eight states transforming as a vector of $SO(8)$,
while the second one is a spinor of $SO(8)$ (the dot over the spinor
index indicates that the vacuum chirality is opposite to that of the
fermionic oscillators). In both kets we understand the eigenvalue $p$
of the momentum. A closed string state at level $n$,  carrying
momentum $p$ with $p^2 = - 4 n/\alpha'$,  is given by the product of a
left moving and a right moving part satisfying the level matching
condition.

Following the previous literature on the subject, we indicate with $\alpha$ and
$\bar{p}_{j}$ the projections of the momentum $p$ respectively along
$e^+$ and along the transverse space (spanned by the vectors $
\hat{w}_j$)
 \begin{equation}
  \label{eq:p+}
    \alpha_r \equiv \sqrt{\frac{\alpha'}{2}}\,2 p^{(r)} e^+ \ , \hspace{0.7cm}
  \bar{p}_{j}^{(r)} \equiv  \sqrt{\frac{\alpha'}{2}} p^{(r)}  \hat{w}_j\ , \hspace{0.7cm} r=1,2,3\ , \hspace{0.7cm} j = 1, 2, \dots 8 \ ,
\end{equation}
where the label $r$ distinguishes the three different states that appear in the
vertex. We shall treat
independently the left and the right moving parts and describe the
interaction in each sector by means of the vertex introduced
in~\cite{Green:1982tc}. The full vertex can be written as a vector
living in the tensor product of three closed string Hilbert spaces, one for each 
 light-cone state involved in the interaction. The chiral part of the vertex is given by
\begin{equation}
  \label{eq:Vs}
  |V_{GS}\rangle = \sqrt{\frac{2}{\alpha'}} \left(P_i-\alpha_1 \alpha_2 \alpha_3
    \frac{n}{\alpha_r} N^r_n A^r_{-n, i} \right) V_b V_f | V_i\rangle + \ldots \ \ ,
 \end{equation}
where we understood the usual delta function imposing momentum
conservation along the ten spacetime directions and the dots stand for terms
that vanish on-shell.   As we shall explain after Eq.~\eqref{eq:PDP}, these terms will not be relevant for the derivation
of the eikonal phase at large $b$.   The zero-mode structure $|V_i\rangle$ of the
vertex is
\begin{eqnarray}
    |V_i\rangle &=& \frac{1}{\alpha_1} |ijj\rangle + \frac{1}{\alpha_2} |jij\rangle  + \frac{1}{\alpha_3} |jji\rangle + \frac{\alpha_1-\alpha_2}{4 \alpha_3} |aai\rangle + \frac{\alpha_1-\alpha_3}{4 \alpha_2} |aia\rangle  \nonumber \\
&+& \frac{\alpha_2-\alpha_3}{4 \alpha_1} |iaa\rangle + \frac{1}{4}
\gamma^{ij}_{ab} \left(|baj\rangle + |bja\rangle + |jba\rangle
\right)~,
\label{Vi}
\end{eqnarray}
where the $\gamma$'s are the $SO(8)$ gamma matrices, all sums over
repeated indices are understood and the kets with three indices are
just the tensor product of the vector or spinor ground states for each
external string state; finally
these kets are normalized as in~\cite{Green:1982tc}
\begin{equation}
  \label{eq:ketnorm}
  {}_r \langle i | j \rangle_{r} = \delta^{ij} \ , \hspace{2cm}
  {}_r \langle a | b \rangle_r = \frac{2}{\alpha_r} \delta^{ab}  \ .
\end{equation}
Most of the complexity of the vertex is in  the
exponentials $V_b$ and $V_f$
\begin{eqnarray}
\label{eq:Vb}
  V_b &=& \exp\left(\frac 12 A^r_{-n, i} N^{rs}_{mn} A^s_{-m, i} + P_i
    N^r_n A^r_{-n, i} \right)~, \\ 
  V_f &=& \exp\left(\frac 12 Q^r_{-n, a} X^{rs}_{mn} Q^s_{-m, a} - S_a \frac{n}{\alpha_r} N^r_n Q^r_{-n, a} \right)~.
 \label{eq:Vf}
\end{eqnarray}
The operators $P_i$ and $S_a$ stand for the following combinations of
the bosonic and fermionic zero-modes
\begin{equation}
  \label{eq:Pi}
  P_i \equiv  \sqrt{\frac{\alpha'}{2}} \left(\alpha_r \bar{p}^{(r+1)}_{i} - \alpha_{r+1} \bar{p}^{(r)}_{i}\right) \  , \hspace{1cm}
  S_a \equiv \alpha_r Q^{(r+1)}_{0 a} - \alpha_{r+1} Q^{(r)}_{0 a} \ ,  
\end{equation}
which, with the cyclic identification between $r=4$ and $r=1$, are
independent of the choice of $r=1,2,3$. Finally, the Neumann
coefficients encoding the actual value of the various couplings are
\begin{gather}
\label{NMX}
N^{rs}_{nm} =  - \frac{nm \alpha_1 \alpha_2 \alpha_3}{n \alpha_s + m
  \alpha_r} N^{r}_{n} N_{m}^{s} \  , \hspace{2cm}
  X^{rs}_{nm} = \frac{n \alpha_s - m\alpha_r}{2\alpha_r \alpha_s}
  N^{rs}_{nm} \ , \\ \label{NMX2}
N_{n}^{r} = - \frac{1}{n \alpha_{r+1}}\left(  \begin{array}{c} - n \frac{\alpha_{r+1}}{\alpha_r} \\ n \end{array} \right) = \frac{1}{\alpha_r n!} \frac{ \Gamma  \left(   - n 
\frac{\alpha_{r+1}}{\alpha_r} \right)  }{ \Gamma  \left(   - n 
\frac{\alpha_{r+1}}{\alpha_r} +1 -n \right) } \ .
\end{gather}
In the kinematic configuration described at the beginning of this section, the above vertex
drastically simplifies if one chooses the direction of the null
vectors as in~\eqref{eq:epm}. In the Regge limit the energy $E$ of the
states $r=1,3$ is much larger than the momentum exchanged which, for a
single graviton exchange, is extremely small and this is reflected in
the asymptotic results for the $\alpha_i$. At leading order in
$E$, we have
\begin{equation}
  \label{eq:specalpha}
  \alpha_1\sim \sqrt{\alpha'}\, {2} {E}\  , \hspace{1.4cm}
  \alpha_2 = - \sqrt{\alpha'} {q_9} \  , \hspace{1.4cm}
  \alpha_3\sim - \sqrt{\alpha'}\, {2} {E} \ , 
\end{equation}
where $q_9\sim {\cal O}(1/E)$, see Eq.~\eqref{p1xxx}. By
using~\eqref{eq:specalpha}, this means that we have to take $\alpha_1$
and $\alpha_3$ large and $\alpha_2$ small\footnote{The states labelled
  here by $r=2$ and $r=3$ have momenta $-q$ and $p_2$ respectively
  according to the notations of Section~\ref{kinsec}.}. Also we set to
zero all the  oscillators of the string labeled by $r=2$, since
we wish to identify that state with the graviton exchanged between the
probe and the target. Then the only surviving Neumann coefficients are
$N^1$ and $N^3$, which become
\begin{equation}
  \label{eq:Nlimit}
   P_i \sim \sqrt{\frac{\alpha'}{2}} \alpha_1 \bar{p}^{(2)}_{i} \to -\sqrt{2}\alpha' E \bar{q}_i \  , \hspace{1cm}
   N_{n}^{1} \to \frac{(-1)^{n-1}}{\alpha_1 n}\  , \hspace{1cm}
   N_{n}^{3} \to -\frac{1}{\alpha_1 n}  \ ,
\end{equation}
where $\bar{q}$ is the momentum introduced\footnote{With a slight
  abuse of language we use the same symbol to indicate both the
  momenta of this section and the 10D vector $\bar{q}$ orthogonal to
  $v$ and $p_2$ introduced in the previous section, because the two
  objects are identical in the non-trivial 8D space.}
in~\eqref{q-v}. Several other simplifications occur in this
limit. We first note that the second and the fifth terms in~\eqref{Vi} dominate
over the others implying that the two energetic external states always
share one index. Then, from the second equation in~\eqref{NMX} we see
that $X^{13}$ is subleading with  respect to $N^{13}$ and similarly
the term proportional to $S_a$ in $V_f$ is subleading with  respect
to the one proportional to $P_i$ in $V_b$. This means that all terms
containing fermionic oscillators $Q^q_a$ can be neglected in the high-energy 
scattering we are interested in. Finally there is a hierarchy
also within $V_b$ and within the prefactor of the full
vertex~\eqref{eq:Vs}: from~\eqref{eq:Nlimit} we see that the
combination $P_i N^{r=1,3}$ is finite in the large $E$ limit, while
$N^{13}$ vanishes in the same limit; similarly in the prefactor, the
term $P_i$ dominates over the other one. Thus, instead of the full
vertex~\eqref{eq:Vs}, we can use the simplified expression
\begin{equation}
  \label{eq:Vss}
  |V_{GS}\rangle \sim  \sqrt{\frac{2}{\alpha'}}   \frac{P_i}{\alpha_2}
   \exp\left\{\sqrt{\frac{\alpha'}{2}} \sum_{n=1}^\infty 
\frac{{\bar{q}}_{\ell}}{n} \Big({A}^3_{-n \ell}+(-1)^n {A}^1_{-n \ell}\Big)\right\} \left[
    |jij\rangle + \frac{\alpha_1-\alpha_3}{4} |aia\rangle \right] \ . 
\end{equation}
We can now easily derive the high-energy scattering amplitude
describing the interaction between a string probe and a stack of
D$p$-branes at large distances. Schematically this process is
described by
\begin{equation}
  \label{eq:BPV}
  |W\rangle = \frac{\kappa_{10} \t_p N}{2} ~ {}_2\langle B| P \left(\kappa_{10}  |V_{GS} \rangle
    |\widetilde{V_{GS}} \rangle \right) \sim \frac{R_{p}^{7-p}
    \pi^{\frac{9-p}{2}}}{\Gamma \left(\frac{7-p}{2} \right)}~
  {}_2\langle B_0| \frac{1}{-t} \left( |V_{GS}
  \rangle |\widetilde{V_{GS}} \rangle \right)~, 
\end{equation}
where $|B\rangle$ is the GS boundary state~\cite{Green:1996um} for the
D$p$-branes written in terms of the oscillators in the Hilbert space
labelled by $r=2$ and $P$ is the string propagator.  The
normalisations on the l.h.s. are the standard gravitational coupling
$\kappa_{10}$, related to the string length and coupling by 
$2 \kappa_{10}^2 = (2 \pi)^7 \ap^4 g^2$, and the tension 
of a single D$p$ brane
\be \t_p^2 = \frac{\pi}{\kappa_{10}^2} \, \si ( 4 \pi^2 \ap \de )^{3-p} \ . \label{dptension} \ee
The final relation in~\eqref{eq:BPV} is obtained by implementing the
high-energy and large impact parameter limits: the boundary state
is truncated to its zero-mode sector $| B_0\rangle$, the string
propagator reduces to the field theory one
and we can use the simplified version of the $3$-string vertex in
Eq.~\eqref{eq:Vss}. The zero-mode structure gives
\begin{equation}
  \label{eq:PDP}
  \frac{2}{\alpha'} \frac{P_h R_{hk} P_k}{\alpha_{2}^{2} (-t)} =
  \frac{\alpha_1^2}{\alpha_2^2} \frac{{\bar{q}}^2}{ t} = -
  \frac{\alpha_1^2}{ \alpha_2^2}  \left( 1 + \frac{q_{9}^{2}}{t}\right)
  \sim - \frac{4E^2}{q_{9}^{2}} - \frac{4E^2}{t}~,
\end{equation}
where $R$ is the reflection matrix~\eqref{MMD}.  Since we are
restricting ourselves to the contribution due to the exchange of the
massless states, we have first to take the impact parameter to be
large and then take the high-energy limit. Then we can neglect the
first term of the final expression in~\eqref{eq:PDP} because it does
not have a pole in $t$. For the same reason, it is possible to neglect
the terms that vanish on-shell in~\eqref{eq:Vs}. These contributions are proportional to $\sum_r P^{-}_r$, where
\begin{equation}
  \label{eq:LCHam}
  P^{-}_r = \frac{2}{\alpha_r} \left[\frac{\alpha'}{2} \frac{\bar{p}_{r}^2}{2} +
 \sum_{n=1}^\infty \si ( A_{-n}^{(r)} A_{n}^{(r)} + n Q_{-n}^{(r)} Q_{n}^{(r)} \de ) \right]~.
\end{equation}
In~\cite{Green:1982tc} it was shown that there are terms that vanish
on-shell and depend on $\tau_0\equiv \sum_r \alpha_r\ln|\alpha_r|$,
while more recently other terms of this type were proposed
in~\cite{Lee:2004cq} (see for instance, Eq.~(5.1) of that
paper). However, in our calculation only the string $r=2$ is kept
off-shell and in the corresponding Hilbert space we focus only on the massless
sector. Thus all the extra terms in~\eqref{eq:Vs} must be proportional
to $-t$, which is the momentum squared of the exchanged
graviton. Again those contributions would cancel the pole
in~\eqref{eq:LCHam} and thus can be neglected.
We then have
\begin{eqnarray}
  |W\rangle & \sim & 
  \frac{R_{p}^{7-p} \pi^{\frac{9-p}{2}}}{\Gamma
    \left(\frac{7-p}{2} \right)}
  \frac{4 E^2}{-t} \exp\left\{\sqrt{\frac{\alpha'}{2}}  \sum_{n=1}^\infty \frac{{\bar{q}}_{\ell}}{n}
    ({A}^3_{-n \ell}+(-1)^n  {A}^1_{-n \ell})\right\} 
  \left[|j\rangle_1 |j\rangle_3 + \frac{\alpha_1}{2} |a\rangle_1
    |a\rangle_3 \right]
  \nonumber \\ \label{eq:zmVD}  && 
   \times ~ \exp\left\{\sqrt{\frac{\alpha'}{2}}  \sum_{n=1}^\infty \frac{{\bar{q}}_{\ell}}{n}
    (\bar{A}^3_{-n \ell}+(-1)^n \bar{A}^1_{-n \ell})\right\} 
  \left[|\bar{j}\rangle_1 |\bar{j}\rangle_3 + \frac{\alpha_1}{2}
    |\bar{a} \rangle_1 |\bar{a}\rangle_3 \right]~.
\end{eqnarray}
It is more natural to write this result as an operator on a single
Hilbert space instead of a product of two kets in two different
spaces. This can be done by taking the adjoint of the objects labelled
with $r=1$, i.e. by transforming $|i\rangle_1$ and $|a\rangle_1$ into
${}_3\langle i|$ and ${}_3\langle a|$, and ${A}^1_{-n \ell}$ into
$(-1)^{n+1} {A}^3_{n \ell}$. After this the square parenthesis
in~\eqref{eq:zmVD} becomes just the identity operator on the zero-mode
sector, as it follows from~\eqref{eq:ketnorm}, and we can rewrite the two
exponentials in terms of an auxiliary string field 
\begin{equation}
  \label{eq:hatX}
  \hat{X}^i(\sigma) = \ii \sqrt{\frac{\alpha'}{2}} \sum_{n\not=0} \left(\frac{A_{ni}}{n} \ex{\ii n \sigma} + \frac{\bar{A}_{ni}} {n} \ex{-\ii n \sigma}\right) ~.
\end{equation}
So we can write~\eqref{eq:zmVD} in an operator form as follows
\begin{equation}
  \label{eq:zmVDop}
  W(\bar{q}) \sim \int \frac{d\sigma}{2\pi} :\ex{\ii \bar{q}
    \hat{X}(\sigma)}:  \left(\frac{R_{p}^{7-p} \pi^{\frac{9-p}{2}}}{\Gamma
    \left(\frac{7-p}{2} \right)}\frac{4E^2}{-t}\right)~,
\end{equation}
where we included an integral over $\sigma$ which is trivial when the matrix elements of $W$ 
are taken between closed string states
satisfying the level matching condition. 

In this derivation we took into account only the
contributions due to the graviton exchange and thus the result
obtained captures reliably only the first term in the small $t$
expansion. 
It should be possible to adapt the derivation discussed
in~\cite{Cremmer:1974ej} and  extend our light-cone calculation of the eikonal phase
to include the contribution of all the string
states exchanged between the string probe  and
the D-branes. We will not perform  this calculation here
because, as already mentioned, the covariant derivation of the
eikonal phase presented in section \ref{covariant} does take into account the full string dynamics 
in the Regge limit. 
As we will show, the complete result can be obtained 
by replacing the round parenthesis in~\eqref{eq:zmVDop}
with the Regge limit of the full elastic tree-level string amplitude ${\cal A}$ in Eq.~\eqref{lep}, not just with the graviton pole, 
\begin{eqnarray}
  \label{eq:eik-phase}
   W(\bar{q}) & = &  {\cal A}(s,\bar{q})~ \int \frac{d\sigma}{2\pi} :\ex{\ii \bar{q} \hat{X}(\sigma)}: \  .  
\end{eqnarray}
Multiplying $W$ by $\ex{\ii {\bar{q}} b}$ and taking the Fourier
transform to write the result in terms of the impact paramenter $b$
instead of the momentum transferred ${\bar{q}}$ we find
\begin{equation}
  \label{eq:Wop}
   W(b)
  = \int\limits \,   \frac{d^{8-p} {\bar{q}}}{(2\pi)^{8-p}}  \, W(\bar{q}) \ex{\ii {\bar{q}} b} 
  \equiv \int\limits_0^{2\pi} \frac{d\sigma}{2\pi}
  : {\cal A}\left(s,b+\hat{X}\right): ~,
\end{equation}
which coincides with Eq.~\eqref{eikgs1}, the  
phase of the leading eikonal operator for string-brane collisions~\cite{D'Appollonio:2010ae}, after including a
factor $2E$ for the relativistic normalization of the external states, $W/2E = 2 \hat \d$. 

This analysis then shows that the exact meaning of the oscillators $a^i_{n}$ in Eq.\eqref{gsmodes} is that of the bosonic
oscillators $A^i_{n}$\footnote{The discussion in this section was based on the Green-Schwarz light-cone formalism but we could have also used the
Ramond-Neveu-Schwarz string quantized in the light-cone gauge. Given that the
eikonal operator is written only in terms of the bosonic oscillators and that the bosonic fields $X$ of the
two formalisms can be identified, we would have obtained exactly the same results. } 
of light-cone quantization, with the light-cone axis aligned with the direction of the large momenta.
As anticipated at the beginning of this section, it is also possible to recast Eq.~(\ref{eq:Wop}) in a covariant form 
(i.e one that does not depend on using a particular  gauge) by exploiting the one-to-one correspondence
between the light-cone states and the DDF operators~\cite{DelGiudice:1971fp}, that we review in App. \ref{app:DDF}.
It is precisely in this latter form that the eikonal phase will be given by the covariant derivation discussed in section \ref{covariant}.

Having identified the Hilbert space on which the eikonal operator acts, 
we can proceed to discuss its main properties and what information it can provide on the string dynamics in the Regge limit,
which will be the subject of the next section.

\section{High-energy inelastic amplitudes from the eikonal phase}
\label{eikonal}

We describe in this section the essential properties of the eikonal operator and what can be learned
 from its simple structure on the high-energy string dynamics. The discussion will be based on the light-cone RNS
superstring so as to make the comparison with the covariant formalism of section \ref{covariant} more direct.
We shall illustrate the general case by deriving from the eikonal operator the inelastic 
amplitudes for transitions from the massless NS-NS sector to the first two massive levels. 
We will also describe a method to provide a covariant characterization of the string states excited by the tidal forces
during a string-brane collision, 
using the DDF operators~\cite{DelGiudice:1971fp} of the NS sector of the 
superstring~\cite{Hornfeck:1987wt} 
to identify the linear combination of covariant string states that corresponds to a given light-cone
state.

As shown in the previous section, the bosonic fields $\hat{X}$ in the leading eikonal operator 
\be {\rm e}^{2 \ii \hat \d(s, b) } \ , \hspace{2cm} \hat \d(s, b) = \frac{1}{4 E} \int\limits_0^{2\pi} \frac{d\s}{2 \pi} 
\, : {\cal A}(s, b +\hat{X}) : \ , \label{eik1}  \ee
can be interpreted as the transverse string coordinates in a light-cone gauge aligned with the direction of the large
momenta.
The most interesting feature of this expression is that the string modes appear
as a simple shift of the impact parameter $b$ by the string position operator $\hat{X}$.
This structure reflects the incoherent scattering of the individual bits of the string
when $\ap s \gg 1$. Another feature of the eikonal operator is that it contains the light-cone modes
of the bosonic fields $\hat{X}$ but not those of the fermionic fields.  It is important to appreciate that 
since we are in the light-cone gauge
the shift $b \mapsto b + \hat{X}$ describes 
the dynamics not only of the string excitations polarized along the directions
transverse to the collision axis but also of the string excitations polarized along the longitudinal 
direction. 
This simple description of the longitudinal polarizations and the absence of the 
fermionic modes are a consequence of the superconformal invariance 
of the covariant worldsheet theory. 

The matrix elements of the eikonal operator between two closed string states give the high-energy  behaviour of the
corresponding two-point amplitudes in the D$p$-brane background. The physical information contained in 
these matrix elements can be most clearly displayed 
by first labeling the light-cone states according to their mass and their representation with respect
to the transverse $SO(8)$  and 
then classifying the independent contractions between the polarization tensors of the external states and the 
momentum transfer that can appear in the amplitudes.

The eikonal operator is the result of an all-loop resummation of string amplitudes, as it is clear from the fact
that the string coupling appears in the exponent. 
Since in this paper we shall only consider tree-level amplitudes, it is  sufficient to study  the matrix
elements of the eikonal phase $\hat \d(s,b)$, which are related
to the string scattering matrix at tree-level by $W(s, b) = 4 E \, \hat \d(s,b)$.
To derive the scattering amplitudes it is more convenient to work in momentum space and write
\be W(s, \bar q) =  {\cal A}(s, t)    \int_0^{2\pi}  \frac{d \s}{2\pi}  \, : {\rm e}^{\ii \bar q \hat X} :
\equiv {\cal A}(s, t)  \sum_{n, m = 0}^\infty \D_{n,m}(\bar q) \, \bar \D_{n,m}(\bar q)
 \label{taylor} \ , \ee
where the operators $\D_{n, m}$ generate by definition  all the transitions between an initial level $m$ and a final level $n$. 
For instance
the inelastic transitions from the ground state to the first two massive levels are due to the operators
\ba \D_{1,0} &=& - \sqrt{\frac{\ap}{2}} \, \bar q^i \,  A_{-1}^i \ , \nb \\
 \D_{2,0} &=&  \frac{\ap}{4}  \bar q^i \bar q^j  \,  A_{-1}^i  A_{-1}^j - \sqrt{\frac{\ap}{8}}
\bar q^i  \,  A_{-2}^i   \ . 
\ea
Level by level we organize the light-cone string spectrum  in irreducible
representations of the transverse $SO(8)$ group.
The irreducible $SO(8)$ representations are traceless tensors of type $(n_1, n_2, ..., n_r)$ which can be represented by  
Young diagrams with $r$ rows of length $n_i$\footnote{Additional details
on our conventions for the irreducible $SO(n)$ tensors can be found in Appendix \ref{app:Young}.}. 
The polarization of a state in the representation corresponding to the Young diagram 
$(n_1, n_2, ..., n_r)$ will be written as follows
\be \w^{(n_1, n_2, ..., n_r)}_{i_1 ...i_{n_1}; j_1 ...j_{n_2}; ...; k_1 ...k_{n_r}} \ , \ee
where the semicolons separate groups of indices in the rows of the diagram. The tensor is
antisymmetric in the indices belonging to the same column  and normalized, $\w \cdot \w = 1$.
To simplify the notation we do not use
the semicolon for totally antisymmetric tensors and often omit the label $(1)$ on vectors. For instance
\be \w^{(2,1)}_{i j; k} \ , \ee
is a tensor antisymmetric in the couple $(i, k)$ and satisfying the relation
\be \w^{(2,1)}_{i j; k}  + \w^{(2,1)}_{j k; i}  + \w^{(2,1)}_{k i; j} = 0 \ . \ee
Let us now analyse in detail the inelastic transitions from the massless states of the NS sector. This will be sufficient 
to understand the general case. 
In the light-cone gauge the massless NS state is a vector of $SO(8)$
\be |\e\pd = \e_i B^i_{-1/2}|0\pd \ . 
\ee
Since the eikonal phase contains only the bosonic oscillators, level by level the states
that can have a non-vanishing matrix element\footnote{As already pointed out in footnote 4, at high energy and large impact parameter the eikonal phase is controlled by soft dynamics: hence, the {\it in} and {\it out} states can be taken to have the same momentum, which we leave understood in the
following equations.}
with the ground state are only those created 
by the action on the vacuum of one $B_{-1/2}$ and any number of bosonic modes $A_{-n}$.
The relevant states in the first level are $64$ and they are displayed together
with their matrix elements in Table 1. 
 \begin{table}[ht] \be 
\begin{tabular}{||l||c||} \hline 
$SO(8)$ representation & Matrix element $\ps \w | \D_{10} | \e \pd$ \\ \hline 
& \\
$ |\w^{(2)}\pd = 
\w^{(2)}_{ij} A_{-1}^i B^j_{-\frac{1}{2}}|0\pd $&  $- \sqrt{\frac{\ap}{2}}  \,
\e^i \, \w^{(2)}_{ij} \bar q^j $\\ & \\
$ |\w^{(1,1)}\pd =  \w^{(1,1)}_{ij} A_{-1}^i B^j_{-\frac{1}{2}}|0\pd$ & $  \sqrt{\frac{\ap}{2}} \,
\e^i \, \w^{(1,1)}_{ij} \bar q^j  $ \\ & \\
$ |\w^{(0)}\pd =  \frac{1}{\sqrt{8}} \, A_{-1}^i B^i_{-\frac{1}{2}}|0\pd$ &  $- \frac{\sqrt\ap}{4}  \, \e \bar q  $ \\ 
& \\ \hline \end{tabular}\label{tab1} 
\ee \caption{Matrix elements of the eikonal phase for transitions from the massless sector to the first level.}  \end{table}

The remaining $64$ NS states of the first level are
\be |\w^{(1,1,1)}\pd = \frac{1}{\sqrt{6}} \w^{(1,1,1)}_{ijk}  B^i_{-\frac{1}{2}} B^j_{-\frac{1}{2}} B^k_{-\frac{1}{2}}|0\pd \ , \hspace{1cm}
 |\w^{(1)}\pd = \w_i B^i_{-\frac{3}{2}}|0\pd \ , \ee
and since they contain either more than one mode $B_{-\frac{1}{2}}$ or the higher mode $B_{-\frac{3}{2}}$
their matrix elements with the eikonal operator vanish. This means that 
the inelastic transitions from the ground state to these states
are subleading in energy. 

The second level contains $352$ states with a non-vanishing inelastic amplitude in the Regge limit.
They transform in the following $SO(8)$ representations
\be \yng(3) \hspace{1cm} \yng(2,1) \hspace{1cm} 2 \times \yng(1)  \hspace{1cm} \yng(2) \hspace{1cm} \yng(1,1) \hspace{1cm}
 \bullet \   \ee
and their explicit form and matrix elements are collected
in Table 2.
\begin{table}[ht] \be
\begin{tabular}{||l||c||} \hline 
$SO(8)$ representation & Matrix element $\ps \w | \D_{20} | \e \pd$ \\ \hline 
& \\
$  | \w^{(3)} \pd = \frac{1}{\sqrt{2}}  \w^{(3)}_{ijk} A_{-1}^i  A_{-1}^j  B^k_{-\frac{1}{2}}|0\pd $& $\frac{\ap}{\sqrt{8}}
\e^i \, \w^{(3)}_{ijk} \bar q^j \bar q^k $\\ & \\
$  | \w^{(2,1)} \pd =  \sqrt{\frac{2}{3}} \w^{(2,1)}_{ij; k} A_{-1}^i  A_{-1}^j B^k_{-\frac{1}{2}}|0\pd$ & $ - \frac{\ap}{\sqrt{6}}
\e^i \, \w^{(2,1)}_{ij; k} \bar q^j \bar q^k$ \\ & \\
$ | \w^{(2)} \pd =  \frac{1}{\sqrt{2}}  \w^{(2)}_{ij} A_{-2}^i B^j_{-\frac{1}{2}}|0\pd$ & $ \frac{\sqrt{\ap}}{2}
\e^i \, \w^{(2)}_{ij} \bar q^j  $ \\  & \\
$ | \w^{(1,1)} \pd = \frac{1}{\sqrt{2}}  \w^{(1,1)}_{ij} A_{-2}^i B^j_{-\frac{1}{2}}|0\pd$ & $- \frac{\sqrt{\ap}}{2}
\e^i \, \w^{(1,1)}_{ij} \bar q^j  $ \\  & \\
$ | \w^{(1)} \pd =  - \frac{ \w_{i}}{4\sqrt{35}} \si [
8  A_{-1}^i  A_{-1}^j B^j_{-\frac{1}{2}} - A_{-1}^j  A_{-1}^j B^i_{-\frac{1}{2}}\de]\! |0\pd $ &
 $- \frac{\ap}{\sqrt{35}} \si ( \e \bar q \w \bar q + \frac{\ap t}{8} \e \w \de )  $ \\ & \\ 
$ | \l^{(1)} \pd =  \frac{ \l_{i}}{4}  A_{-1}^j  A_{-1}^j B^i_{-\frac{1}{2}}|0\pd$ & $ - \frac{\ap t}{8} \e \l   $ \\ & \\ 
$ | \w^{(0)} \pd = \frac{1}{4} \, A_{-2}^i B^i_{-\frac{1}{2}}|0\pd$ & $- \frac{\sqrt{\ap}}{4\sqrt{2}}\e \bar q   $ \\ 
& \\ \hline \end{tabular} \label{tab2}
\ee \caption{Matrix elements of the eikonal phase for transitions  from the massless sector  to the second level.}   \end{table}
We see that the set of the representations that can be reached from the ground state in a high-energy collision comprises
the same representations present at level one, created now by the action of the modes $A^i_{-2}$, together with two
new rank three tensors and two vectors. 
The remaining $800$ states of the $NS$ sector  
transform in the following representations of $SO(8)$
\be \yng(1,1,1,1,1) \hspace{0.8cm} \yng(1,1,1,1) \hspace{0.8cm} \yng(2,1,1)  \hspace{0.8cm}\yng(1,1,1)  \hspace{0.8cm}
\yng(2,1) \hspace{0.8cm} 2 \times \yng(1,1) \hspace{0.8cm} \yng(2)  \hspace{0.8cm} 2 \times \yng(1)  \hspace{0.8cm} \bullet  
\ee
Their explicit form in terms of the string modes is easily derived and always involves more than one $B_{-\frac{1}{2}}$
or the higher modes $B_{-\frac{3}{2}}$ and $B_{-\frac{5}{2}}$. As in the previous case, this implies that 
the inelastic amplitudes for the transitions from the ground state to any
of these states are subleading in energy.

It is in the second level that we find the first example of a degenerate $SO(8)$ representation, the two vectors 
$|\w^{(1)}\pd$ and $|\l^{(1)}\pd$. The degeneracy of the representation is matched
by the presence in  the amplitudes of two independent contractions between the momentum transfer and the polarization
tensors 
\be \e \bar q \,\,  \w \bar q \ , \hspace{2cm} \e \w \,\, t \ . \ee
The basis
chosen for the vectors in Table 2 has the property
that only the state $ | \w^{(1)} \pd$  can be produced at large values of the impact parameter
since its amplitude contains a term $\e \bar q \,\, \w \bar q$, without powers of $t$ that would cancel the graviton
pole.

The allowed representations and couplings for the transitions from the massless
sector to the higher levels follow a similar pattern. 
At level $l$ one obtains  all the $SO(8)$ representations and couplings present at level $l-1$ together with two new
rank-$(l+1)$ $GL(8)$ tensors of symmetry type $(l+1)$ and $(l,1)$
\be \yng(1)\hspace{-0.08cm}\overbrace{\yng(6)}^\text{l boxes} \hspace{2cm} 
\hspace{-0.08cm}\overbrace{\yng(6,1)}^\text{l boxes} \  \ee
These two tensors are not traceless and they generate a series of lower rank irreducible $SO(8)$ tensors
when the traceless and the trace part are separated. The resulting pattern of $SO(8)$ representations in the
light-cone gauge can be compared with the pattern followed by the covariant $SO(9)$ representations
derived in the next section and displayed in Eq.~\eqref{so9irreps}.

In general two string states in the $SO(8)$ representations $r_1$ and $r_2$ can be connected by the eikonal phase 
if they are created by exactly the same fermionic modes and
if there is at least one non-vanishing contraction between a subset (including the empty set) of the bosonic indices of their polarization
tensors such that the two Young diagrams obtained from the original ones by
removing the boxes corresponding to the  contracted fermionic and bosonic indices
consist of a single row. 

Whenever an $SO(8)$ representation $r$ appears in a given level with multiplicity
$c_r$, as it is the case for the two vectors in the second level,  
there will also be $c_r$ linearly independent contractions of the polarization tensors of the initial and final state
and the momentum transfer 
$\bar q$. The possible inequivalent couplings can be read from Eq.~\eqref{taylor}
and have a very simple form. In the case of a transition from the massless sector they are for instance 
\be \e^k \w_{k i_1...1_{n-1}} \bar q^{i_1} ...  \bar q^{i_{n-1}} t^a\ , \hspace{1cm}
\e^k \bar q_k \w_{i_1...1_{n}} \bar q^{i_1} ...  \bar q^{i_n} t^b \ , \hspace{1cm} a, b \in \mathbb{N} \ . \label{r28} \ee
The higher powers of $t$ appear whenever we take the trace in a couple of transverse indices to decompose 
the $GL(8)$ tensors into irreducible $SO(8)$ tensors.

It is worth recalling that, so far, we have only discussed the exponent appearing in the eikonal operator.
When the full operator is considered also 
transitions produced by the repeated action of $\hat \d(s, b)$ should be taken into account.
Moreover, since the eikonal operator is the exponential of a normal-ordered operator, 
to put the exponential itself in normal-ordered form one has to apply the Baker-Campbell-Hausdorff 
formula. As discussed in detail in \cite{Amati:1987uf}, this produces the exponential damping of each exclusive 
transition that is necessary in order to ensure unitarity.

In order to complete our discussion of the properties of the eikonal operator it remains to illustrate 
one more feature, the fact that not all the light-cone states obtained by acting
on an initial state only with the bosonic modes have non-vanishing transition
amplitudes in the Regge limit. In fact the opposite is true, as the number of the partitions of the level increases 
there are more and more linear combinations of light-cone states that decouple at high energy.
This is a consequence of the restricted number of inequivalent amplitudes allowed by the eikonal operator,
cfr. Eq.~\eqref{r28}. 

The simplest example of this decoupling occurs at level three.
The inelastic transitions from the ground state are in this case determined by the operator
\ba
 \D_{3,0} &=& - \frac{1}{6} \si ( \frac{\ap}{2} \de )^{\frac{3}{2}}  \bar q^i \bar q^j   \bar q^k \,  A_{-1}^i  A_{-1}^j A_{-1}^k 
 + \frac{1}{4}\frac{\ap}{2}   \bar q^i \bar q^j \,  A_{-1}^i  A_{-2}^j 
- \frac{1}{3} \sqrt{\frac{\ap}{2}}  \bar q^i  \,  A_{-3}^i \ . \ea
There are two linearly independent $(2,1)$ tensors that can be formed using the states
\be  A_{-1}^i A_{-2}^j B^k_{-\frac{1}{2}}|0 \pd \ . \ee
Consider the following orthonormal basis
\ba |\w^{2,1}\pd_1 &=& \frac{1}{\sqrt{2}} \w^{(2,1)}_{ij;k}  \si ( A_{-1}^i A_{-2}^j  - A_{-1}^j A_{-2}^i \de )  B^k_{-\frac{1}{2}}|0 \pd \ , 
\nb \\
 |\w^{2,1}\pd_2 &=& \frac{1}{\sqrt{6}} \w^{(2,1)}_{ij;k}  \si ( A_{-1}^i A_{-2}^j  + A_{-1}^j A_{-2}^i \de ) 
B^k_{-\frac{1}{2}}|0 \pd \ .  \ea
The inelastic high-energy amplitudes are 
 \be  {}_1\ps \w^{2,1}| \D_{3,0} |\e\pd = 0 \ , \hspace{1cm} 
{}_2\ps \w^{2,1} | \D_{3,0} |\e\pd =  \frac{\ap}{2\sqrt{6}} \qb^i \qb^j \w^{(2,1)}_{ij;k} \e^k \ , \ee
and therefore only the state $ |\w^{2,1}\pd_2$ is produced at high energy, consistently with the fact
that at level three there is only one independent contraction for tensors of type $(2,1)$.

The examples discussed so far show that
the inelastic scattering amplitudes in the Regge limit can be easily derived using the eikonal operator. 
What is still missing in order to fully characterize a given transition is the
knowledge of the linear combinations of covariant states that corresponds to the light-cone states
 taking part in the collision.
One way of deriving  this information is to study the behaviour of the light-cone states, which are 
labeled by their transformation properties under the
transverse rotation group $SO(8)$, with respect to  the action of the
full ten-dimensional Lorentz group. In this section we shall present a different method which is based on the
DDF operators~\cite{DelGiudice:1971fp}, reviewed in App. \ref{app:DDF}, since this method
makes the relation between the description of the high-energy dynamics
in the light-cone gauge and the covariant description discussed in the next section more direct. 

The method works in the following way. Let us assume that all the covariant vertex operators 
 in the string spectrum at level $l$ are known and let us decompose them with respect to the transverse
$SO(8)$.
Consider now a light-cone state at level $l$. Using the one-to-one correspondence between 
the light-cone modes and the DDF operators, we interpret it as a covariant state created by the operators 
$A_{-n, j}$ and $B_{-r, j}$. Performing then the integrals in Eq.~\eqref{ABigen} we  obtain an explicit expression 
for the state in terms of the modes of the worldsheet fields $X^\m$ and $\psi^\m$. From this expression 
we can identify the linear combination of $SO(8)$ components of the covariant vertices that corresponds to the
given light-cone state. We will discuss some explicit examples of this 
method at the end of section \ref{covariant-eikonal}.

\section{The eikonal phase from the covariant dynamics}
\label{covariant}

The string eikonal operator gives a remarkably compact description of
the high-energy dynamics in the Regge limit. It constrains the class
of states that can be excited in a string-string or string-brane
collision as well as the form of the interaction vertices. As shown in
section \ref{gssec}, the eikonal operator becomes very simple when
written in  a light-cone gauge
adapted to the high-energy process under study. The aim of this
section is to describe the covariant dynamics that gives rise to such
a simple operator.

From the covariant point of view the dynamics simplifies because in
the Regge limit the dominant interactions among strings and D-branes
are those mediated by the exchange in the $t$-channel of the states of
the leading Regge trajectory. The exchange of this set of states leads
to the characteristic Regge behaviour $(\ap s)^{a(t)}$ displayed by
the tree-level string scattering amplitudes in the limit of large
energy $s$ and fixed momentum transfer $t$.  This universal variation
of the amplitudes with a $t$-dependent power of the energy can be
understood as due to the exchange in the $t$-channel of an effective
string state, the Reggeon.

In the Regge limit the integrals that define the tree-level string
amplitudes are dominated by the region of small worldsheet
distances. This fact has the important consequence that it is possible
to describe the emission of a Reggeon in terms of a local vertex operator ${\cal V}_R$
\cite{Ademollo:1989ag, Ademollo:1990sd, Brower:2006ea} and to write
any scattering amplitude for a process involving two sets of states at
large relative boost in a simple factorized form.  Consider for
instance a four-particle process with particles $1$ and $2$ in the
initial state and particles $3$ and $4$ in the final state.
In the Regge limit, defined as $|t'|=|p_1+p_3|^2\ll |s'|=|p_1+p_2|^2$,
this amplitude is
\be 
{\cal A}(s', t') \sim \Pi_R \, C_{1 3
  R}(s', t') \, C_{2 4 R}(s', t') \ , \label{r0} 
\ee 
where $\Pi_R$ denotes the Reggeon propagator, which is process 
independent, while $C_{1 3 R}$ and $C_{2 4 R}$ denote the couplings 
of the Reggeon to the two sets of states, which can be easily 
derived by evaluating the corresponding
three-point functions with the vertex ${\cal V}_R$.  Similarly, for a
two-point amplitude in the background of a collection of Dp-branes,
the main focus of this paper, one has 
\be {\cal A}(s, t) \sim
\Pi^{D_p}_R \, C_{1 2 R}(s, t) \, \bar C_{1 2 R}(s, t) 
\ . \label{r1} 
\ee 
In this case the Reggeon is absorbed by the D-brane and the function
$\Pi^{D_p}_R$ represents the Reggeon tadpole in the D-brane
background, which is again independent of the external states, while
$C_{1 2 R}$ and $\bar C_{1 2 R}$ are the holomorphic and
the antiholomorphic parts of the three-point couplings of the two
external states to the Reggeon.

In the following we will review the construction of the Reggeon vertex operator for
the superstring. 
We will then apply this formalism to derive the Regge limit of string amplitudes with massive states.
When evaluating the high-energy limit, it is important to remember
that S-matrix elements involving the longitudinal polarizations of a
massive particle contain, for purely kinematic reasons, additional
powers of the energy, as shown in Eq.~\eqref{eq:vi}.
We will explain how these factors of the energy are taken into proper account 
by the contractions of the Reggeon vertex with the tensor part of the physical vertex operators
of the massive string states. 

Since the  high-energy scattering at fixed momentum transfer 
is always dominated by the exchange of the Reggeon, all tree-level amplitudes will share the same $\Pi_R$
or $\Pi^{D_p}_R$ and therefore the same energy dependence.
The different inelastic processes will be distinguished by the couplings of the Reggeon to 
the external states which 
can be expressed as the contraction of the external polarizations with a tensor formed using the metric,
the momentum transfer $q$ and the longitudinal vector $v$. We will give explicit examples of these
tensors for all the transitions from the massless NS sector to the first two massive levels of the
superstring.

Once the high-energy dynamics has been formulated in terms of Reggeon exchange, it is possible to
provide a simple covariant derivation of the eikonal phase.  This derivation shows in a 
clear and direct way how the operator $\hat \d(s, b)$ emerges from
the full covariant dynamics.
The covariant equivalent of the 
eikonal phase is the operator that associates to every couple of physical string states 
their three-point function with the Reggeon. While these couplings have a somewhat complex
form when the external states are chosen to transform in irreducible representations of the covariant little group,
they become elementary if one chooses a basis adapted to the kinematics of the high-energy scattering,
the basis of the DDF operators. In this basis the tree-level scattering matrix in the Regge limit takes
a compact form that, as we will show, coincides with $\hat \d(s, b)$ in Eq.~\eqref{eikgs1}, with the modes
of the fields $\hat{X}$ identified with the bosonic DDF operators.

\subsection{The Reggeon vertex  operator}

The general form of a string amplitude in the Regge limit displayed in
Eqs.~(\ref{r0})--(\ref{r1}) can be derived by first noticing that the
worldsheet integral is dominated by the region of small distances
\cite{Ademollo:1989ag, Ademollo:1990sd, Brower:2006ea} 
and then by analysing the energy dependence of the relevant
factorization channel. The relation between the Regge limit and the
limit of short worldsheet distances is a consequence of the fact that
the essential dependence of a string amplitude on the external momenta
is contained in the correlation function of the exponential part
(${\rm e}^{\ii k X}$) of each vertex operator.
For a four-particle process the dependence
of this correlator on the Mandelstam variables is for instance 
\be {\rm e}^{-\frac{\ap t'}{4} \ln |z|^2 - \frac{\ap s'}{4} \ln | 1 - z|^2} \ , \hspace{1.4cm} z = \frac{z_{13}z_{24}}{z_{14}z_{23}} \ . \ee
When $s'$ is large and $t'$ finite the integral 
over $z$ is dominated by a neighbourhood of the origin of size $ \ap s' |z|^2 \le 1$ .

The class of states that contribute at leading order in energy can
then be easily identified and shown to be process independent by
factorizing the amplitude in the $t$-channel on a complete set of
string states. The contribution of the states of level $l$ are
suppressed by a factor of $|z|^{2 l}$, where $z$ is the parameter
controlling the factorization in the $t$-channel. As $|z|^2$ is of
order $1/(\alpha' s)$ in the Regge limit, we require that the increase
in the power of $z$ with the mass level is compensated by an equal
increase in the power of $s$ in the three-point couplings. The final
step is to show that the sum over the intermediate states together
with the integral over the worldsheet are equivalent to the insertion
of a single
local operator, the Reggeon, in the three-point couplings with the external states.

Let us focus on the case of a two-point amplitude on the disc, but the
result, as it should be clear from the discussion, is
general~\cite{Brower:2006ea}.  We start from the following
representation of the disc amplitude with two closed strings\footnote{In our conventions $d^2 z= 2  d{\rm Re}z d{\rm Im}z$.} 
\be
A_{12} = \frac{\ap}{8\pi} \int d^2z \, \ps 0 | {\cal V}^{(-1,
  -1)}_{(S_1, \bar S_1)}\, {\cal V}^{(0, 0)}_{(S_2, \bar S_2)}
z^{L_0-1} \bar z^{L_0-1} |D_p \pd \ , \label{r2} \ee 
where on the left
there is the $SL(2,\mathbb{C})$-invariant vacuum state and on the
right the boundary state corresponding to the collection of $N$ D$p$-branes.\footnote{We consider here explicitly the case of a transition between states
of the NS sector. The discussion for the transitions between states of the
R sector is similar, with the external vertices both in the $-\frac{1}{2}$ picture.
Transitions between the NS sector and the R sector are subleading in energy since
at  level $l$ the highest spin that can be exchanged in the $t$-channel   in the R sector
is equal to $l+\frac{1}{2}$ and then lower than the spin $l+1$ of the states of the leading Regge trajectory. }
As usual, the vertex operators
${\cal V}$ factorize in a holomorphic and an antiholomorphic component
and, following a standard notation, the superscripts on the vertices
refer to the picture of each component. For a closed string state at
level $l$ labeled by the left and right $SO(9)$ representations $(S,
\bar S)$ we write 
\ba {\cal V}_{(S, \bar S)} &=&  \frac{\kappa_{10}}{2 \pi} \, V_S 
\, \bar V_{\bar S} = \frac{\kappa_{10}}{2 \pi}   \, \e_{\m_1...\m_r} \, V_S^{\m_1...\m_r}   \ \
 \bar \e_{\n_1...\n_s} \,  V_{\bar S}^{\n_1...\n_s}   \ ,  \label{vo} 
\ea
where we extracted from the normalization of the vertex operators an explicit factor $\frac{\kappa_{10}}{2 \pi}$.
The polarization tensors are normalized
as $\e_{\m_1...\m_r} \e^{\m_1...\m_r} = \bar\e_{\m_1...\m_r} \bar\e^{\m_1...\m_r} = 1$ and
the overall normalization of the vertices is fixed by the requirement that the corresponding state
has unit norm. Finally left-right symmetric states will be simply written as ${\cal V}_{S}$.

Let us insert a complete set of string states  in the picture $(-1, -1)$, labeled by a triple $(l, n_l, \bar n_l)$ where
$l$ is the level and $n_l$, $\bar n_l$ the set of all the remaining left and right
quantum numbers needed to identify the state, including its momentum. 
The disc amplitude then becomes a sum of products of 
 three-point couplings  on the sphere and  one-point functions on the disc with D$p$-brane boundary conditions
\be  A_{12} = \frac{\ap}{8\pi}   \sum_{(l, n_l,  \bar n_l)} \int d^2z \, \si (z \bar z \de)^{l-1-\frac{\ap t}{4}} 
\si \ps  {\cal V}^{(-1, -1)}_{(S_1, \bar S_1)} \,  {\cal V}^{(0, 0)}_{(S_2, \bar S_2)} {\cal V}^{(-1, -1)}_{l, n_l, \bar n_l}   \de \pd_{{\cal S}}
\si \ps  {\cal V}^{(-1, -1)}_{l, n_l, \bar n_l}   \de \pd_{D_p} \
. \label{r13}  
\ee
In order to identify the states that give the leading contribution in the Regge limit we need to determine the
scaling with the energy of  the three-point correlators. A 
state can contribute to the Regge limit only if the power of the energy in its
three-point couplings with the external states matches its level. The only operators 
with this property are the operator ${\cal Q}_l$ whose  holomorphic part is
\be Q_l = \frac{1}{\sqrt{l!}} 
\psi^+ \si (\ii\sqrt{\frac{2}{\ap}} \p X^+ \de )^l
{\rm e}^{\ii k X }\ , \hspace{1cm} l \ge 0 \ , \label{r14} 
\ee
where $X(z,\bar{z}) = X(z)+\bar X(\bar{z})$ and $X^{+} = e^+ X$, $\psi^{+} = e^+ \psi$ are the components of the string fields along the light-cone directions defined in Eq.~\eqref{eq:epm}.
In order to see this note that there are only two possible sources for the
factors of the energy in the three-point couplings. The first one is
related to the contractions of the operators $\p^r X^+$ in the vertex
of the intermediate state with the exponential part of the external states 
\begin{eqnarray}
\sqrt{\frac{2}{\alpha'}} \frac{\p^r X^+(z)}{\sqrt{\alpha'} E} \;
{\rm e}^{\ii p_1 X_1 (w)}  \sim \p^{r-1}_z \left(\frac{1}{z-w}\right) 
{\rm e}^{\ii p_1 X_1 (w)} ~, \\
\sqrt{\frac{2}{\alpha'}} \frac{\p^r X^+(z)}{\sqrt{\alpha'} E} \;
{\rm e}^{\ii p_2 X_2(w)}  \sim -\p^{r-1}_z \left(\frac{1}{z-w}\right) 
{\rm e}^{\ii p_2 X_2(w)}~.
\end{eqnarray}
The second possibility follows from the contractions of the operators
$\p^r X^+$ or $\p^r \psi^+$ with the tensor part of the external states 
\begin{align}
\ii \sqrt{\frac{2}{\alpha'}} \frac{\p^r X^+(z)}{\sqrt{\alpha'}E}\, 
\ii \sqrt{\frac{2}{\alpha'}} \frac{\p^s X^\r(w)}{\sqrt{\alpha'}E} 
& \sim \sqrt{\frac{2}{\alpha'}} \frac{v^\r}{m} \p^{r-1}_z \p^{s-1}_w \left(\frac{1}{z-w}\right)^2 ~\ ,  \label{vm1}
\\   
\frac{\p^r \psi^+(z)}{\sqrt{\alpha'}E}  \p^s \psi^\r(w) & \sim \sqrt{\frac{2}{\alpha'}}  \frac{v^\r}{m} \p^r_z \p^s_w \left(\frac{1}{z-w}\right) \ , 
\label{vm2}
\end{align}
where $v$ is the longitudinal polarization and $m$ the mass of the
state. This second class of contractions is very important for all the
massive states of the string spectrum.  Since in both cases one can
only obtain one factor of $E$, all the operators that contain
derivatives of $X^+$ higher than the first or fermionic fields in
addition to $\psi^+$ give a subleading contribution, because the power
of $z$ in Eq.~\eqref{r13} grows faster than the power of the energy in
the coupling. The result of this analysis is that the class of
operators that contribute to the Regge limit is universal, independent
on the external states and is given by Eq.~\eqref{r14}.

Since all other contractions yield subleading contributions, we can
then restrict the sum over the intermediate states in Eq.~\eqref{r13}
to the states of the leading Regge trajectory with all the
polarization indices in the $x^+$ direction. Notice that the amplitude
for their exchange has the largest power of the energy compatible with
helicity conservation. For this class of states the one-point function
on the disc in~\eqref{r13} is independent of $l$
\be \label{TpkR}
 \, \ps {\cal Q}_l \pd_{D_p} = \frac{\t_p \kappa_{10} N}{2} = \frac{1}{\kappa_{10}}  \frac{\pi^{\frac{9-p}{2}}}{\Gamma(\frac{7-p}{2})}
\, R_p^{7-p} \ , \ee
where $R_p$ is the scale of the D$p$-brane background, given in Eq.~\eqref{Rp}, and $\t_p$ the tension
of a single D$p$-brane, given in Eq.~\eqref{dptension}. 
In conclusion, we find
\ba &&  {\cal A}_{12} =\frac{\ap}{8\pi \kappa_{10}} \frac{\pi^{\frac{9-p}{2}}}{\Gamma(\frac{7-p}{2})}
\, R_p^{7-p} \,
\sum_{l=0}^{\infty} \int d^2z \, \si (z \bar z \de)^{l-2-\frac{\ap t}{4}} \, \,
\si \ps  {\cal V}^{(-1, -1)}_{(S_1, \bar S_1)} {\cal V}^{(0, 0)}_{(S_2, \bar S_2)} 
{\cal Q}_l  \de \pd_{{\cal S}} \label{eq:515a}
 \ . \ea
It is precisely  the inclusion of this infinite series of operators  that gives rise to the Regge behaviour of the amplitude
$(\ap s)^{\a(t)}$ \cite{Ademollo:1989ag, Ademollo:1990sd,
  Brower:2006ea}, while the  leading singularity in $t$ accounts only
for the graviton pole. 
The final step in the derivation is to express the previous sum over the three-point couplings 
of the intermediate states as a single three-point coupling with an effective vertex operator.
This can be achieved by first summing the exponential series and then
performing the integral over $z$. The result can be written as follows
\be  \label{eq:515}
{\cal A}_{12} = \Pi^{D_p}_R(t) 
 \si \ps  V^{(-1)}_{S_1}  V^{(0)}_{S_2}
 V^{(-1)}_R   \de \pd \, \si \ps   \bar V^{(-1)}_{ \bar S_1}  \bar V^{(0)}_{ \bar S_2}
\bar V^{( -1)}_R \de \pd   \ , 
\ee
where the Reggeon vertex for the superstring in the $-1$ picture is
\ba \label{eq:rv-1}
V^{(-1)}_R = \frac{\psi^+}{\sqrt{\alpha'}E}
\si (  \sqrt{ \frac{2}{\ap}}  \frac{\ii \p X^+}{\sqrt{\alpha'}E} 
\de )^{\frac{\ap t}{4}} \!\!  {\rm e}^{-\ii q X} \ .
\ea 
A factor of $\sqrt{\alpha'} E$ for each string field along $e^+$ has been included in $V_R$ so as to make the sphere correlator in 
Eq.~\eqref{eq:515} energy independent. Then the dependence on the energy, which is universal, is included in the Reggeon tadpole which, in our case, coincides 
with the function ${\cal A}$ introduced in Eq.~\eqref{lep}
\be  
\Pi^{D_p}_R = \frac{\pi^{\frac{9-p}{2}}}{\Gamma(\frac{7-p}{2})}
\, R_p^{7-p} \, \Gamma \si (-\frac{\ap t}{4} \de ) {\rm e}^{-\ii \pi \frac{\ap
    t}{4}} (\alpha' s)^{1+\frac{\alpha' t}{4}}\ . 
\ee
In writing Eq.~\eqref{eq:515}, we also extracted for convenience 
from the three-point couplings the factor that gives the normalization of the sphere amplitude setting
\be \si \ps {\cal O} \de \pd \equiv \frac{\ap \kappa_{10}^2}{32 \pi^3} 
\si \ps {\cal O} \de \pd_{{\cal S}} \ . \ee
As anticipated, the two-point function on the disc has  been expressed as the product of a single three-point
coupling on the sphere and the Reggeon tadpole on the brane.

The OPEs with the energy-momentum tensor and with the supercurrent
show that the Reggeon vertex~\eqref{eq:rv-1}, although it carries an
off-shell momentum $q$, behaves as a superconformal primary of
dimension one half at high energies \cite{Ademollo:1990sd} (i.e. terms
in the OPEs violating the superconformal invariance exist, but are
suppressed by powers of $1/E$).  This guarantees that, for our
purposes, the correlation functions of the Reggeon with physical
string states are well-defined and in particular invariant under
global superconformal transformations. Also we can use the OPE with
the supercurrent to obtain the Reggeon vertex in the $0$ picture
\begin{equation}
V^{(0)}_R \!= \! \si [  - 
\frac{2}{\ap} \frac{\p X^+ \p X^+}{\alpha' E^2} - \ii q \psi 
\frac{\psi^+ \p X^+}{\alpha' E^2} 
- \frac{\ap t}{4}  \frac{\psi^+  \p \psi^+}{\alpha' E^2} \de ]
\si (\sqrt{ \frac{2}{\ap}} \frac{\ii \p X^+}{\sqrt{\alpha'}E} 
\de )^{\frac{\ap t}{4}-1} \!\!\! {\rm e}^{-\ii q X_L} \,.  
\label{rv0}  
\end{equation}
In explicit calculations of the 3-point correlators in~\eqref{eq:515},
it is convenient to move the superghost charge from the Reggeon to the
$(S_2,\bar{S}_2)$ vertex so as to have both external states in the
simpler $(-1,-1)$ picture and use Eq.~\eqref{rv0} for the Reggeon
vertex.

The factorization of the scattering amplitudes and the form of the Reggeon vertex result in some simple selection 
rules for the inelastic transitions in the Regge limit, which represent the covariant version of the selection rules 
discussed in section \ref{eikonal} for the eikonal phase. 
In order to identify the $SO(9)$ representations that can be related by the exchange of the Reggeon
and the structure of the corresponding couplings, let us first analyse the possible contractions of the
fermionic fields. The first term in Eq.~\eqref{rv0} can only connect components of 
the vertex operators of the external states with the same number of fermionic fields,
while the second and the third also components containing two additional fermionic fields. 
When the variation in the number of fermionic fields is due to the second term, one of the
two additional indices of the polarization tensor must be longitudinal. When it is due to the third term, both 
additional indices  must be longitudinal. 
In order to have a non-vanishing three-point coupling all the remaining fermionic fields must be contracted
among the two external states, which results in the contraction of the corresponding indices of the two
polarization tensors. 

As for the bosonic fields, there are three types of contractions. They can be contracted between the two external states,
giving the contraction of the indices of the external polarizations, or with the exponentials,
giving the contraction of the polarizations with the momentum transferred $q$,
or with the $\p X^+$ operators in the Reggeon vertex, giving the contraction of the polarization with the longitudinal vector $v$.

As a result, two external states transforming in the $SO(9)$ representations  with Young diagrams 
$Y_1$ and $Y_2$
can be connected by Reggeon exchange if the following two conditions are fulfilled.
First of all they should respect the selection rules described above for the fermionic indices.
Moreover there should be at least one non-vanishing contraction between a subset (including the empty set)
of the bosonic indices of their polarizations such that, when the boxes corresponding to the contracted bosonic
indices and to the fermionic indices are removed from the original diagrams,  
the two resulting Young diagrams have at most two rows. 

Let us apply these selection rules to the case of the inelastic transitions from the ground state to the massive states
at level $l$. The exchange
of a Reggeon in the $t$-channel allows to excite only the following three classes 
of $SO(9)$ representations with one, two or three rows
\be  \begin{array}{lll}  \yng(1)\hspace{-0.08cm}\overbrace{\yng(6)}^\text{n boxes} & \hspace{2cm}
\yng(1,1)\hspace{-0.08cm}\overbrace{\yng(6,3)}^\text{n boxes} &  \hspace{2cm}
\yng(1,1)\hspace{-0.08cm}\overbrace{\yng(6,3)}^\text{n boxes}  \vspace{-0.3cm} \\ 
&  \hspace{2cm} \hspace{0.27cm}\underbrace{\hspace{0.9cm}}_\text{m boxes} &
 \hspace{2cm}  \hspace{0.27cm}\underbrace{\hspace{0.9cm}}_\text{m boxes} \vspace{-0.72cm} \\
& &  \hspace{2cm} \yng(1)  \\ & & \end{array} \label{so9irreps}  \ee
For the first class there are no constraints
on the indices of the massive polarization tensor and the integer $n$ can take all the values from zero to $l$. 
For the second and the third class the integers $n$ and $m$, with $n \ge m$, can take all the values from zero to $l-1$
such that $n+m \le l-1$.  The polarization index in the third row and all the polarization
indices in the second row, except the first one, must be longitudinal.
The values of the integers $n$ and $m$ that actually appear at level $l$ depends on the physical spectrum
of that level, which can be read for instance from the generating function discussed in \cite{Hanany:2010da}.

As a final remark, let us emphasize that the discussion above is limited to tree level and therefore
to the exchange of a single
Reggeon. The leading high-energy contributions of the higher-genus surfaces which are
resummed by the eikonal operator take into account the exchange of an arbitrary number 
of Reggeons. As already mentioned in section \ref{eikonal} this has two main consequences:
on one hand one should take into account transitions mediated by multiple Reggeon exchange and therefore
given by the repeated application of the previous rules,
on the other hand every exclusive transition will be exponentially suppressed as required by unitarity.

\subsection{Regge limit of the covariant string amplitudes with massive states}

We now show how the Reggeon vertex considerably simplifies the evaluation
of the Regge limit of the string amplitudes and discuss how to take into proper account
the factors of the energy carried by the longitudinal polarizations of the massive states.

In the evaluation of the three-point couplings it is convenient to take both external states in the $-1$
picture and the Reggeon in the $0$ picture.
We write the three-point couplings as the product of an holomorphic and anti-holomorphic part
\ba C_{(S_1, \bar S_1), (S_2, \bar S_2), (R, R)}  &=& \frac{\kappa^3_{10}}{(2 \pi)^3} \, 
C_{S_1, S_2, R}    \bar  C_{\bar S_1, \bar S_2, R}     \ , \ea
where 
\be  C_{S_1, S_2, R} = \ps V_{S_1}(p_1) V_{S_2}(p_2) V_R \pd \ , \ee
and similarly for the anti-holomorphic part. We will also isolate the tensor structure which is specific to a given process by writing 
\be C_{S_1, S_2, R} = 
 \e_{\m_1...\m_r} \z_{\n_1...\n_s} T^{\m_1...\m_r;\n_1...\n_s}_{S_1,
   S_2, R}  \, \ . 
\ee 
Here and in the following the polarizations of the initial and of the final states
will be denoted by $\e$ and $\z$, respectively. For compactness, we will often write directly the tensors $T_{S_1, S_2, R}$ 
rather then the couplings $C_{S_1, S_2, R}$. However, the contraction with the external polarizations
 will   always be understood and we will make use of
momentum conservation, of the mass-shell condition and of the physical state conditions to simplify our 
equations. 

In the following we will derive the inelastic amplitudes for transitions from the NS-NS ground state to the first two massive
levels of the string spectrum, which is equivalent to evaluating the holomorphic couplings
with the Reggeon of one massless and one massive state. 
Once we know the holomorphic couplings, we know all the amplitudes for a given level
due to the factorized form of the scattering amplitudes
\be 
{\cal A}_{g, (S,\bar S)} =  \Pi^{D_p}_R C_{g, S, R} \bar C_{g, \bar S, R} \ . 
\ee
Let us begin with  the elastic amplitude. The massless vertex in the $-1$ picture is
\be V_g^\mu(p) =   \psi^\m {\rm e}^{\ii p X} \, {\rm e}^{- \f} \
, \label{mvg} 
\ee
and we need to evaluate
\ba && T^{\m;\r}_{g, g, R} = 
\ps V_g^\mu(p_1) V_g^\r(p_2) V_R \pd  \ .\ea
In this case only  the contractions of the
Reggeon with the exponential part  of the vertex operators give a non-vanishing result and we find
\be  T^{\m;\r}_{g, g, R}  =  \h^{\m\r} \ . \ee
The Regge limit of the elastic amplitude is therefore
\ba &&  {\cal A}_{g, g} =  \Pi^{D_p}_R C_{g, g, R} \bar C_{g, g, R}
 =  \e_{\m\n} \z^{\m\n} \,  \frac{\pi^{\frac{9-p}{2}}R^{7-p}}{\Gamma\left(\frac{7-p}{2}\right)} 
 \,  {\rm e}^{-\ii \pi \frac{\ap t}{4}} (\ap s)^{1+\frac{\ap t}{4}} \Gamma \si (-\frac{\ap t}{4} \de ) 
   \ . 
\ea
The three-point functions for transitions to the first and to the second level will display a more interesting 
structure.  They are given by  tensors $T_{g, S, R}$  formed from the metric $\h$ and the vectors $q$ and $v$ and
reflecting the symmetry properties of the external states. 
It is worth noticing that the covariant amplitudes
depend on the masses of the external states also in the high energy limit, 
as a consequence of the contractions in \eqref{vm1} and \eqref{vm2} and the kinematic relations
in \eqref{eq:q-vHE} and \eqref{eq:epsexp}. The masses can be eliminated and the results expressed 
only in terms of $\ap$ 
using the relation $m^2 = \frac{4l}{\ap}$ for states at level $l$.

Due to the physical state conditions, the
tensors $T_{g, S, R}$ can be rewritten in terms of $SO(8)$ tensors living in the space orthogonal to the collision axis.
If the polarizations of the external states are similarly decomposed into irreducible $SO(8)$
components, each covariant amplitude generates a class of amplitudes labeled by $SO(8)$ representations.
It is after this decomposition that 
the selection rules and the simple structure of the high-energy limit of the S-matrix elements  predicted by the eikonal 
operator and discussed in section \ref{eikonal} become evident, as we will explain in detail in section \ref{covariant-eikonal}.

\subsubsection{Transitions to the first level}

The simplest  inelastic transitions are those from the massless sector to the first massive level.
The $NS$ sector of the first massive level  contains $128$  physical states: a rank-2 traceless totally symmetric tensor 
$S_2$ ($44$ components) and a rank-3
totally antisymmetric tensor $A_3$ ($84$ components). 
The corresponding  vertex operators in the $-1$ picture are
\ba V^{\r\a}_{S_2} &=& 
\ii \sqrt{ \frac{2}{\ap}} \,  \, \psi^{\r} 
 \p X^{\a }  {\rm e}^{\ii p X} \, {\rm e}^{- \f} \ , \nb \\ 
V^{\r\a\g}_{A_3} &=&  \, \frac{1}{\sqrt{3!}} \,  \,
 \psi^\r \psi^\a \psi^\g {\rm e}^{\ii p X} \, {\rm e}^{- \f }  \ . \label{mvfl} \ea
Let us evaluate 
\ba && T^{\m;\r\a}_{g, S_2, R} =   
\ps V_g^\mu(p_1) V^{\r\a}_{S_2}(p_2) V_R \pd \ .\ea
Since this is the first example where it is important to take into account
the longitudinal polarizations of a massive state, we describe this calculation
in some detail. In this case not only the contraction of the Reggeon with the exponential part of the vertex operators 
are relevant at high energy but also the contractions with their tensor part. We find\footnote{In this subsection $m$ indicates the mass of the string states at the first massive level, $m^2 =4/\alpha'$.}
\ba  && T^{\m;\r\a}_{g, S_2, R} =  - \sqrt{\frac{\ap}{2}} \,  
\si [ \h^{\m\r} \si ( q^\a - \frac{2}{\ap} \si ( 1 +
\frac{\ap t}{4}  \de ) \frac{v^\a}{m} \de ) + \frac{q^\m}{m} v^\r 
 \si ( q^\a - \frac{t}{2 m} v^\a \de ) \de ] \ . 
\ea
The first term in the previous expression comes from the first term in the Reggeon vertex, contracted both
with the exponentials and with the $\p X^\r$ of the massive state. The second term 
comes from the second term in the Reggeon vertex, the $q \psi$ part contracted
with the massless vertex  and the $\psi^+ \p X^+$ part with the massive vertex. The third term in the
Reggeon vertex clearly does not contribute to this transition. 
We can write
\ba  && T^{\m;\r\a}_{g, S_2, R} = - \sqrt{\frac{\ap}{2}} \,  
\si [ \si ( \h^{\m\r}+ \frac{q^\m}{m} v^\r  \de ) \si ( q^\a - \frac{m}{2}\si (  1 +
\frac{\ap t}{4}\de ) \, v^\a  \de )+ \frac{q^\m}{2} v^\r v^\a \de ] \ . \label{s2c}\ea
We can simplify further this expression defining 
\be Q^\a = q^\a - 
\frac{ t}{2 m} \, v^\a \ , \hspace{1cm} \bar q^\a = q^\a - \frac{m}{2}\si (  1 +
\frac{ t}{m^2}\de ) \, v^\a \ , \hspace{1cm}
\d_\perp^{\m\r} = \h^{\m\r}+ \frac{q^\m}{m} v^\r  \ . \label{tten} \ee
When contracted with the physical polarization tensor of the massive state, the vector $\bar q$ coincides with the momentum
transfer in the directions transverse to the collision axis, see Eq.~\eqref{eq:q-vHE}; similarly Eq.~\eqref{eq:epsexp} ensures that the tensor $\d_\perp$ reduces to the Kronecker delta in the transverse directions when the first index
is contracted with the massless polarization and the second with the massive one. In the following we will use the same symbol
$\d_\perp$ for both 
the Kronecker delta in the transverse directions and the tensor defined
in Eq.~\eqref{tten}.
By using the symmetry properties of the polarization $S_2$, the final result can be written 
as~\footnote{Groups of indices are symmetrized or antisymmetrized always with weight one.
For instance  $A^{(\rho} B^{\alpha)} = \frac{1}{2} \left(A^{\rho} B^{\alpha} + A^{\alpha} B^{\rho}   \right)$ 
and $A^{[\rho} B^{\alpha]} = \frac{1}{2} \left(A^{\rho} B^{\alpha} - A^{\alpha} B^{\rho}   \right)$.}
\ba  && T^{\m;\r\a}_{g, S_2, R} = -  \sqrt{\frac{\ap}{2}} \,  
\si [ \d_\perp^{\m(\r}\bar q^{\a)} + \frac{q^\m}{2} v^\r v^\a \de ] \ . \label{s2t} \ea
The three-point coupling with the state $A_3$ can be derived following the same steps.
In this case only the second term in the Reggeon vertex contributes and the result is
\ba  && T^{\m;\r\a\g}_{g, A_3, R} = \frac{\sqrt{6}}{m} \h^{\m[\r} q^\a v^{\g]} \ , \label{a3c} \ea
which can be rewritten, using the antisymmetry in the indices $\r, \a, \g$ , as follows 
\ba  && T^{\m;\r\a\g}_{g, A_3, R} = \frac{\sqrt{6}}{m} \d_\perp^{\m[\r} \bar q^\a  v^{\g]}\ . \label{a3t} \ea

\subsubsection{Transitions to the second level}

The $NS$ sector of the second massive level  contains  $1152$ bosonic
physical states in the following six irreducible representations of $SO(9)$ 
\be \yng(3) \hspace{1cm} \yng(2,1,1) \hspace{1cm} \yng(2,1) \hspace{1cm} \yng(1,1,1,1,1) \hspace{1cm}
\yng(1,1) \hspace{1cm} \yng(1)  \ .  \ee
The corresponding normalized vertex operators in the $-1$ picture are \cite{Bianchi:2010es} 
\footnote{In the list given in \cite{Bianchi:2010es} there are typos in the vertices for $U$ and $V$.}
 \ba
V^{\r\a\g}_{Z} &=& - \frac{1}{\sqrt{2}}  \frac{2}{\ap} \p X^\r \p X^\a
\psi^\g \, {\rm e}^{\ii p X} \, {\rm e}^{-\f} \ , \nb \\
V^{\r\a\g\w}_{Y} &=&  \ii \sqrt{\frac{3}{8}} \sqrt{\frac{2}{\ap}} \p
X^{(\r}  \psi^{\a)} \psi^\g \psi^\w \, {\rm e}^{\ii p X} \, {\rm e}^{-\f} \ , \nb \\
V^{\r\a\g}_{U} &=& - \frac{1}{\sqrt{6}} \si [\frac{2}{\ap} \p X^\r \p
X^\a \psi^\g   + 2 \p \psi^{(\r} \psi^{\a)} \psi^\g \de ] \, {\rm e}^{\ii p X}
 \, {\rm e}^{-\f}\ , \nb \\ \label{mvsl} 
V^{\r\a\g\w\xi}_{X} &=& \frac{1}{\sqrt{5!}}  \, \psi^\r \psi^\a
\psi^\g \psi^\w \psi^\xi \, {\rm e}^{\ii p X}  \, {\rm e}^{-\f}\ , \\ 
V^{\g\w}_{V} &=&  \frac{\ii}{5}\sqrt{\frac{2}{7}} \sqrt{\frac{2}{\ap}}
\si [ \si ( \hat \h_{\r\a} \p X^\r  \psi^\a \de ) \psi^\g \psi^\w   -
\frac{7}{2} 
\si ( \p^2 X^\g \psi^\w  - 2 \p X^\g \p \psi^\w \de )  \de ] \, {\rm
  e}^{\ii p X}  \, {\rm e}^{-\f} \ , \nb \\ 
V^\g_{W} &=&  \frac{1}{8\sqrt{22}} \hat \h_{\a\r} \si [- \frac{2}{\ap}
\p X^\g \p X^\a \psi^\r   +  5 \frac{2}{\ap} \p X^\r \p X^\a \psi^\g
+ 11 \p \psi^\r \psi^\a \psi^\g \de ]  \, {\rm e}^{\ii p X}  \, {\rm
  e}^{-\f} \ ,  \nb
\ea 
where $\hat\eta$ is the transverse metric as defined in~\eqref{eq:Minma} and the overall normalizations have been chosen so as to normalize to one the string states $\zeta^{Z}_{\rho\alpha\gamma} V^{\r\a\g}_{Z}$, $\zeta^{Y}_{\rho\alpha;\gamma;\omega} V^{\r\a\g\w}_{Y}$ etc. In this subsection $m$ indicates the mass of the string states at the second massive level, $m^2 = 8/\alpha'$.

In the second level we find the first examples  of tensors of mixed symmetry in the holomorphic superstring
spectrum, namely the states $Y$ and $U$. The vertex $V_Y$ describes a tensor of type $(2, 1, 1)$ 
which can be obtained from a generic tensor by first symmetrizing in
$\r\a$ and then antisymmetrizing in $\r\g\w$. Similarly the vertex $V_U$ describes a tensor of type $(2, 1)$ 
which can be obtained from a generic tensor by first symmetrizing in
$\r\a$ and then antisymmetrizing in $\r\g$. In the list above we showed explicitly only the symmetrization in $\r\a$
of the tensor part of the physical vertices, since the antisymmetrization can be left understood 
due to our convention of choosing the polarization tensors as manifestly
antisymmetric in the column indices. 

The three-point couplings with one massless state, one state $S$ of the second level and the Reggeon can be derived 
following the same steps as for the first level and are completely characterized by the tensors $T_{g, S, R}$. In
the following we will
display for each state of the second level the corresponding tensor, first in terms of the metric $\h$ and the
momentum transfer $q$ and then in terms of the transverse tensors $\d_\perp$ and $\bar q$ defined in 
Eq.~\eqref{tten}. 

According to the selection rules for the transitions mediated by
Reggeon exchange, the totally antisymmetric rank-five tensor $X$ is
not produced at high energy in transitions from the ground state.  
The coupling for the state $Z$ is
\be \sqrt{2} T_{g, Z, R}^{\m;\r\a\g} =   \, \si ( \h^{\m\g} + 
\frac{q^\m}{m} v^\g \de ) \si [  \frac{\ap}{2} Q^\r Q^\a 
  -  \frac{ t}{2 m^2} v^\r v^\a\de ] - \frac{1}{   m}  \h^{\m\g} \si ( Q^\r v^\a + Q^\a v^\r \de ) \ , \label{zc} \ee
and in terms of $\d_\perp$ and $\bar q$
\ba  \sqrt{2} T_{g, Z, R}^{\m;\r\a\g} &=& 
\d_\perp^{\m\g} \si [  \frac{\ap}{2} \bar q^\r \bar q^\a + \frac{1}{m } \si ( \bar q^\r v^\a + \bar q^\a v^\r \de )
  -   \frac{ t}{2 m^2} v^\r v^\a \de ]\nb\\  &+& 
 \frac{\bar q^\m}{m^2}  \si ( \bar q^\r v^\a + \bar q^\a  v^\r \de ) v^\g 
+   \frac{\bar q^\m}{m}  v^\r v^\a v^\g   \ .  \label{zt} \ea
The coupling for the state $Y$  is 
\ba  \label{eq:541}
\sqrt{\frac{8}{3}} T_{g, Y, R}^{\m;\r\a\g\w} 
=   \sqrt{\frac{\ap}{2}} \frac{3}{m} \h^{\m[\a}  q^\w v^{\g]} Q^\r +
\sqrt{\frac{\ap}{2}} \frac{3}{m} \h^{\m[\r}  q^\w v^{\g]} Q^\a \,,
\ea 
which, after using the symmetry properties of $\zeta^Y_{\rho\alpha;\gamma;\omega}$, can be written in terms of $\d_\perp$ and $\bar q$
\be
\sqrt{\frac{8}{3}} T_{g, Y, R}^{\m;\r\a\g\w} =  
 \sqrt{\frac{\ap}{2}} \frac{4}{m}  \d_\perp^{\m[\r}  \bar q^\w v^{\g]}  \bar q^\a 
+  \sqrt{\frac{\ap}{2}} \, 2 \, \d_\perp^{\m[\r} \bar q^\w v^{\g]} v^\a  \ . \label{yt} 
\ee
The coupling for the state $U$  is 
\ba \label{eq:543}
\sqrt{6} T_{g, U, R}^{\mu;\rho \alpha \gamma} & =&   \si ( \h^{\m\g} + 
\frac{q^\m}{m} v^\g \de ) \si [  \frac{\ap}{2} Q^\r Q^\a 
  - v^\r v^\a \frac{ t}{2 m^2}\de ] -  \h^{\m\g}\frac{1}{m} \si ( Q^\r v^\a + Q^\a v^\r \de )  \nb \\
 &-&  \frac{1}{m} \h^{\m\r} \si ( q^\a v^\g - q^\g v^\a\de ) +
 \frac{1}{m}  \h^{\m\a} \si ( q^\g v^\r - q^\r v^\g\de ) \nb \\
&+& \frac{t}{2m^2}  \si (\h^{\m\a} v^\g v^\r + \h^{\m\r} v^\g v^\a - 2 \h^{\m\g} v^\a v^\r \de) \ ,  \ea
where the first line comes from the first term in the vertex $V_U$ and the following two lines from the second term.
Using the symmetry properties of $\zeta_{\rho\alpha;\gamma}^U$ and $m^2=8/\alpha'$, this result can be written as follows 
in terms of $\d_\perp$ and $\bar q$
\ba   
\sqrt{6} T_{g, U, R}^{\m;\r\a\g} &=& \frac{\ap}{4} \qb^\a \si ( \d_\perp^{\m\g} \qb^\r - \d_\perp^{\m\r} \qb^\g \de )
+ \frac{\ap m}{8}\qb^\a\si (\d_\perp^{\m\g}  v^\r -\d_\perp^{\m\r}  v^\g \de )
+ \frac{\ap m}{8}\d_\perp^{\m\a}\si ( \qb^\g v^\r -\qb^\r v^\g \de ) \nb \\
&-& \frac{\ap }{4} \qb^\m v^\a\si ( \qb^\g v^\r -\qb^\r v^\g \de )
- \frac{\ap t}{8} v^\a \si (  \d_\perp^{\m\g} v^\r - \d_\perp^{\m\r} v^\g \de ) 
\label{ut} \ . \ea
Finally let us give the results for the last two vertices in Eq~\eqref{mvsl} 
\ba \sqrt{\frac{7}{2}} \, T_{g, V, R}^{\m;\g\w} &=&  
\frac{1}{2} \sqrt{\frac{\ap}{2}} \h^{\m[\g} q^{\w]} 
- \frac{1}{m} \sqrt{\frac{2}{\ap}} \h^{\m[\g} v^{\w]} \si ( 1 + \frac{\ap t}{2} \de ) + \frac{\ap}{8} q^\m q^{[\g} v^{\w]} \ , \ea
\ba && 8 \sqrt{22} \, T_{g, W, R}^{\m;\g} = q^\m \si ( \frac{5\ap}{8} Q^\g - \frac{v^\g}{2m} \de )
+ 5 \si [ \frac{9 t}{2 m^2} \h^{\m\g} - \frac{q^\m v^\g}{m}\si ( 1 - \frac{9 t}{2 m^2}
\de )   \de ] \nb \\ &+& \frac{ 11 t}{2 m^2} \si ( \h^{\m\g} + \frac{q^\m v^\g}{m} \de )
+  11 \frac{q^\m}{m^2} \si ( Q^\g + \frac{m}{2} v^\g \de ) \ , \ea
and in terms of the transverse quantities  $\d_\perp$ and $\bar q$
\ba\sqrt{\frac{7}{2}} \, T_{g, V, R}^{\m;\g\w} &=& 
 \frac{ \ap}{4} \bar q^\m \bar q^{[\g}  v^{\w]} +
\frac{1}{2} \sqrt{\frac{\ap}{2}}  \d_\perp^{\m[\g}  \bar q^{\w]}
- \frac{3 \ap t}{16} \d_\perp^{\m[\g} v^{\w]}  \label{vt}
 \ , \ea
\ba && 8 \sqrt{22} \, T_{g, W, R}^{\m;\g} = 2\ap \bar q^\m \bar q^\g + \frac{28 t}{m^2} \d^{\m\g}_\perp
+ \frac{8}{m} \bar q^\m v^\g  \ . \label{wt} \ea

\subsection{Covariant derivation of the eikonal phase}

As we have seen, the high-energy behaviour of the
tree-level two-point  amplitudes between arbitrary string states in the background
of a collection of D$p$-branes can be described in a simple and elegant way 
in terms of Reggeon exchange. We can summarize this dynamical information in 
an operator $W_R(s,  q)$ defined as follows 
\be W_R(s,  q) = \Pi^{D_p}_R \, \sum_{i, \bar i, j, \bar j}  \,
|S_i, S_{\bar i}\pd \, C_{(S_i, S_{\bar i}), (S_j, S_{\bar j}), R}  \ps S_j, S_{\bar j} | \ , \label{coveik} \ee
where the sum is over the complete physical spectrum of the string. 

As discussed in the previous subsection, if we choose a basis of physical states which transform in irreducible representations of the
covariant little group, the couplings $C_{S_i, S_j, R}$ 
can be expressed in terms of tensors $T_{S_i, S_j, R}$. These tensors, which are
written in terms of the metric, the momentum transfer and the longitudinal polarization
vector, have an interesting structure reflecting
the symmetry properties  of the external states.  
However their explicit form becomes 
more and more complex as the mass of the string states  increases
and in order to evaluate them one also needs to know level by level  the covariant spectrum of the
string.

There is another choice of basis, the basis provided by the DDF operators \cite{DelGiudice:1971fp}, 
which allows
to easily enumerate the physical states\footnote{Although in \cite{Brower:1973iz, Schwarz:1974ix,
Hornfeck:1987wt} only the DDF operators for the NS sector were explicitly discussed,
the construction can be easily extended to the R sector by acting with the operators defined in Eq.\eqref{ABigen} on the Ramond ground 
state and taking $r \in \mathbb{Z}$ in the definition of the fermionic operators.}
and in which the couplings to the Reggeon become elementary, although 
in this basis  only the $SO(8)$ symmetry group of the
space transverse to the collision axis is manifestly realized. 
The simple couplings of the DDF operators to the Reggeon  make
it possible to represent the formal sum in
Eq.~\eqref{coveik} in a compact operator form. We will show that the result coincides with $\hat \d(s, b)$, 
the phase of the eikonal operator.
Using the Reggeon vertex we can then 
derive the eikonal phase from the full covariant dynamics
and identify the
modes of the string coordinates $X$ in $\hat \d(s, b)$ with the bosonic DDF
operators.

In order to derive the eikonal phase, we need to consider transitions between two generic NS states or two
generic R states, the transitions from the NS to the R sector being subleading in energy. 
Let us describe explicitly 
the couplings of two NS states to the Reggeon. As reviewed in Appendix \ref{app:DDF}, 
a generic physical state of the NS sector at level $l$ and carrying momentum $p$  is generated by the action of 
a finite collection of bosonic and fermionic modes
\be \{ A_{-n_{a}, i_{a}}, B_{-r_{b}, i_{b}} \}_{a, b} \ , \hspace{1cm} l =  N_a + N_b - \frac{1}{2} \hspace{1cm}
N_a = \sum_a n_a \ , \hspace{0.4cm} N_b =  \sum_b r_b\ , \ee
on a vacuum state carrying a momentum $p_T = p + \si ( N_a + N_b \de ) k$ which satisfies $p_T^2 = \frac{2}{\ap}$. 
The null vector $k$ has the property that $k p_T =  \frac{2}{\ap}$. 
Finally the GSO projection requires that $N_b$ is
a half-integer. 

To represent the external states $S_1$ and $S_2$ in a string-brane collision we need two collections of modes labeled by two couples
on indices $(a_1, b_1)$ and $(a_2, b_2)$, two momenta $p_{T, i}$ 
and two null vectors $k_i$, $i = 1, 2$. We choose the two null vectors proportional to $e^+$, $k_i = \l_i e^+$.
In the high-energy limit they can be identified since the boost parameters coincide, $\l_i \sim \frac{\sqrt{2}}{\ap E}$. 
The transverse polarization vectors of the initial and final state will be denoted by $\e$ and $\z$, respectively.
Consider now the coupling of two generic DDF vertices of the NS sector to the Reggeon 
\be C_{S_1, S_2, R} = \ps V^{(-1)}_{S_1}(z_1) V^{(-1)}_{S_2}(z_2) V^{(0)}_R(z_3) \pd \ . \ee
Since the DDF operators do not contain $X^-$ and $\psi^-$, in evaluating the correlation
functions one can simplify the vertices of both the Reggeon and the external states
by retaining only the following terms
\ba
V^{(0)}_R(z) \! &\sim& \! 
\si (\sqrt{ \frac{2}{\ap}} \frac{\ii \p X^+(z)}{\sqrt{\alpha'}E} 
\de )^{\frac{\ap t}{4}+1} \!\!\! {\rm e}^{-\ii q X(z)} \ , \nb \\
A_{-n,j}(z) &\sim& -\ii \sqrt{\frac{2}{\ap}} \oint_{z} dw \,\, (\epsilon_j)_{\mu} \partial X^{\mu}
{\rm e}^{-\ii n k X } \ , \nb  \\
B_{-r,j}(z) &\sim& -\ii  \oint_{z} dw \,\, (\epsilon_j)_{\mu}  \psi^{\mu}
(\ii k \partial X)^{\frac{1}{2}} 
{\rm e}^{-\ii r k X}   \ . \label{crddf}
\ea
From Eq.~\eqref{crddf} we see that
since there are no transverse fermions in the Reggeon vertex,  for a non-vanishing result the two states must be created by the action of exactly
the same fermionic DDF operators. It is easy to verify that the contour integrals in the definition
of the $B_{-r, j}$ reduce to the ones appearing in the scalar product between the two states
and then give simply the contraction between the corresponding polarization vectors, $\e \z$. 
The action of  the
operator $W_R(s, q)$ on the fermionic modes $B_{-r, j}$ then reduces to the action of the identity operator.

As for the bosonic modes, it is always possible to contract all of them with the exponential part of the Reggeon vertex,
with the only constraint imposed by momentum conservation in the $x^+$ direction.
The contractions with the Reggeon of the transverse bosons $\e_{j_a} \p X$ in the definition of the modes $A_{-n_a, j_a}$  give
\be  \sqrt{\frac{\ap}{2}}   \frac{\e_{j_{a_1}} \bar q}{w_1-z_3} \, {\rm e}^{-\ii n_{a_1} k X(w_1)} \ , 
\hspace{2cm} -  \sqrt{\frac{\ap}{2}}   \frac{ \z_{j_{a_2}} \bar q }{w_2-z_3} \, {\rm e}^{\ii n_{a_2} k X(w_2)} \ , \ee
respectively for the initial and the final state. Evaluating the contour integrals  
the dependence on the $z_i$ disappears, as required since all the vertices are conformal primaries.
The result is then simply to replace every bosonic mode in the initial state with
$ - \sqrt{\frac{\ap}{2}} \e_{i_{a_1}} \bar q$ and every bosonic mode
 in the final state with $\sqrt{\frac{\ap}{2}} \z_{j_{a_2}} \bar q$.
A similar substitution holds for the antiholomorphic part and one should also  impose the constraint 
\be \sum n_{a_1} - \sum n_{a_2} - \sum \bar n_{\bar a_1} + \sum \bar n_{\bar a_2} = 0 \ . \label{sigmac} \ee
Finally, there are also contractions between the DDF operators whenever some of the bosonic modes that create the initial and the final state
coincide. As it was the case for the fermionic
modes, these contractions simply give the scalar product between the bosonic DDF operators, $\e \z$, and
clearly respect the constraint in Eq.~\eqref{sigmac}. 

A similar analysis can be performed for the transitions between two generic states of the Ramond sector,
taking the external states in the $-\frac{1}{2}$ picture and the Reggeon vertex in the $-1$ picture, and 
it leads to the same conclusions.

The action of $W_R(s, q)$ in the DDF basis is therefore extremely simple. It acts like the identity
on the fermionic modes $B_{-r, i}$.
Its action on the bosonic modes is non-trivial and consists in replacing any number of $A_{-n, i}$ 
satisfying the condition in Eq.~\eqref{sigmac}
with the momentum transfer $\bar q$. This operator can then be written as follows
in terms of the DDF operators
\ba W_R(s, q) &=&  {\cal A}(s, t)  \,  \int_0^{2\pi}  \, \frac{d \s}{2\pi} \, : {\rm e}^{\ii \bar q X} : 
= {\cal A}(s, t)   \int_0^{2\pi} \, \frac{d \s}{2\pi} \,
{\rm e}^{\ii \bar q X^{<}} {\rm e}^{\ii \bar q X^{>}} {\rm e}^{\ii \bar q \bar X^{<}} {\rm e}^{\ii \bar q \bar X^{>}}
 \label{taylor2} \ , \ea
where  the integral over $\s$ enforces the constraint  in Eq.~\eqref{sigmac} and 
\be X^{>} = \ii \sqrt{\frac{\ap}{2}} \sum_{n = 1}^\infty \frac{A_n}{n} {\rm e}^{\ii n \s} \ , 
\hspace{1cm} X^{<} = - \ii \sqrt{\frac{\ap}{2}} \sum_{n = 1}^\infty \frac{A_{-n}}{n} {\rm e}^{- \ii n \s} \ , \ee
with similar expressions for $\bar X$.
This is precisely the  operator $\hat \d(s, \bar q)$ in \eqref{taylor}
which, as this derivation shows, can be interpreted as a covariant operator expressed in the basis
of the DDF operators.

\section{The eikonal operator and the covariant amplitudes}
\label{covariant-eikonal}

The derivation of the eikonal phase given in the previous section shows that, although the individual 
amplitudes with covariant external states may have a somewhat complex tensor structure, all the dynamical
information, including the longitudinally polarized states, can be summarized in a simple operator. 
The aim of this section is to understand in detail the relation between the scattering amplitudes
of the covariant states and the matrix elements of the eikonal operator.

The first step is to decompose the covariant tensors
with respect to the $SO(8)$ group used to define the light-cone gauge, namely
the symmetry group of the space transverse to the collision axis.
Each $SO(9)$ representation then breaks into several $SO(8)$ components
whose couplings to the Reggeon have the simple structure of the basic couplings in Eq.~\eqref{r28}. 
We shall call the set of all the $SO(8)$ components
of the covariant physical states at a given level the covariant basis for that level.

To show that the covariant and the light-cone calculations precisely match it is necessary to
perform a change of basis 
from the covariant basis to a high-energy basis. The latter is characterized by the 
following dynamical property: every subspace of states transforming in the same
representation of $SO(8)$ is decomposed 
into two orthogonal sets, containing states having a vanishing or a non-vanishing coupling to the Reggeon
respectively. 

Let us define the high-energy basis more precisely.
Generically a given $SO(8)$ representation $r$ will appear in the decomposition of several
covariant states and with several linearly independent couplings. 
If $d_r$ is the degeneracy of the representation and $c_r$ the number of the inequivalent couplings,
there will be $d_r-c_r$ linear combinations of elements of the covariant basis that decouple at high energy.
The states in the $c_r$-dimensional orthogonal subspace are those with a non-vanishing coupling to the Reggeon.
For each $SO(8)$ representation we choose an orthonormal basis in the  
$d_r-c_r$ and $c_r$-dimensional subspaces. The set of all these states forms the high-energy basis.

The covariant and the high-energy basis are related by a unitary transformation. Once rewritten
in the high-energy basis,
the dynamical information derived from the covariant amplitudes
reproduces the list of $SO(8)$ representations that can be excited
at high-energy and the corresponding transition amplitudes
derived from the eikonal phase.
Since the high-energy basis is given explicitly in terms
of linear combinations of $SO(8)$ components of covariant string states,  we obtain in this way
a covariant  characterization of the light-cone states that are created by the 
action of the eikonal phase on a given initial state.

We will perform in detail the comparison between the eikonal phase and the covariant amplitudes
for the inelastic transitions from the massless NS-NS sector to the first two massive levels of the superstring,
finding perfect agreement with the results derived in section \ref{eikonal}. 
It is interesting to note that this precise agreement is found only after the explicit dependence
of the covariant amplitudes on the masses of the external states is rewritten in terms of $\ap$ using the relation 
$m^2 = \frac{4l}{\ap}$ for states at level $l$. 

The comparison is straightforward for the first level, since in this case there is no degeneracy in the $SO(8)$ 
representations that appear in the decomposition of the covariant states and therefore
the covariant and the high-energy basis coincide. The first degenerate 
representations appear in the second level, which will be studied in detail since it clearly exemplifies 
all the generic features of the relation between the covariant amplitudes and the matrix elements of the eikonal operator.

When decomposing the $SO(9)$ representations with respect to the transverse $SO(8)$,
we will write  the polarization 
tensors of the covariant states as products of longitudinal vectors $v$
and of $SO(8)$ polarization tensors $\w$ 
having non vanishing components only in the transverse directions, $v^\a \w_{\a...} = 0$. 
We will use the following notation for the component of the polarization of a covariant state $S$ 
transforming in the $(n_1,n_2,..., n_r)$ representation of $SO(8)$
\be \z^{S, (n_1,n_2,..., n_r) } \ . \ee
The decomposition with respect to the transverse $SO(8)$ followed by the unitary transformation
just described simplifies the form of the covariant amplitudes and reduces them to the matrix elements of the eikonal phase.
It is also interesting to proceed in the opposite direction and to provide a covariant characterization
of the light-cone states of a given matrix element. This is the well-known problem 
of finding the covariant representations which can be formed by combining  the physical states in the light-cone gauge. 
As discussed at the end of section
\ref{eikonal}, a possible way to derive this information 
is to first identify a light-cone state with a covariant state using the DDF operators and then to find
to which linear combination of $SO(8)$ components $\z^{S, (n_1,n_2,..., n_r) }$
of physical $SO(9)$ states it corresponds. This procedure will be illustrated
with a few specific examples at the end of this section.

\subsection{$SO(8)$ decomposition of the covariant amplitudes: first massive level}

The two $SO(9)$ representations in the first level have the following decomposition with respect to the transverse $SO(8)$
\be \yng(2)\ \  \mapsto \  \ \yng(2) \ \ + \ \ \yng(1) \ \ + \ \ \bullet \ \ , \ee
and \be \yng(1,1,1)\ \  \mapsto \  \ \yng(1,1,1) \ \ + \ \ \yng(1,1) \ \ . \ee
We decompose the polarization tensors into irreducible $SO(8)$ components. For $S_2$  we have
\ba \z^{S_2,(2)}_{\r\a} =  \w^{(2)}_{\r\a} 
\ , \hspace{0.7cm} \z^{S_2,(1)}_{\r\a} = \sqrt{2} \, \w_{(\r} v_{\a)} 
\ , \hspace{0.7cm} \z^{S_2,(0)}_{\r\a} =  \frac{1}{3\sqrt{8}}
\si ( - \d_\perp^{\r\a} + 8 v^\r v^\a \de )  \ , \label{so8s2} \ea
and  for $A_3$  we find 
\ba \z^{A_3, (1,1,1)}_{\r\a\g} =  \w^{(1,1,1)}_{\r\a\g} \ , \hspace{1.4cm} 
\z^{A_3, (1,1)}_{\r\a\g} =  \sqrt{3} \, \w^{(1,1)}_{[\r\a} v_{\g]} \ . \label{so8a3}  \ea
The decompositions above 
can be easily derived by writing a tensor with the correct symmetry 
properties using $\w$, $v$ and the Kronecker delta in the transverse directions $\d_\perp$ 
and requiring that it is traceless and normalized. 
The couplings of the covariant states to the Reggeon are given by the tensors in Eq.~\eqref{s2t} and Eq.~\eqref{a3t}.
When contracted with the  $SO(8)$ polarizations we find
\ba  T_{g, S_2, R}^{\m,\r\a}  \z^{S_2,(2)}_{\r\a} &=& - \sqrt{\frac{\ap}{2}} \d_\perp^{\m\r} \w^{(2)}_{\r\a} \bar q^\a \ , \nb \\
T_{g, S_2, R}^{\m,\r\a}  \z^{S_2,(1)}_{\r\a} &=& 0 \ , \nb \\
T_{g, S_2, R}^{\m,\r\a}  \z^{S_2,(0)}_{\r\a} &=& - \frac{\sqrt{\ap}}{4} \bar q^\m \ , \ea
and
\ba 
 T_{g, A_3, R}^{\m,\r\a\g} \z^{A_3, (1,1,1)}_{\r\a\g} &=& 0 \ , \nb \\
 T_{g, A_3, R}^{\m,\r\a\g}  \z^{A_3, (1,1)}_{\r\a\g}  &=&   \sqrt{\frac{\ap}{2}} \d_\perp^{\m\r} \w^{(1,1)}_{\r\a} \bar q^\a\ . \ea
Note that the vector and the rank-3 antisymmetric tensor of $SO(8)$ decouple at high energy.
The remaining representations and couplings are in perfect agreement with those derived using the eikonal operator and given in
Eq.~\eqref{tab1}. 
The states in the first massive level that can be excited in the high energy scattering of a massless NS-NS state
on a stack of D$p$-branes are therefore 
\be  S = \z^{{\cal S}_2,(2)} S_2 \ , \hspace{1cm} A = \z^{{\cal A}_3, (1,1)} A_3  \ , \hspace{1cm} I = \z^{{\cal S}_2,(0)}  S_2 \ , \ee
for a total of $64$ degrees of freedom. The covariant and the high-energy basis coincide for the first level.

\subsection{$SO(8)$ decomposition of the covariant amplitudes: second massive level}

We perform now the same analysis for the second level, since it neatly displays all the generic features 
of the relation between the covariant amplitudes and the matrix elements of the eikonal phase.
The explicit expressions for the polarization tensors that correspond to the various $SO(8)$ components of the covariant states
of the second level are listed in Appendix \ref{app:Young}.

The state $Z$ in 
the $(3)$ of $SO(9)$ has the following decomposition with respect to $SO(8)$
\be \yng(3)\ \  \mapsto \  \ \yng(3) \ \ + \ \ \yng(2) \ \ + \ \ \yng(1) \ \ + \ \ \bullet \ \ . \ee
The coupling of the covariant state $Z$ to the Reggeon is given by the tensor in Eq.~\eqref{zt}.
Contracting this tensor with the polarizations in the list in Eq.~\eqref{zpol} 
we find 
\ba T_{g, Z, R}^{\m;\r\a\g}\z^{Z, (3)}_{\r\a\g} &=& \frac{\ap}{\sqrt{8}} \d_\perp^{\m\g} \w^{(3)}_{\r\a\g} \bar q^\r \bar q^\a \ , \nb \\
T_{g, Z, R}^{\m;\r\a\g}\z^{Z, (2)}_{\r\a\g} &=& \frac{2}{\sqrt{6}m} \d_\perp^{\m\g} \w^{(2)}_{\g\r} \bar q^\r 
=  \sqrt{\frac{\ap}{12}} \d_\perp^{\m\g} \w^{(2)}_{\g\r} \bar q^\r   \ , \nb \\
T_{g, Z, R}^{\m;\r\a\g}\z^{Z, (1)}_{\r\a\g} &=&\frac{1}{2\sqrt{165}} \si ( - \frac{3}{2} \ap \bar q^\m \bar q \w 
+ \frac{\ap t}{8} \w^\m \de )  \ , \nb \\
T_{g, Z, R}^{\m;\r\a\g}\z^{Z, (0)}_{\r\a\g} &=&-  \frac{1}{2\sqrt{11}} \frac{3}{m} \bar q^\m \ . \label{zp} \ea
Note the presence of two inequivalent couplings for the vector component. 
The state $Y$ in the $(2,1,1)$  of $SO(9)$ has the following decomposition with respect to $SO(8)$
\be \yng(2,1,1)\ \  \mapsto \  \ \yng(2,1,1) \ \ + \ \ \yng(1,1,1) \ \ + \ \ \yng(2,1) \ \ + \ \ \yng(1,1)   \label{so9so8ybis}  \ . \ee  
From the coupling of the covariant state $Y$ to the Reggeon  in Eq.~\eqref{yt}
we derive the following couplings for its $SO(8)$ components
\ba  \sqrt{\frac{8}{3}} T_{g, Y, R}^{\m;\r\a\g\w} \z^{Y, (2,1,1)}_{\r\a;\g;\w} &=& 0 \ , \nb \\ 
 \sqrt{\frac{8}{3}} T_{g, Y, R}^{\m;\r\a\g\w} \z^{Y, (1,1,1)}_{\r\a;\g;\w} &=& 0 \ , \nb \\ 
 \sqrt{\frac{8}{3}} T_{g, Y, R}^{\m;\r\a\g\w} \z^{Y, (2,1)}_{\r\a;\g;\w} &=& \sqrt{\frac{\ap}{2}} \frac{4}{\sqrt{3}m} \d_\perp^{\m\g}
\w^{(2,1)}_{\r\a;\g} \bar q^\a \bar q^\r = \frac{\ap}{\sqrt{3}} \d_\perp^{\m\g}
\w^{(2,1)}_{\r\a;\g} \bar q^\a \bar q^\r \ , \nb \\ 
 \sqrt{\frac{8}{3}} T_{g, Y, R}^{\m;\r\a\g\w} \z^{Y, (1,1)}_{\r\a;\g;\w} &=& - \sqrt{\frac{4\ap}{7}} \d_\perp^{\m\a}
\w^{(1,1)}_{\a\g} \bar q^\g \ .  \label{yp}\ea
Note that all the tensors with more than two antisymmetric indices decouple. 
The state $U$ in the $(2,1)$  of $SO(9)$ has the following decomposition with respect to $SO(8)$
\be \yng(2,1)\ \  \mapsto \  \ \yng(2,1) \ \ + \ \ \yng(2) \ \ + \ \ \yng(1,1) \ \  + \ \ \yng(1)  \  \ . \label{so9so8u}  \ee
Using the coupling of the covariant state $U$ to the Reggeon in Eq.~\eqref{ut} and the polarization tensors in Eq.~\eqref{up}
we find
\ba  \sqrt{6} T_{g, U, R}^{\m; \r\a\g}\z^{U, (2,1)}_{\r\a;\g} &=&  \frac{\ap}{2} \d_\perp^{\m\g} \w^{(2,1)}_{\r\a; \g} \bar q^\a \bar q^\r \ , \nb \\
 \sqrt{6} T_{g, U, R}^{\m; \r\a\g}\z^{U, (2)}_{\r\a;\g} &=& -  \frac{2\sqrt{2}}{m} \d_\perp^{\m\g} \w^{(2)}_{\g\r} \bar q^\r 
= -  \sqrt{\ap} \d_\perp^{\m\g} \w^{(2)}_{\g\r} \bar q^\r \ , \nb \\
 \sqrt{6} T_{g, U, R}^{\m; \r\a\g}\z^{U, (1,1)}_{\r\a;\g} &=& 0  \ , \nb \\
 \sqrt{6} T_{g, U, R}^{\m; \r\a\g}\z^{U, (1)}_{\r\a;\g} &=& - \frac{\sqrt{7}}{28} 
\si ( 3 \ap \bar q^\m \w \bar q + \frac{ 5}{4} \ap t \w^\m \de )   \ . \ea
Note the decoupling of the vector component of $U$, similar to the decoupling of the vector component
of the state $S_2$ in the first level.
The state $V$ in the $(1,1)$ of $SO(9)$ gives a two-form and a vector of $SO(8)$.
Using the coupling in Eq.~\eqref{vt} we find
\ba T_{g, V, R}^{\m; \g\w} \z^{V, (1,1)}_{\g\w}  &=& 
\frac{\sqrt{\ap}}{2 \sqrt{7}} \d_\perp^{\m\g} \w_{\g\w} \bar q^\w \ ,  \nb \\
T_{g, V, R}^{\m; \g\w} \z^{V, (1)}_{\g\w} &=& \frac{\ap}{4\sqrt{7}}  \bar q^\m \bar q \w
- \frac{3 \ap t }{16\sqrt{7}} \w^\m  \ . \ea
Finally the state $W$ in the vector representation of $SO(9)$ gives a vector and a scalar of $SO(8)$.
The coupling in Eq.~\eqref{wt} gives
\ba  8 \sqrt{22} T_{g, W, R}^{\m; \g}\z^{W, (1)}_{\g} &=& 2 \ap \bar q^\m \w \bar q + \frac{28}{m^2} t \w^\m \ ,  \nb \\
8 \sqrt{22} T_{g, W, R}^{\m; \g}\z^{W, (0)}_{\g} &=& \frac{8}{m} \bar q^\m  \ . \ea

\subsection{From the covariant to the high-energy basis}

We now compare the $SO(8)$ tensors and  couplings 
given by the covariant amplitudes for the second level with
those given by the eikonal phase, deriving the unitary transformation that
connects the covariant and the high-energy basis.

Let us start with the tensors of rank $3$, namely the $(3)$ and the $(2,1)$ representations. 
The $(3)$ representation of $SO(8)$ appears only in the covariant state $Z$ 
\be  F = \z^{Z, (3)} \, Z \ , \ee
and it is produced with the same amplitude as given by the eikonal operator, cfr. Eq.~\eqref{zp} and Eq.~\eqref{tab2}.
Consider now the $(2,1)$ representation of $SO(8)$. This representation appears in the covariant states $Y$ and $U$
and provides the simplest example of a degenerate $SO(8)$ representation. Since there is only one independent coupling
to the Reggeon, there is a linear combination of the $(2,1)$ components of $Y$ and $U$
that decouples at  high energy.
The new basis is easily identified as well as
the corresponding couplings to the Reggeon\footnote{We leave
understood that the polarizations $\z^{S, (r)}$ in the definition of a state of the high-energy basis transforming 
in the representation $r$ of $SO(8)$
are all formed using the same transverse polarization $\w^{(r)}$.}  
\ba H_1 &=&  \frac{1}{2} \si (  \z^{Y, (2,1)} \, Y - \sqrt{3}  \z^{U, (2,1)} \, U \de ) \ ,  \hspace{1cm} C_{g,H_1,R} = 0 \ ,  \nb \\
H_2 &=& \frac{1}{2} \si ( \sqrt{3}  \z^{Y, (2,1)} \, Y  +  \z^{U, (2,1)} \, U \de )\ ,  \hspace{1cm}  C_{g,H_2,R} = -
 \frac{\ap}{\sqrt{6}} \epsilon^\a  \w^{(2,1)}_{\a\r; \g} \bar q^\r \bar q^\g \ . \label{uncino} \ea
The coupling to the Reggeon of the state $H_2$, which is the one produced at high energy  in the $(2,1)$ 
representation, coincides with the coupling given for this representation by the eikonal operator, cfr. Eq.~\eqref{tab2}.

Let us now turn to the rank-two tensors. The $(2)$ appears in $Z$ and $U$ and the high-energy basis is
\ba S_1 &=&  \frac{1}{\sqrt{3}} \si (  \sqrt{2} \z^{Z, (2)} \, Z +  \z^{U, (2)} \, U \de ) \ , \hspace{1cm}  C_{g,S_1,R} = 0 \ ,  \nb \\
S_2 &=& \frac{1}{\sqrt{3}} \si (  \z^{Z, (2)} \, Z  -  \sqrt{2} \z^{U, (2)} \, U\de ) \ , \hspace{1cm}  C_{g,S_2,R} = 
 \frac{\sqrt{\ap}}{2} \epsilon^\mu  \w^{(2)}_{\m\r} \bar q^\r \ , \ea
again in agreement with the eikonal operator, cfr. Eq.~\eqref{tab2}.
The $(1,1)$ appears in $Y$ and $V$ and the high-energy basis is
\ba A_1 &=&\frac{1}{\sqrt{7}} \si (  \z^{Y, (1,1)} \, Y +   \sqrt{6}  \z^{V, (1,1)} \, V \de ) \ ,   \hspace{1cm}  C_{g,A_1,R} = 0 \ ,  \nb \\
A_2 &=& \frac{1}{\sqrt{7}} \si (  \sqrt{6} \z^{Y, (1,1)} \, Y  -   \z^{V, (1,1)} \, V \de ) \ , \hspace{1cm} C_{g,A_2,R} = 
- \frac{\sqrt{\ap}}{2} \epsilon^\mu  \w^{(1,1)}_{\m\r}  \bar q^\r \ . 
\ea
Similarly the $SO(8)$ scalar appears in $Z$ and $W$ and the high-energy basis is
\ba I_1 &=&\frac{1}{\sqrt{11}} \si (  \sqrt{2} \z^{Z, (0)} \, Z + 3  \z^{W, (0)} \, W \de ) \ ,  \hspace{1cm}  C_{g,I_1,R} = 0 \ ,  \nb \\
I_2 &=& \frac{1}{\sqrt{11}} \si ( 3 \z^{Z, (0)} \, Z -   \sqrt{2} \z^{W, (0)} \, W  \de ) \ ,  \hspace{1cm}  C_{g,I_2,R} = 
-\frac{\sqrt{\ap}}{4\sqrt{2}} \epsilon_\mu \bar q^\m 
\ . \ea
In both cases we find agreement with the eikonal operator, cfr. Eq.~\eqref{tab2}. 
Finally the vector of $SO(8)$ appears in $Z$, $U$, $V$ and $W$. This is the first case in which there are degenerate $SO(8)$ representations 
both in the covariant and 
in the high-energy basis, the two vectors in Eq.~\eqref{tab2}. 
Since for the vector there are two inequivalent couplings, we will find two linear combinations
that decouple at high energy and two linear combinations with a non-vanishing coupling to the Reggeon, 
related by a unitary transformation to the states in Eq.~\eqref{tab2}.
A basis  for the two-dimensional space of vectors of $SO(8)$ that  do not couple to the Reggeon is 
\ba B_1 &=&
 \frac{5\sqrt{15}}{22}\z^{Z, (1)} \, Z+ \frac{\sqrt{77}}{22}  \z^{V, (1)} \, V
+ \frac{4 \sqrt{2}}{22}  \z^{W, (1)} \, W \ ,  \nb \\
B_2 &=& - \frac{1}{11}\sqrt{\frac{35}{3}}\z^{Z, (1)} \, Z +  
\frac{1}{2} \sqrt{\frac{11}{6}} \z^{U, (1)} \, U + \frac{1}{\sqrt{11}}  \z^{V, (1)} \, V
- \frac{7}{22} \sqrt{\frac{7}{2}}  \z^{W, (1)} \, W \ .  \ea
In the two-dimensional space of vectors  of $SO(8)$  that couple to the Reggeon we choose the basis that corresponds to the one in 
Eq.~\eqref{tab2}
\ba
 B_3 &=& \frac{1}{4} \si ( \sqrt{\frac{21}{11}}\z^{Z, (1)} \, Z +  \sqrt{\frac{15}{2}} \z^{U, (1)} \, U -\sqrt{5}  \z^{V, (1)} \, V
- \sqrt{\frac{35}{22}}  \z^{W, (1)} \, W\de ) 
\ , \nb \\
B_4 &=& - \frac{1}{4} \si ( \sqrt{\frac{5}{33}}\z^{Z, (1)} \, Z -  \sqrt{\frac{7}{6}} \z^{U, (1)} \, U - \sqrt{7}  \z^{V, (1)} \, V
+\frac{13}{ \sqrt{22}}  \z^{W, (1)} \, W\de ) 
\ . \ea
It is easy to verify that the couplings of these two vectors to the Reggeon are  indeed
\be   C_{g,B_3,R} = - \frac{\ap}{\sqrt{35}} \epsilon_\mu \si (  \bar q^\m \bar q^\g + \frac{\ap t}{8} \d_\perp^{\m\g}   \de ) \w_{\gamma} \ , 
\hspace{1cm}  C_{g,B_4,R} = - \frac{\ap t}{8} \epsilon_\mu \d_\perp^{\m\g} \l_{\gamma}  \ , \ee
in agreement with Eq.~\eqref{tab2}. 

The analysis of the second level is now complete.
The states in the second massive level that can be excited in the high-energy scattering of a massless state
on a stack of Dp-branes are therefore 
\be F \ , \hspace{0.8cm} H_2 \ , \hspace{0.7cm}  B_3 \ ,  \hspace{0.7cm}
 B_4 \ ,  \hspace{0.7cm} S_2 \ , \hspace{0.7cm}  A_2   \ , \hspace{0.7cm} I_2   \ , \ee
in the following irreducible representations of $SO(8)$
\ba \yng(3) \hspace{1cm} \yng(2,1)  \hspace{1cm} \yng(1)  \hspace{1cm}\yng(1)  \hspace{1cm}\yng(2) 
 \hspace{1cm} \yng(1,1) \hspace{1.2cm} \bullet \ ,  \ea
for a total of $352$ degrees of freedom. All the other $SO(8)$ components
of the covariant states at level two decouple from the ground state in the Regge limit.
All the covariant amplitudes for the $SO(8)$ tensors in the high-energy basis agree with the
matrix elements of the eikonal phase. 

\subsection{From the light-cone to the covariant states} \label{lc2c}

 Having described in detail how to relate the covariant dynamics to the light-cone dynamics
 encoded in the eikonal operator, we now discuss a method to proceed in the opposite
 direction. This method, based on the DDF operators introduced in Appendix~\ref{app:DDF},
 allows to connect level by level
 the light-cone states with linear
 combinations of the $SO(8)$ components of the covariant states.

 The DDF operators are in one-to-one correspondence
 with the light-cone states and the $SO(8)$
 symmetry rotating the transverse polarization vectors $\epsilon_j$ is the only part of the Lorentz
 group manifestly realized. All massive states will then appear decomposed in $SO(8)$ representations. 
 The fact that the DDF operators are constructed using the worldsheet fields $X^\m$ and $\psi^\m$
 of the covariant theory will allow us to identify for any $SO(8)$ representation in the light-cone gauge
 the corresponding linear combination of the $SO(8)$ components of the covariant states.

 The spectrum of the physical states of the NS sector of the superstring 
 is generated by the action  on the vacuum state in Eq.~\eqref{eq:PT} of the DDF operators
 in Eq.~\eqref{ABigen}, followed by the
 GSO projection on states
 with definite worldsheet fermion number. 
 The GSO projection preserves only the states containing an odd number
 of $B_{-r,j}$, so the first non trivial physical state is obtained by
 applying the operator $B_{- \frac{1}{2},j}$ to the vacuum in
 Eq.~\eqref{eq:PT}
 \begin{equation}
 \label{eq:state0lc}
 B_{- \frac{1}{2},j} |p_T;0 \rangle = -  (\epsilon_j
 \psi_{-\frac{1}{2}}) | p_T - \frac{1}{2} k;0 \rangle \ , \hspace{1cm}
 \left( p_T - \frac{1}{2} k \right)^2 = p^2_T - p_T k =0 \ .
 \end{equation}
 These are the eight physical polarizations of the massless states corresponding
 to the covariant vertex operator in Eq.~\eqref{mvg} with momentum $p = p_T -k/2$.

 At the first massive level we have three types of states: those
 created by the action of $B_{-3/2}$, those created by the action of
 three $B_{-1/2}$ and finally the states with one $B_{-1/2}$ oscillator
 and one $A_{-1}$ oscillator. 
 In this case the correspondence between
 the light-cone and the covariant states is unambiguous since there is
 a unique way to combine the $SO(8)$ representations into $SO(9)$
 representations. The first two sets of states can be immediately
 identified with the covariant states with polarizations $\z^{S_2,
   (1)}$ and $\z^{A_3, (1,1,1)}$.  Similarly the symmetric,
 antisymmetric and trace part of the states created by the action of
 $A_{-1, j} B_{-1/2, k}$ correspond to the covariant states with
 polarizations $\z^{S_2, (2)}$, $\z^{A_3, (1,1)}$ and $\z^{S_2, (0)}$.

 Let us show how the previous identifications can be derived using the DDF operators, focusing
 on the states $A_{-1, j} B_{-1/2, k}$, which are the only ones in the first level produced in a high-energy 
 collision of a massless state with a D$p$-brane. This analysis will  allow us
 to introduce all the tools that are necessary to apply 
 the method to a general light-cone state in the higher massive levels, where the
 correspondence between light-cone and covariant states is not already completely fixed by the
 decomposition of the $SO(9)$ representations. We have
 \begin{equation}
 | \zeta_{jk}  \rangle \equiv  A_{-1, j} B_{-\frac{1}{2},k} | p_T;0\rangle
 =  - A_{-1, j} (\epsilon_k \psi_{-\frac{1}{2}})| p_T -
 \frac{1}{2} k ;0 \rangle  \ , 
 \end{equation}
 where we have performed the contour integral present in the definition
 of the $B$ DDF oscillator; by carrying out explicitly the integral in
 the definition of $A$ we get
 \ba
 | \zeta_{jk} \rangle &=&  \epsilon^{j}_{\mu} \epsilon^{k}_{\nu} 
  \left[ \alpha_{-1}^{\mu} \psi_{-\frac{1}{2}}^{\nu} - 
 (k \psi_{- \frac{1}{2}})  \psi_{- \frac{1}{2}}^{\mu} \psi_{-
   \frac{1}{2}}^{\nu}   -  (k \psi_{-\frac{3}{2}}) \eta^{\mu \nu}
 + 
 (k \psi_{- \frac{1}{2}})(k \alpha_{-1})  \eta^{\mu \nu}
 \right]  | p_T - \frac{3}{2} k ;0 \rangle \ ,
\nb \\ &=&
  \left[ \alpha_{-1}^{j} \psi_{-\frac{1}{2}}^{k} - 
 (k \psi_{- \frac{1}{2}})  \psi_{- \frac{1}{2}}^{j} \psi_{-
   \frac{1}{2}}^{k}  
 + \delta^{jk} \left( (k \psi_{- \frac{1}{2}})(k \alpha_{-1}) 
  - (k \psi_{-\frac{3}{2}}) \right)
 \right]  |p^{(1)} ;0 \rangle\ ,
 \label{zetajk}
 \ea
 where  $p^{(1)} = p_T - \frac{3}{2} k$ and for simplicity we defined $\alpha_{-1}^{j} = (\epsilon_j)_{\mu}
 \alpha_{-1}^{\mu}$ and
 similarly for the $\psi$ oscillators. This approach can
 obviously be used for any other light-cone state and, as explained at the end
 of section \ref{eikonal}, provides an
 algorithm to map the light-cone spectrum in the NS formalism into the
 covariant spectrum. In order to make this
 mapping fully explicit we need however some extra work.  Consider first  the
 antisymmetric combination
 \begin{equation}
 \label{antisym}
 | \zeta_{jk}^a \rangle =  \frac{1}{2} \left(  |\zeta_{jk} \rangle -
   | \zeta_{kj} \rangle  \right) =
 \left[ \frac{1}{2} \si ( \alpha_{-1}^{j} \psi_{-\frac{1}{2}}^{k} -
 \alpha_{-1}^{k} \psi_{-\frac{1}{2}}^{j} \de ) -  (k \psi_{- \frac{1}{2}}) \psi_{-\frac{1}{2}}^{j} \psi_{-\frac{1}{2}}^{k}    \right] |p^{(1)};0 \rangle \ .
 \end{equation}
It is straightforward to show
 that the following state
 \begin{eqnarray}
 |\Sigma^{[MN]} \rangle = \left[ 
 \alpha_{-1}^{M} \psi_{-\frac{1}{2}}^{N} -  \alpha_{-1}^{N} \psi_{-\frac{1}{2}}^{M} +
 (p^{(1)} \psi_{-\frac{1}{2}})
 \psi_{-\frac{1}{2}}^{M}   \psi_{-\frac{1}{2}}^{N} \right] |p^{(1)};0 \rangle  \ ,
 \label{G12kk}
 \end{eqnarray}
 is spurious when the indices $M,N$ run over the space orthogonal to the momentum $p$. This means that it is physical and at the same time it
 can be written as $G_{-1/2}$ acting on another state\footnote{It is a zero norm state that is decoupled from the physical spectrum. This kind of states were first considered in~\cite{DelGiudice:1970dr}.} 
 \begin{equation}
   \label{eq:spur}
  |\Sigma^{[MN]}\rangle =  G_{-\frac{1}{2}} \psi_{-\frac{1}{2}}^{M}
  \psi_{-\frac{1}{2}}^{N} |p^{(1)};0 \rangle ~.
 \end{equation}
 Then by using~\eqref{G12kk} in~\eqref{antisym}, we can rewrite the 
 antisymmetric state as follows
 \begin{eqnarray}
 | \zeta_{jk}^a \rangle & = & \frac{1}{2} 
 G_{-\frac{1}{2}} \psi_{-\frac{1}{2}}^{j}
 \psi_{-\frac{1}{2}}^{k}  |p^{(1)};0 \rangle 
 \label{zetab}
 - 
 \frac{1}{2} 
 \left[\left(p_T+\frac{1}{2} k\right) \psi_{-\frac{1}{2}} \right] 
 \psi_{-\frac{1}{2}}^{j}   \psi_{-\frac{1}{2}}^{k}  |p^{(1)};0 \rangle~.
 \end{eqnarray}
 Of course we can neglect the first line because it is a spurious
 state; then let us focus on the combination $p_T+k/2$: it is
 orthogonal  to both  the momentum of the state and the  eight
 polarizations $\epsilon_i$. Then this combination must be proportional
 to the unit vector  $v^\mu$ describing the ninth physical polarization present
 in the description of the massive state. It  is straightforward to
 generalize this relation and the corresponding expression of the momentum 
 of the massive state  to the $n^{\rm th}$ massive level 
 \begin{equation}
   \label{eq:PKv}
   v = -\frac{1}{\sqrt{2n}} \left(p_T  +\left(n-\frac{1}{2}\right)
     k\right) \ , \hspace{0.7cm} p^{(n)} = p_T - \left(n+\frac{1}{2} \right) k \ , \hspace{0.7cm}  p^{(n)} v =0 \ . 
 \end{equation}
 Thus we can rewrite the antisymmetric state~\eqref{zetab} as
 \begin{equation}
   \label{eq:zetab2}
   | \zeta_{jk}^a \rangle = \frac{1}{\sqrt{2}} 
   \psi_{-\frac{1}{2}}^v \psi_{-\frac{1}{2}}^{j}
 \psi_{-\frac{1}{2}}^{k}  |p^{(1)};0 \rangle~.
 \end{equation}
 where $\psi_{-\frac{1}{2}}^{v} = v_{\mu} \psi_{-\frac{1}{2}}^{\mu}$. 
 In this form it is clear that $\zeta_{jk}^a$ corresponds to $\z^{A_3, (1,1)}$, the
 $SO(8)$ part of the covariant state $A_3$ in Eq.~\eqref{mvfl} and Eq.~\eqref{so8a3}, where one of the Lorentz indices
 is along the $v$ direction and the remaining two are in the eight-dimensional space
 perpendicular to the light-cone directions $e^\pm$.

 The same approach can be followed to rewrite the symmetric and the trace parts of $|\zeta_{ij}\rangle$. 
 By using~\eqref{eq:PKv} it can be checked that 
 \begin{eqnarray}
  | \zeta_{jk}^{st} \rangle \!\! & \equiv & \!\! \frac{1}{2}
  \left(|\zeta_{jk}\rangle + |\zeta_{kj}\rangle\right) =
 \left[  \delta^{jk} \left( \frac{1}{2
      \sqrt{2}} |s_2\rangle + \frac{1}{3} |s_1 \rangle  \right) +
 \frac{1}{2} \left(  \alpha_{-1}^{j} \psi_{-\frac{1}{2}}^{k}  +
   \alpha_{-1}^{k} \psi_{-\frac{1}{2}}^{j} \right) 
 \right.  \nonumber \\ \label{eq:syma} & & \left. 
  ~ - \frac{1}{6} \eta^{jk}  \left( \eta_{\rho \sigma} - \frac{p^{(1)}_\rho p^{(1)}_\sigma}{(p^{(1)})^2} 
 - 3 v_{\rho} v_{\sigma} \right)\alpha_{-1}^{\rho} \psi_{-\frac{1}{2}}^{\sigma}
 \right] |p^{(1)};0 \rangle ,
 \end{eqnarray}
 where $|s_i\rangle$ are spurious states whose explicit expression can
 be found at the end of the Appendix~\ref{app:conv}. Then
 Eq.~\eqref{eq:syma} becomes
 \begin{equation}
   \label{eq:symb}
   | \zeta_{jk}^{st} \rangle = \frac{1}{2} \left[  \alpha_{-1}^{j}
     \psi_{-\frac{1}{2}}^{k} +  \alpha_{-1}^{k} \psi_{-\frac{1}{2}}^{j} 
     - \frac{\delta^{jk}}{3} \left(\sum_{i=1}^{8}  \alpha_{-1}^{i}
       \psi_{-\frac{1}{2}}^i - 2 \alpha_{-1}^v \psi_{-\frac{1}{2}}^v
     \right)\right] |p^{(1)};0 \rangle~.
 \end{equation}
 Separating the traceless and the trace part of the previous tensor, one can see
 that these states indeed correspond to the covariant $SO(8)$ components
 $\z^{S_2, (2)}$ and $\z^{S_2, (0)}$ in Eq.~\eqref{mvfl} and Eq.~\eqref{so8s2}.

 A similar analysis can be performed for the higher massive levels. 
 As an example let us consider here the two $\yng(2,1)$ of $SO(8)$ 
 present at level two in the light-cone spectrum.
 It is not difficult to
 build two  light-cone states transforming as tensor of type $(2,1)$ of $SO(8)$.
 The first one is 
 \begin{eqnarray}
    \sqrt{2} A_{-1, \ell} | \zeta_{jk}^a \rangle & = & \frac{1}{\sqrt{2}}
   \left[ (k \psi_{- \frac{1}{2}})\,\psi_{- \frac{1}{2}}^{\ell}  \left(
     \alpha_{-1}^{j} \psi_{ -\frac{1}{2}}^{k}  -  \alpha_{-1}^{k}
     \psi_{ -\frac{1}{2}}^{j}  \right)\right. 
   \label{eq:U'} \\ \nonumber
  &-& \left. \alpha_{-1}^{\ell} \left(\alpha_{-1}^{j} \psi_{
         -\frac{1}{2}}^{k}  -  \alpha_{-1}^{k} \psi_{ -\frac{1}{2}}^{j}
       - 2 (k \psi_{-\frac{1}{2}})  \psi_{ -\frac{1}{2} }^{j}
 \psi_{-\frac{1}{2}}^{k}   \right) \right]|p^{(2)};0 \rangle ~,
 \end{eqnarray}
 where $p^{(2)} = p_T - \frac{5}{2} k$ and, for the sake of simplicity,
 we assumed that $\ell\not=j\not=k$.
 Notice that the state \eqref{eq:U'} has unit norm and can be written as a linear combination of the states $| \w^{(2,1)}\pd$ introduced in the second line of~\eqref{tab2} 
 \begin{align}
 \nonumber
   \sqrt{2} A_{-1, \ell} | \zeta_{jk}^a \pd & = \frac{1}{\sqrt{3}} \left( | \w^{(2,1)}_\a\pd - | \w^{(2,1)}_\b\pd\right) ~, \\
   \label{eq:U'2tab}
 (\w^{(2,1)}_\a)_{\ell'j'|k'} & = \frac{1}{2} \left( \delta^\ell_{\ell'} \delta^j_{j'} \delta^k_{k'} +\delta^\ell_{j'} \delta^j_{\ell'} \delta^k_{k'} -  \delta^\ell_{k'} \delta^j_{j'} \delta^k_{\ell'} - \delta^\ell_{j'} \delta^j_{k'} \delta^k_{\ell'} \right) \ , \\
 (\w^{(2,1)}_\b)_{\ell'j'|k'} & = \frac{1}{2} \left( \delta^\ell_{\ell'} \delta^k_{j'} \delta^j_{k'} +\delta^\ell_{j'} \delta^k_{\ell'} \delta^j_{k'} -  \delta^\ell_{k'} \delta^k_{j'} \delta^j_{\ell'} - \delta^\ell_{j'} \delta^k_{k'} \delta^j_{\ell'} \right) \ ,
 \nonumber
 \end{align}
 where the state $|\w^{(2,1)}_\a\pd$ is obtained from the polarization
 $\w^{(2,1)}_\a$ and, of course, $\w^{(2,1)}_\b$ is just $\w^{(2,1)}_\a$
 with $j$ and $k$ exchanged.

 The relation between~\eqref{eq:U'} and the $SO(8)$ components $\z^{Y,
   (2,1)}$ and $\z^{U, (2,1)}$ of the covariant states $Y$ and $U$ in
 Eq.~\eqref{so8y} and Eq.~\eqref{up} is far from obvious, but this
 state must be a linear combination of them.
 Again, in order to make this connection manifest, one needs to
 eliminate a spurious state from the expression obtained by using the
 DDF operators
 \begin{equation}
   \label{eq:psi2new}
  |\hat{\psi} \rangle = \sqrt{2} \left[
  A_{-1, \ell} | \zeta_{jk}^a\pd 
   + \frac 14 \Bigl( |S_{jl;k} \rangle - |S_{kl;j}
   \rangle \Bigr) \right]  ~, 
 \end{equation}
 where 
 \begin{equation}
   \label{eq:S3}
    |S_{jl;k} \rangle = G_{-\frac{1}{2}} \alpha_{-1}^{(\ell}
   \psi_{-\frac 12}^{j)} \psi_{-\frac 12}^k |p^{(2)};0 \rangle \ , 
 \end{equation}
 is  explicitly given in Eq. (\ref{Sjellk}). The state $|{\hat{\psi}}\rangle$ has norm equal to 1. Using Eqs. (\ref{eq:PKv}) for $n=2$, one can rewrite $k$ in 
 (\ref{eq:U'}) in terms of  $p^{(2)}$ and $v$. The  terms with $p^{(2)}$ in (\ref{eq:psi2new})
 cancel  and then 
 it is  straightforward to check that
 \begin{equation}
   \label{eq:H1YU}
 |{\hat{\psi}}\rangle =  \sqrt{2} \left[-\frac 12 \left(|Y_\a \rangle -
      |Y_\b \rangle \right) - \frac{1}{12} \left(|U_\a \rangle
      - |U_\b \rangle \right) \right] ~,
 \end{equation}
 where the kets on the r.h.s. represent the covariant states $Y$ and $U$
 with a specific choice of the polarization. Their explicit
 expressions are\footnote{As mentioned above, we are assuming that all indices in $Y$ and $U$ are different; the full expressions, including the terms necessary to ensure the tracelessness condition, can be found in~\eqref{Yc}--\eqref{Uc}.}
 \begin{equation}\label{eq:symMN2}
    |U_\a\rangle = \left(
      \alpha_{-1}^{\ell} \alpha_{-1}^{j} \psi_{- \frac{1}{2}}^{k} -
      \alpha_{-1}^{k} \alpha_{-1}^{(j} \psi_{- \frac{1}{2}}^{\ell)} -
      3 \psi_{-\frac{3}{2}}^{(\ell} \psi_{-\frac{1}{2}}^{j)} \psi_{-\frac{1}{2}}^{k}
    \right) |p^{(2)};0 \rangle = \sqrt{6} |\zeta^{U,(2,1)}_\a\pd \ , 
 \end{equation}
 and similarly for $ |U_\b\rangle$. The states $|\zeta^{U,(2,1)}_{\a,\b}\pd$ are obtained from the state, corresponding to the vertex operator $V_U$ in Eq.~\eqref{mvsl}, with the same 
 polarizations introduced in~\eqref{eq:U'2tab} and  
 with the indices restricted to those of $SO(8)$. Similarly
 \begin{equation}
   \label{eq:symMN}
   |Y_\a \rangle = \alpha_{-1}^{(\ell} \psi_{-\frac{1}{2}}^{j)} 
  \psi_{-\frac{1}{2}}^k  \psi_{-\frac{1}{2}}^v  \, |p^{(2)};0 \rangle =
  \frac{1}{\sqrt{2}} |\zeta^{Y,(2,1)}_\a \pd ~,
 \end{equation}
 is obtained from the  state corresponding to the vertex operator $V_Y$ in~\eqref{mvsl}
contracted with  the $(2,1)$ tensor 
in~\eqref{so8y} where the $\omega$'s are again those introduced in~\eqref{eq:U'2tab} and the indices are restricted to those of $SO(8)$. 
It is easy to check that
 \begin{equation}
   \label{eq:H264}
   \sqrt{2} A_{-1, \ell} | \zeta_{jk}^a \pd = -\frac{1}{\sqrt{3}} \left(|H_{2\a}\rangle - |H_{2\b}\rangle \right)~,
 \end{equation}
 where $H_2$ is the state introduced in~\eqref{uncino}, while the orthogonal state within the same $SO(8)$ representation is
 \begin{equation}
   \label{eq:H2YU}
    \frac{1}{\sqrt{6}} \left[- \left(|Y_{\a} \rangle -
      |Y_{\b} \rangle \right) + \frac{1}{2} \left(|U_{\a} \rangle
      - |U_{\b} \rangle \right)\right] = 
    - \frac{1}{\sqrt{3}} \left(|H_{1\a}\rangle - |H_{1\b}\rangle \right)~,
 \end{equation}
 where again $H_1$ is the state introduced in~\eqref{uncino}. In terms
 of the DDF oscillators Eq.~\eqref{eq:H2YU} corresponds to
 \begin{equation}
   \label{eq:H1DDF}
 \sqrt{\frac{2}{3}} \left(
   B_{-\frac{3}{2}}^{ (\ell} B_{-\frac{1}{2}}^{ j)} B_{-\frac{1}{2}}^{k} - 
   B_{-\frac{3}{2}}^{ (\ell} B_{-\frac{1}{2}}^{ k)} B_{-\frac{1}{2}}^{j} \right) |p_T ;0 \rangle ~.
 \end{equation}

 \section{Conclusions}
 \label{Conclusions}

 The leading eikonal operator represents one
 of the rare examples of resummation of the complete perturbative
 series of string theory. Its simple and somewhat intuitive structure,
 which generalizes to an extended object the eikonal phase of a point
 particle, encodes a wealth of information on the high-energy string
 dynamics. To make this information accessible in every detail, we
 completed the definition of the eikonal operator showing that it acts
 on the Hilbert space of physical string states in the light-cone
 gauge, with the spatial direction determined by the collision
 axis. Once this is established, it is possible to evaluate and give
 the correct interpretation to its matrix elements which capture the
 asymptotic behaviour at high energy of arbitrary four-point or
 arbitrary two-point amplitudes, in the case of a
 string-string or a string-brane collision respectively.

 From the covariant point of view, the high-energy dynamics described
 in the light-cone gauge by the eikonal operator is the multiple
 exchange of effective Reggeon states. Using the Reggeon operator we 
 were able to provide a simple and fully covariant derivation of the
 eikonal phase. 

We also discussed the asymptotic high-energy behaviour of the
 covariant string amplitudes with massive states, explaining how to take into 
 account the longitudinal polarizations in a simple way.  We illustrated our methods by
 calculating all the transition amplitudes from the massless NS-NS
 sector to the first two massive levels of the superstrings. In this
 way we could show in detail how the simple properties of the matrix
 elements of the eikonal operator emerge from the covariant amplitudes.

 The covariant dynamics of the massive string spectrum is an
 interesting topic on its own.  The second massive level provides
 several useful examples, including states transforming as 
 tensors of mixed symmetry. In the Regge limit the external massive string
 states can be considered approximately massless and the amplitudes
 discussed in our paper may help to understand the consistent
 interactions of massless fields of higher spin.

 Equipped with a detailed understanding of the leading eikonal
 operator, it is possible to generalise the analysis of the string high-energy scattering  in several directions. An
 interesting problem is to derive an eikonal operator which includes the
 classical corrections in $R_p/b$. It is possible that new qualitative
 features arise: for instance, the shift $b\to b+\hat{X}$, which gives
 the string eikonal operator starting from the leading eikonal phase, might not
 be enough to capture the full dynamics already at the first subleading
 order.

 Another interesting extension of our results is to study the
 high-energy string-brane scattering in the stringy regime $b\ll
 l_s$. In this case, the effects of the $t$-dependent phase in
 Eq.~\eqref{lep} are not exponentially suppressed and, in order to
 restore unitarity, one should enlarge the Hilbert space of the eikonal
 operator to include also the open string oscillators. 

Finally it
 should be possible to extend our analysis to more complicated D-brane
 bound states that represent the microstates of macroscopic black
 holes. In this case, it would be interesting to generalise  the
 approach of~\cite{Veneziano:2012yj} and see whether and under what
 conditions the tidal forces present in the string high-energy
 scattering can distinguish different microstates.

 \vspace{7mm}
 \noindent {\large \textbf{Acknowledgements} }

 \vspace{5mm} 

We would like to thank Massimo Bianchi for discussions.  
This research is partially supported by STFC (Grant ST/J000469/1, {\it String theory, gauge 
 theory \& duality}) and by INFN. G.D. and P.D. gratefully acknowledge the hospitality
of the {\it Coll\'ege de France}, where part of this work was carried out. R.R. wishes to thank the {\it Institut Lagrange de Paris} for hospitality 
and support during the completion of this work. G.D. would also like to thank Chiara Luciana D'Appollonio, whose company and 
curiosity were invaluable to him whilst working on this research project.

 \begin{appendix}

 \section{Conventions}
 \label{app:conv}

 In this appendix we collect our conventions for the description of type II string
 theories in flat space in the RNS formalism. The standard bosonic
 coordinates describing the embedding of the string in spacetime are
 indicated by ${X}^\mu(z,\bar{z})$, with $z=\ex{\tau + \ii \sigma}$
 where $\tau$ and $\sigma$ are the usual variables parametrising the
 (Euclidean) worldsheet. We use the greek letters $\alpha, \beta, \mu,
 \nu,\dots$ for the $10$-dimensional indices and a mostly plus metric
 $\eta=(-\,+\,\dots\,+)$. When describing a massive string state, we
 use the capital latin letters $I,J,\dots$ for the $9$-dimensional
 indices that live in the space orthogonal to the momentum $p^\mu$ and
 the small latin letters $i,j,\dots$ for the $8$-dimensional indices
 that live in the subspace orthogonal to both $p^{\mu}$ and the
 longitudinal polarization $v$, see for instance
 Eq.~(\ref{eq:long}). The 2D equations of motions imply that the
 ${X}^\mu$ are a combination of a holomorphic and an
 anti-holomorphic part and similarly the worldsheet spinors
 ${\psi}^\mu$ have a holomorphic and an anti-holomorphic component
 \begin{equation}
   \label{eq:hol-split}
   {X}^\mu(z,\bar{z}) ={X^\mu(z)+\bar{X}^\mu(\bar{z})} \ , \hspace{1cm}
   \psi^\mu(z,\bar{z}) = \left(
     \begin{array}{c}
       \bar\psi^\mu(\bar{z}) \\ \psi^\mu(z)
     \end{array}\right) \ . 
 \end{equation}
 The OPE's of these fields are
 \begin{equation}
   \label{eq:OPEs}
    X^\mu(z) X^\nu(w) \sim - \frac{\alpha'}{2} \eta^{\mu\nu} \log(z-w) \ , \hspace{1cm}
   \psi^\mu(z) \psi^\nu(w) \sim \frac{\eta^{\mu\nu}}{z-w} \ , 
 \end{equation}
 and similarly for the anti-holomorphic fields, while the corresponding
 mode expansions and commutation relations (in the NS sector) are
 \begin{gather}
   \label{eq:modeexp}
   X^\mu(z) = q^\mu -\ii \sqrt{\frac{\alpha'}{2}} \alpha_0^\mu \log z + \ii
   \sqrt{\frac{\alpha'}{2}}\sum_{n\neq 0}\frac{\alpha_n^\mu}{n} z^{-n} \ , \hspace{1cm}
   \psi^\mu(z) = \sum_{r\in {\mathbb Z}+\frac{1}{2}}\! \psi_{r}^\mu
   z^{-r-\frac{1}{2}} \ , \\ \nonumber
   [q^\mu,\alpha_0^\nu]= \ii \eta^{\mu\nu} \sqrt{\frac{\alpha'}{2}} \ , \hspace{1cm}
   [\alpha^\mu_n,\alpha^\nu_m]=\eta^{\mu\nu} n\, \delta_{m+n} \ , \hspace{1cm}
   \{\psi^\mu_r,\psi^\nu_s\} = \eta^{\mu\nu} \delta_{r+s} \ , 
 \end{gather}
 where tilded (left) and untilded (right) operators are completely
 independent and thus (anti)commute. The left part of the Fock space is
 built on the vacuum state $|p;0\rangle$ defined by the following relations
 \begin{equation}
   \label{eq:vacuum}
   \alpha_0^\mu |p;0\rangle =\sqrt{\frac{\alpha'}{2}} p^\mu |p;0\rangle \ , \hspace{1cm}
   \alpha_n^\mu |p;0\rangle = \psi_r^\mu |p;0\rangle = 0 \ , \hspace{1cm} \mbox{if} ~~~
   n\geq 1  , \hspace{0.2cm} r\geq \frac{1}{2} \ , 
 \end{equation}
 and  the right part is built on a vacuum state
 $|p;\bar{0}\rangle$ defined in a similar way. The physical NS-NS spectrum is obtained
 by taking the tensor product of a left and a right state annihilated
 separately by $L_0 - 1/2$, $G_r$ and $\bar{L}_0 - 1/2$,
 $\bar{G}_r$, with $r\geq 1/2$, where
 \begin{gather}
   \label{eq:superVir}
   G_r =\sum_{n\in {\mathbb Z}} \alpha_{-n} \psi_{r+n} \ , \hspace{1cm}
   L_m =   L_m^{(\alpha)} +  L_m^{(\psi)} \ , \\ \nonumber
   L_m^{(\alpha)} = \frac{1}{2} \sum_{n\in {\mathbb Z}} :\alpha_{-n}
   \alpha_{n+m}: \ , \hspace{1cm}
   L_m^{(\psi)} = \frac{1}{2} \sum_{r\in {\mathbb Z}+\frac{1}{2}} \!
   \left(r + \frac{m}{2}\right) :\psi_{-r} \psi_{m+r}: \ . 
 \end{gather}
 Finally physical states $|ph\rangle$ should obey the level-matching
 condition $L_0 |ph\rangle = \bar{L}_0 |ph\rangle$ and the GSO
 projection selects the states with an odd number of $\psi$ oscillators
 both in the left and in the right part.

 For completness we list the standard $N=1$ super-Virasoro algebra
 satisfied by the generators introduced above 
 \begin{eqnarray}
 \{G_r, G_s\} &=& 2 L_{r+s} + \frac{d}{2}\left(r^2 -
   \frac{1}{4}\right)\, \delta_{r+s,0}~,  \nonumber \\
 \left[L_n, G_r \right] &=& \left( \frac{1}{2}m -r \right)G_{r+n}~,
 \label{GrGs} \\ 
 \left[L_n, L_m \right]  &=& (n-m) L_{n+m} + \frac{d}{8} m (m^2-1)
 \delta_{n+m,0}~,  \nonumber  
 \end{eqnarray}
 where $d=10$ is the space-time dimension. 

 We conclude this appendix, by giving the explicit expressions of the states used in Section~\ref{lc2c}. With the definitions given above, it is
 straightforward to check that the states below are physical (here $p^{(2)} = p_T - \frac{5}{2} k$)
 \begin{eqnarray}
 |Y_{(I[J)HK]} \rangle &=& \frac{1}{2} \left[  
 \left(\alpha_{-1}^{I} \psi_{-\frac{1}{2}}^J +\alpha_{-1}^{J} \psi_{-\frac{1}{2}}^I \right) 
  \psi_{-\frac{1}{2}}^H  \psi_{-\frac{1}{2}}^K  \right. \nonumber \\
 &-&  \frac{\alpha_{-1, L} \psi_{-\frac{1}{2},L} }{d-3} 
  \left( 2 \hat{\eta}^{IJ} \psi_{-\frac{1}{2}}^H - \hat{\eta}^{IH} \psi_{-\frac{1}{2}}^J  - \hat{\eta}^{JH} \psi_{-\frac{1}{2}}^I \right)    \psi_{-\frac{1}{2}}^K  \nonumber \\
  &-& \left.  \frac{ \alpha_{1,L} \psi_{-\frac{1}{2},L}}{d-3} \left(\hat{\eta}^{IK} \psi_{-\frac{1}{2}}^{J} + \hat{\eta}^{JK} \psi_{-\frac{1}{2}}^{I}  \right) \psi_{-\frac{1}{2}}^{H}   \right]|0, p^{(2)} \rangle \ , 
 \label{Yc}
 \end{eqnarray}
 which reduces to~\eqref{eq:symMN} when all indices are different and $K=v$,
 \begin{eqnarray}
 |U_{(I[J)K]} \rangle &=& \left[ \frac{3}{2} \left( \psi_{-\frac{3}{2}}^{I}   \psi_{-\frac{1}{2}}^{J} + 
  \psi_{-\frac{3}{2}}^{J}  \psi_{-\frac{1}{2}}^{I} \right) \psi_{-\frac{1}{2}}^{K} - 
   \alpha_{-1}^{I} \alpha_{-1}^{J} \psi_{- \frac{1}{2}}^{K} + \frac{1}{2} \alpha_{-1}^{K} \left(
   \alpha_{-1}^{J} \psi_{- \frac{1}{2}}^{I} +  \alpha_{-1}^{I} \psi_{- \frac{1}{2}}^{J} \right) \right.    \nonumber \\
  &- &  \frac{3}{2}
  \frac{ \psi_{-\frac{3}{2}, H}  \psi_{-\frac{1}{2}, H}}{d-2}\left( 2 \hat{\eta}^{IJ}   \psi_{-\frac{1}{2}}^{K}
 - \hat{\eta}^{IK}  \psi_{-\frac{1}{2}}^{J} -\hat{\eta}^{JK}  \psi_{-\frac{1}{2}}^{I} \right)  \nonumber \\
 &+&  \frac{\alpha_{-1H}\, \alpha_{-1 H}}{2(d-2)}\left(2 \hat{\eta}^{IJ}  \psi_{- \frac{1}{2}}^{K} - \hat{\eta}^{KJ} \psi_{- \frac{1}{2}}^{I} - \hat{\eta}^{KI} \psi_{- \frac{1}{2}}^{J} \right) 
 \nonumber \\ 
 &-& \left. \frac{ \alpha_{-1 H}\, \psi_{- \frac 12, H} }{2(d-2)} \left(2 \hat{\eta}^{IJ}  \alpha_{-1}^{K} - \hat{\eta}^{KJ} \alpha_{-1}^{I} - \hat{\eta}^{KI} \alpha_{-1}^{J} \right) \right]  |0,  p^{(2)} \rangle \ , 
 \label{Uc}
 \end{eqnarray}
 which reduces to~\eqref{eq:symMN2} when all indices are different, and
 \begin{eqnarray}
 | Z_{(MNP)} \rangle  & = & {\hat{\alpha}}_{-1}^{M}  {\hat{\alpha}}_{-1}^{N} 
 {\hat{\psi}}_{-\frac{1}{2}}^{P} +
  {\hat{\alpha}}_{-1}^{M}  {\hat{\alpha}}_{-1}^{P} {\hat{\psi}}_{-\frac{1}{2}}^{N} +  {\hat{\alpha}}_{-1}^{N}  {\hat{\alpha}}_{-1}^{P} {\hat{\psi}}_{-\frac{1}{2}}^{M}  - \frac{1}{d+1} \nonumber \\
 & \times & \left[ \hat{\eta}^{MN}  \hat{\eta}^{QR} \left(\alpha_{-1, R } \alpha_{-1Q} \psi_{-\frac{1}{2}} ^{P}+ 2 \alpha_{-1,R} \psi_{-\frac{1}{2}, Q} \alpha_{-1}^{P}  \right)    \right. \nonumber \\
 &  + &  \hat{\eta}^{MP}  \hat{\eta}^{QR} \left(\alpha_{-1, R } \alpha_{-1Q}
  \psi_{-\frac{1}{2}}^{N}+ 2 \alpha_{-1,R} \psi_{-\frac{1}{2}, Q} \alpha_{-1}^{N}  \right) 
  \nonumber \\
  & + & \left. \hat{\eta}^{NP}  \hat{\eta}^{QR} \left(\alpha_{-1, R } \alpha_{-1Q}
  \psi_{-\frac{1}{2}}^{M}+ 2 \alpha_{-1,R} \psi_{-\frac{1}{2}, Q} \alpha_{-1}^{M}  \right)   \right]  |0 , p^{(2)} \rangle  \ , 
 \label{Zc}
 \end{eqnarray}
 where ${\hat{\alpha}}_{-1}^{M} = \alpha_{-1}^{M} - \frac{p^M (p \alpha_{-1})}{p^2}$ and analogously  for $\psi_{-\frac{1}{2}}^{M}$. 
 Also the following states are physical (here $p^{(1)} = p_T - \frac{3}{2} k$)
 \begin{eqnarray}
   \label{eq:s1s2}
   |s_1 \rangle &=& \left( G_{-\frac{1}{2}} p^{(1)} \alpha_{-1} +
     \frac{1}{2} G_{-\frac{3}{2}} \right) | p^{(1)}; 0\rangle   \\ \nonumber
     &=&  \left( \frac{3}{2} p^{(1)}  \psi_{-\frac{3}{2}} + p^{(1)}  \psi_{-\frac{1}{2}} \
 p^{(1)}  \alpha_{-1} + \frac{1}{2} \alpha_{-1}  \psi_{-\frac{1}{2}} \right) |0, p^{(1)} \rangle\ , 
 \\ 
 |s_2\rangle &=& -G_{-\frac{1}{2}} \left( (v \psi_{-\frac{1}{2}}) \, (p^{(1)}
   \psi_{-\frac{1}{2}}) - 2 v \alpha_{-1} \right)|p^{(1)};0\rangle \\ \nonumber
   &=&  \left(2 v  \psi_{-\frac{3}{2}} + (v \psi_{-\frac{1}{2}}) (p^{(1)} \alpha_{-1}) +
 (v \alpha_{-1}) (p^{(1)} \psi_{-\frac{1}{2}}) \right)|0, p^{(1)} \rangle~,
 \\  |S_{jl;k} \rangle & = &  
 G_{-\frac{1}{2}} \alpha_{-1}^{(\ell}
   \psi_{-\frac 12}^{j)} \psi_{-\frac 12}^k |0,p^{(2)} \rangle
 \label{Sjellk}  
 \\ \nonumber  &=& 
 \left[ (p^{(2)} \psi_{-\frac{1}{2}} ) \alpha_{-1}^{(\ell}  \psi_{-\frac{1}{2}}^{j)} \psi_{-\frac{1}{2}}^{k} +
 \psi_{-\frac{3}{2}}^{(\ell} \psi_{-\frac{1}{2}}^{j)} \psi_{-\frac{1}{2}}^{k}  + \alpha_{-1}^{\ell} \alpha_{-1}^{j}
 \psi_{-\frac{1}{2}}^{k}  - \alpha_{-1}^{k} \alpha_{-1}^{(\ell} \psi_{-\frac{1}{2}}^{j)} \right]
 | 0, p^{(2)} \rangle \ . 
 \end{eqnarray}
 Since they are explicitly written as super-descendant, these states are spurious.

 \section{Explicit formulae for the kinematics}
 \label{app:kine}

 In this appendix we collect some formulae relevant for the kinematics
 discussed in Sect.~\ref{kinsec} in a particular reference frame. In order
 to simplify the comparison with the light-cone computation discussed
 in Sect.~\ref{gssec}, it is convenient to choose the spatial momentum of
 the outgoing massive particle to be aligned along the same direction
 of $\vec{e}^\pm$; then we have
 \begin{equation}
  p_{2}^{\mu} = \left( -E , 0_p ; 0_{8-p} , - \sqrt{E^2 -M^2} \right) \ , 
 \label{p2v}
 \end{equation}
 where the first $p+1$ directions are parallel to the (Neumann
 directions of the) D$p$-branes and the entries after the semicolon are
 along the Dirichlet directions. Then the light-cone directions defined
 in~\eqref{eq:epm} read
 \begin{equation}
   \label{eq:lico1}
  ( e^+ )^{\mu} = \frac{1}{\sqrt{2}} (-1,0,\ldots,0,1) \ , \hspace{1cm}
   (e^-)^{\mu} = \frac{1}{\sqrt{2}} (1,0,\ldots,0,1) \ .
 \end{equation}
 The most direct way to describe the physical polarization of massive
 particles is to introduce 9 vectors perpendicular to their
 momentum. For instance, in the case of the outgoing state~\eqref{p2v}
 we have the unit vectors $\hat{w}^i$
 \begin{equation}
   \label{eq:versorw}
   \hat{w}_1 = \left(0,1, 0_{p-1}; 0_{8-p}, 0 \right)\ , 
   \ldots \ , \hat{w}_8 =\left(0,0_p; 0_{7-p},  1, 0 \right)  \ , 
 \end{equation}
 and, as the ninth one, $v^{\mu}$ corresponding to the longitudinal
 polarization 
 \begin{equation}
   \label{eq:long}
  v^{\mu}_2 =  \left( \frac{\sqrt{E^2 -M^2}}{M} , 0_p ; 0_{8-p} ,
    \frac{E}{M}  \right) \ .   
 \end{equation}
 In terms of the DDF construction reviewed in Appendix~\ref{app:DDF}, these massive state can be generated by choosing
 \begin{eqnarray}
 p_T = \left(\sinh \alpha, 0_p ; 0_{8-p},\cosh \alpha\right) \ , \hspace{1cm} k = {\rm e}^{- \alpha} \sqrt{2}  e^+ 
 = {\rm e}^{- \alpha} (-1,0_p;0_{8-p},1) \ ,
 \label{pTk}
 \end{eqnarray}
 where $k^2=0$ and $k p_T=1$ for any $\alpha$ (this parameter can be
 used to obtain the desired energy in~\eqref{p2v}).
 The possible momenta of the ingoing particle take the following form
 \begin{gather}
   p_1^\mu = \left( E,0_p;{\bar{p}}_1, \sqrt{E^2-M^2} + q_9 \right) \ , 
 \label{p1xxx}
 \\
 q^{\mu=9} =\frac{t+M^2}{2 \sqrt{E^2-M^2}}\ , \hspace{1cm}
 ({\bar{p}}_1)^{2} + (q^{\mu=9})^2 = -t \equiv (p_1+p_2)^2 \ , 
 \label{pbarq}
 \end{gather}
 where $p_1^2=0$.  Finally $\epsilon^{\mu}_k$ indicates the (left) part of the
 polarisation of the massless NS-NS state. Since we are focusing on a
 massless state we have eight independent polarisations. 
 It is convenient to choose them as follows
 \begin{equation}
 \epsilon^{\mu}_{k} = \left(\frac{{\bar{p}}_{1}^{k}}{E+\sqrt{E^2-M^2} +
     q^9},  \delta^{i}_{k} , -\frac{{\bar{p}}_{1}^{k}}{E+\sqrt{E^2-M^2} +
     q^9} \right) \ . 
 \label{emuK}
 \end{equation}
 This implies that we can neglect the $p_1^\mu$ part in $q^\mu$ and
 write $p_2^\mu$ in place of the latter
 \begin{equation}
   \label{eq:eetaq}
   \epsilon_k q = \epsilon_k p_2 = \epsilon^{\mu=0} \left(E
     +\sqrt{E^2-M^2}\right) \sim \bar{p}_1^k~,
 \end{equation}
 where in the final step we kept only the leading term in the high
 energy expansion.

 We now wish to decompose the indices $\alpha$ and $\rho$ of the
 massive states in 8D part ($i$, $j$) and the component along $v\equiv
 v_2$, see Eq.~\eqref{eq:versorw} and~\eqref{eq:long}. Then it is
 convenient to rewrite $\epsilon^\mu$ in terms of their $v$-component
 and 8D part
 \begin{equation}
   \label{eq:emuiv}
   \epsilon_k v = \frac{\epsilon^{\mu=0}}{M^3} 
   \left[(E^2-M^2)^{\frac{3}{2}} + (E^2-M^2) E - E^2 \sqrt{E^2-M^2} -
     E^3 \right]  \sim - \frac{\bar{p}_1^k}{M}~,
 \end{equation}
 where again in the last step we implemented the high energy limit. 
 The polarisation of the massive state (of momentum $p_2$) can be
 written in terms of tensor products of the vectors $\hat{w}_i$ and
 $v$.

 \section{Reminder of the DDF construction}
 \label{app:DDF}

 In this Appendix we collect a few known results about the DDF operators and states \cite{DelGiudice:1971fp}  to be used in the main body of the paper.
 We do so by using their generalization to the
 NS sector of the superstring presented in~\cite{Hornfeck:1987wt}~\footnote{In this Appendix, except for the expansion of the fermionic coordinate $\psi$ in terms of the oscillators,  we follow the notation of Ref.~\cite{Hornfeck:1987wt} where dimensionless variables are used.  In this notation  the  string coordinate is given in  Eq. (\ref{eq:modeexp}) for $\alpha' =2$.}. As usual we will discuss explicitly the construction for the left movers. Identical considerations apply to the right movers.

 One first introduces an auxiliary tachyon-like  momentum $p_T$ and the corresponding tachyonic state
 \begin{equation}
   \label{eq:PT}
   |p_T;0\rangle  \ , \hspace{1cm} p_T^2=1 \ . 
 \end{equation}
 The physical states in the NS sector of the superstring can be
 constructed by acting on this ground state with the DDF oscillators~\footnote{The fermionic DDF oscillators $B_{-r,j} $ were originally constructed in Refs.~\cite{Brower:1973iz,Schwarz:1974ix}.}
 \begin{gather}
 \label{ABigen}
 A_{-n,j} = -\ii \oint_{0} dz \,\, (\epsilon_j)_{\mu} \left(\partial X^{\mu} +\ii n (k \psi)  \psi^{\mu}   \right)
 {\rm e}^{-\ii n k X (z)}\;, \\
 B_{-r,j} = \ii \oint_{0} dz \,\, (\epsilon_j)_{\mu} \left( \partial X^{\mu} \, (k \psi) - \psi^{\mu} 
 (k \partial X) + \frac{1}{2} \psi^{\mu} (k \psi) \frac{(k \partial \psi)}{(k \partial X) } \right)
 \frac{ {\rm e}^{-\ii r k X (z)}  }{(\ii k \partial
   X)^{\frac{1}{2}}} \;,
 \nonumber
 \end{gather}
 where $n$ ($r$) is a positive integer (half-integer), $k$ is an
 arbitrary null vector whose scalar product with $p_T$ is one ($p_T k
 =1$), and $\epsilon_j$ is any one of the eight unit vectors
 perpendicular to $\vec{k}$ and with a trivial time component.
 The spectrum of the physical states of the NS sector of the superstring 
 is generated by the action  on the vacuum state in Eq.~\eqref{eq:PT} of the DDF operators
 in Eq.~\eqref{ABigen}, followed by the
 GSO projection on states
 with definite worldsheet fermion number. 
 The GSO projection preserves only the states containing an odd number
 of $B_{-r,j}$, so the first non trivial physical state is obtained by
 applying the operator $B_{- \frac{1}{2},j}$ to the vacuum in
 Eq.~\eqref{eq:PT}
 \begin{equation}
 \label{eq:state0lcx}
 B_{- \frac{1}{2},j} |p_T;0 \rangle = -  (\epsilon_j
 \psi_{-\frac{1}{2}}) | p_T - \frac{1}{2} k;0 \rangle 
\ , \hspace{1cm} \left( p_T - \frac{1}{2} k \right)^2 = p^2_T - p_T k =0 \ .
 \end{equation}
 These are the eight physical polarizations of the massless states corresponding
 to the covariant vertex operator in Eq.~\eqref{mvg} with momentum $p = p_T -k/2$.
 We shall be interested in the massive states obtained by acting on the above massless one by applying any number of $A_{-n,i}$ DDF operators. Since each one of them carries a momentum $-n k$ the outcome will be a state of total momentum
 \be
 p =  p_T - (n+\frac12) k \ , \hspace{1cm} n = \sum n_k  \ , \hspace{1cm}  p^2  = - 2n \ . 
 \ee
 The generic (right moving component of a) state can thus be seen as a collection of  photons moving in the direction of $k$ together with a single tachyon  moving in a different direction.
 In a convenient Lorentz frame we can take the latter to move in the opposite direction to $k$, in other words we can restrict the kinematics to the 2-dimensional space given by $k$. In this frame, defined modulo a Lorentz boost $\alpha$ along this axis, we can write the different vectors as follows
 \bea
 k &=& e^{-\alpha} \left(-1, 0_p; 0_{8-p}, 1\right) \ ,  \nonumber \\ 
 p_T &=& \left(\sinh \alpha, 0_p; 0_{8-p},  \cosh \alpha \right) \ , \nonumber \\ 
 p &=& \frac12  \left(e^{\alpha} + 2ne^{-\alpha}, 0_p; 0_{8-p}, e^{\alpha} - 2n e^{-\alpha} \right) \ . 
 \label{kpTp}
 \eea
 The latter expression 
 coincides with the 4-momentum $p_2$ given in (\ref{p2v}) provided we identify
 \beq
 \sqrt{\frac{\alpha'}{2}} E = - \frac{1}{2} ( e^{\alpha} + 2ne^{-\alpha}) \ , \hspace{1cm}
 \frac{\alpha'}{2}  M^2  = 2 n \ , 
 \label{E2} 
 \eeq
 the overall minus sign being there because this is an outgoing string~\footnote{We have reinserted $\alpha'$ because $E$ and $M$ of Appendix \ref{app:kine} 
 have dimension of an energy.}.

 We shall be interested in  a high-energy process where, in the rest frame of the branes, the components of $p$ are large for the two external, generic, closed string states. This corresponds to the limit  of large boost parameter, $\alpha \gg 1$, in which the momenta carried by the ``photons'' of the DDF operators are, instead, very small (of order $e^{-\alpha}$). 
  In this regime  $p_1$, as given in (\ref{p1xxx}), is also of the same form as in (\ref{kpTp}) and in the limit we have
 \beq
 p_1 \rightarrow - p_2 \rightarrow \frac{1}{2}  e^{\alpha} ( 1, 0_p; 0_{8-p}, 1) \ . 
 \eeq
 If we now impose  on the string coordinates
 in~\eqref{ABigen} the particular light-cone gauge corresponding to setting to zero the string oscillations of
 $k\partial X$ and $k\psi$, the contour integrals become simple (since $n k p = n$ for any $p$) 
 and  the DDF bosonic oscillators reduce to the modes of the transverse bosonic
 coordinates in that particular light-cone gauge. From Eq. (B.4) we see that, at  large $\alpha$, this is precisely the gauge choice that allows us to obtain our simple eikonal operator.

 \section{Polarizations of the massive string states}
 \label{app:Young}

 The massive string states transform in irreducible representations of the little group
 $SO(9)$ and can be described in terms of irreducible tensors $\z$ of the Lorentz group which satisfy
 a transversality condition of the form $p^{\m_i}\z_{\m_1...\m_i...\m_n} = 0$ (where $p$
 is the momentum of the state) in all their indices. 
 We restrict our discussion to the left movers and
  use the following standard notation for the symmetrization and the antisymmetrization 
  of a group of $n$ indices 
 \be T_{(i_1...i_n)} = \frac{1}{n!} \ \sum_{\s \in S_n} T_{\s(i_1)...\s(i_n)} \ ,
 \hspace{1cm} T_{[i_1...i_n]} = \frac{1}{n!} \ \sum_{\s \in S_n}  {\rm sign}(\s) T_{\s(i_1)...\s(i_n)} \ , \ee
 where the sum is over all the elements $\s$ of the symmetric group $S_n$.

 To every irreducible tensor one can associate a Young diagram
 which specifies its symmetry type. We write the vertex operators and the corresponding string states
 as the product of a polarization tensor $\z$ and a polynomial in the string fields or modes with the same symmetry properties
 as the polarization tensor. A generic rank-$n$ tensor can be projected onto its irreducible components by 
 the action of the Young symmetrizer corresponding to a Young diagram with $n$ boxes. According to our
 conventions the Young symmetrizer first symmetrizes the indices in the rows
 and then antisymmetrizes the indices in the columns. The polarization tensors are 
 therefore totally
 antisymmetric in the indices corresponding to the columns of the diagram.  Moreover if we denote by
 $\s_1..\s_k$ the indices of a given column and by $\n$ any index of the column to its right,
 the polarization tensors satisfy  the following identity
 \be \z_{...[\s_1...\s_k \n] ...} = 0 \ . \ee
 Since we are considering the irreducible representations of an orthogonal group, every tensor is also
 traceless in every couple of indices. 
 Finally the polarization tensor of a generic string state is normalized to one \be \z_{\m_1...\m_n} \z^{\m_1...\m_n} = 1 \ . \ee
 Using the previous identities the normalization coefficients of the massive vertex operators, as for instance
 those in Eq.~\eqref{mvsl}, can be easily derived. 

 As discussed in section \ref{covariant-eikonal}, to relate the covariant amplitudes and the matrix elements of the eikonal phase it is necessary
 to decompose the covariant tensors with respect to the $SO(8)$ symmetry group of the space
 transverse to the collision axis. The covariant polarization $\z^C$ of a state $C$ in a given
 representation of the little group $SO(9)$ then decomposes into several polarizations
 $\z^{C, (n_1,...,n_r)}$ transforming  as irreducible tensors of $SO(8)$ of type $(n_1,...,n_r)$.
 Taking into account the transversality condition satisfied by the polarization tensors $\z^C$, the 
 $SO(8)$ components $\z^{C, (n_1,...,n_r)}$ can be written explicitly in terms of the longitudinal vector
 $v$, the Kronecker delta $\d_{\perp}$ in the transverse space and an $SO(8)$
 polarisation tensor of the required symmetry type $\w^{(n_1,...,n_r)}$ and satisfying $\w \cdot \w = 1$.

 In section \ref{covariant-eikonal} we gave the explicit form of the $\z^{C, (n_1,...,n_r)}$ 
only  for the states of the first massive level.
 Here we collect the formulae for the reduction of the polarization tensors of the states  of the second massive level.
 The state $Z$ in 
 the $(3)$ of $SO(9)$ has the following decomposition with respect to $SO(8)$
 \be \yng(3)\ \  \mapsto \  \ \yng(3) \ \ + \ \ \yng(2) \ \ + \ \ \yng(1) \ \ + \ \ \bullet \ \ , \ee
 or in terms of the polarization tensor
 \ba  \z^{Z, (3)}_{\r\a\g} &=& \w^{(3)}_{\r\a\g} \ , \nb \\
 \z^{Z, (2)}_{\r\a\g} &=&  \frac{1}{\sqrt{3}}\si ( \w^{(2)}_{\r\a} v_{\g}+\w^{(2)}_{\a\g} v_{\r}
 +\w^{(2)}_{\g\r} v_{\a} \de ) \ , \nb \\
 \z^{Z, (1)}_{\r\a\g} &=&    \frac{1}{\sqrt{330}}\si ( \si (\d_{\perp\r\a} - 10 v_\r v_\a \de )  \w_{\g}
 +\si (\d_{\perp\a\g} - 10 v_\a v_\g \de )  \w_{\r}
 +\si (\d_{\perp\g\r} - 10 v_\g v_\r \de )  \w_{\a} \de )  \ , \nb \\
 \z^{Z, (0)}_{\r\a\g} &=&\frac{1}{\sqrt{88}}\si ( \d_{\perp\r\a}  v_{\g}+\d_{\perp\a\g} v_{\r}
 +\d_{\perp\g\r} v_{\a} - 8 v_\r v_\a v_\g\de )   \label{zpol}  \ . \ea
 The state $Y$ in the $(2,1,1)$  of $SO(9)$ has the following decomposition with respect to $SO(8)$
 \be \yng(2,1,1)\ \  \mapsto \  \ \yng(2,1,1) \ \ + \ \ \yng(1,1,1) \ \ + \ \ \yng(2,1) \ \ + \ \ \yng(1,1)   \label{so9so8y}  \ . \ee  
 We decompose the polarization tensor accordingly
 \ba  \z^{Y, (2,1,1)}_{\r\a;\g;\w} &=& \w^{(2,1,1)}_{\r\a;\g;\w} \ ,  \nb \\
  \z^{Y, (1,1,1)}_{\r\a;\g;\w} &=& \frac{\sqrt{3}}{2} \si ( \w^{(1,1,1)}_{\r\g\w} v_\a 
 + \w^{(1,1,1)}_{\a[\g\w} v_{\r]}  \de ) \ , \nb \\
  \z^{Y, (2,1)}_{\r\a;\g;\w} &=&
 \frac{1}{\sqrt{3}} \si ( \w^{(2,1)}_{\r\a\g} v_{\w} + \w^{(2,1)}_{\g\a\w} v_{\r}+ \w^{(2,1)}_{\w\a\r} v_{\g} \de ) \ , \nb \\
  \z^{Y, (1,1)}_{\r\a;\g;\w} &=&\sqrt{\frac{2}{7}} \, 3 \si (
  \w^{(1,1)}_{[\r\g} v_{\w]}v_\a -  \frac{1}{6} \w^{(1,1)}_{[\r\g} \d_{\perp \w]\a} \de )  \label{so8y}  \ . \ea
 The state $U$ in the $(2,1)$  of $SO(9)$ has the following decomposition with respect to $SO(8)$
 \be \yng(2,1)\ \  \mapsto \  \ \yng(2,1) \ \ + \ \ \yng(2) \ \ + \ \ \yng(1,1) \ \  + \ \ \yng(1)  \  \ , \label{so9so8ubis}  \ee
 or in terms of the polarization tensor  
 \ba  \z^{U, (2,1)}_{\r\a;\g} &=& \w^{(2,1)}_{\r\a;\g} \ , \nb \\
  \z^{U, (2)}_{\r\a;\g} &=& \frac{1}{\sqrt{2}} \si ( \w^{(2)}_{\r\a} v_\g - \w^{(2)}_{\g\a} v_\r \de ) \ , \nb \\
  \z^{U, (1,1)}_{\r\a;\g} &=& \frac{1}{\sqrt{6}} \si (2  \w^{(1,1)}_{\r\g} v_\a +
 \w^{(1,1)}_{\a\g} v_\r -  \w^{(1,1)}_{\a\r} v_\g \de )   \ , \nb \\
  \z^{U, (1)}_{\r\a;\g} &=& \frac{\sqrt{7}}{4} \si ( \w_\g v_\r v_\a - \w_\r v_\g v_\a 
 - \frac{1}{7} \w_\g \d_{\perp\r\a} + \frac{1}{7} \w_\r \d_{\perp\g\a} \de ) \ .  \label{up} \ea
 The state $V$ in the $(1,1)$ of $SO(9)$ gives a two-form and a vector of $SO(8)$.
 Decomposing the polarization tensor we find
 \ba  \z^{V, (1,1)}_{\g\w} &=& \w^{(1,1)}_{\g\w} \ , \hspace{1cm}
  \z^{V, (1)}_{\g\w} =  \frac{1}{\sqrt{2}}\si ( \w_\g v_\w - \w_\w v_\g \de )  \ . \ea
 Finally the state $W$ in the vector representation of $SO(9)$ gives a vector and a scalar of $SO(8)$.
 The corresponding polarization tensors are
 \ba  \z^{W, (1)}_{\g} &=& \w_{\g} \ , \hspace{1cm}
  \z^{W, (0)}_{\g} =  v_\g \ . \ea

 \end{appendix}

\newpage


\providecommand{\href}[2]{#2}\begingroup\raggedright\endgroup

\end{document}